\documentclass{article}
\usepackage{amsfonts}
\usepackage{amsmath}
\usepackage{amssymb}

\newtheorem{theorem}{Theorem}

\newtheorem{remark}[theorem]{Remark}

\input{tcilatex}
\begin{document}

\title{The Dynamical Behavior of \\
Detected vs. Undetected Targets}
\author{Ronald Mahler, Eagan MN, USA, EML: mahlerronald@comcast.net}
\maketitle

\begin{abstract}
This paper is a sequel of the 2019 paper \cite{MahSensors1019Exact}. \ It
demonstrates the following: \ a)\ the Poisson multi-Bernoulli mixure (PMBM)
approach to detected vs. undetected (U/D) targets cannot be rigorously
formulated using either the two-step or single-step multitarget recursive
Bayes filter (MRBF); b)\ it can, however, be partially salvaged using a
novel single-step MRBF; c) probability hypothesis density (PHD) filters can
be derived for both the original \textquotedblleft S-U/D\textquotedblright\
approach in \cite{MahSensors1019Exact} and the novel \textquotedblleft
D-U/D\textquotedblright\ approach; d) important U/D formulas in \cite%
{MahSensors1019Exact} can be verified using purely algebraic methods rather
than the intricate statistical analysis employed in that paper; and e) the
claim, that PMBM\ filters can propagate detected and undetected targets
separately in parallel, is doubtful.
\end{abstract}

\section{Introduction \label{A-Intro}}

This paper is a sequel of the 2019 paper \cite{MahSensors1019Exact}. \ The
primary content of that paper had two main parts:

\begin{enumerate}
\item In Section 4, a critique of the first three versions of the Poisson
multi-Bernoulli mixture (PMBM) filter.

\item In Section 5, a statistical theory of a peripheral multitarget
tracking concept that is central to PMBM filters: \textquotedblleft
undetected\textquotedblright\ vs. \textquotedblleft
detected\textquotedblright\ targets (hereafter referred to as
\textquotedblleft U-targets\textquotedblright\ vs. \textquotedblleft
D-targets\textquotedblright ). \ 
\end{enumerate}

A summary of Section 4, along with a short critique of a fourth
(\textquotedblleft trajectory\textquotedblright )\ PMBM filter incarnation,
subsequently appeared in \cite{Mah-TRFS-arXiv2024}. \ \ 

In this paper it will be demonstrated that: \ a) the PMBM approach to
U/D-targets is seriously theoretically erroneous, but b) can be partially
salvaged by adopting a novel Bayesian theoretical reformulation of it.

This Introduction is organized as follows: \ PMBM\ filters (Section \ref%
{A-Intro-AA-PMBM}), the static model for U/D targets (Section \ref%
{A-Intro-AA-SUD}),the dynamic model for U/D targets (Section \ref%
{A-Intro-AA-DUD}), the five purposes of this paper (Section \ref%
{A-Intro-AA-Purpose}), and the organization of the paper (Section \ref%
{A-Intro-AA-Org}).

\subsection{PMBM Filters \label{A-Intro-AA-PMBM}}

Like most multitarget algorithms, PMBM\ filters presume detection-type
sensors and the \textquotedblleft small target\textquotedblright\ model. \
That is, a) at any given time \ $t_{k}$ \ a detection process is applied to
a sensor signature, resulting in a finite set \ $Z_{k}\subseteq \mathbb{Z}$
\ of point measurements (\textquotedblleft detections\textquotedblright )
that are generated by targets or by clutter/false alarms; and b) single
point targets can generate at most a single point detection, and target
detections can be generated by only a single point target \cite[p. 433]%
{Mah-Artech}.

PMBM filters are also based on the random finite set (RFS) two-step
multitarget recursive Bayes filter (MRBF) \cite[p. 484]{Mah-Artech}, which
recursively alternates between time-update and measurement-update steps%
\begin{eqnarray}
f_{k|k-1}(X|Z_{1:k-1}) &=&\int
f_{k|k-1}(X|X_{k-1})\,f_{k-1|k-1}(X_{k-1}|Z_{1:k-1})\delta X_{k-1}
\label{eq-MF-1} \\
f_{k|k}(X|Z_{1:k}) &=&\frac{f_{k}(Z_{k}|X)\,f_{k|k-1}(X|Z_{1:k-1})}{\int
f_{k}(Z_{k}|X)\,f_{k|k-1}(X|Z_{1:k-1})\delta X}  \label{eq-MF-2}
\end{eqnarray}%
with initial distribution \ $f_{0|0}(X)$; \ and where \ $%
f_{k|k-1}(X|X_{k-1}) $ \ is the multitarget Markov density and \ $%
f_{k}(Z|X)=L_{Z}(X)$ \ is the multitarget measurement density; and where \ $%
\int \cdot \delta X_{k-1}$ \ and \ $\int \cdot \delta X$ \ are RFS set
integrals (Eq. (\ref{eq-SetInt})).

\subsection{The Static Model for U/D Targets \label{A-Intro-AA-SUD}}

The purpose of Section 5 of \cite{MahSensors1019Exact} was to devise a
theoretically rigorous basis for U/D targets based on their \textit{%
instantaneous statistical} behavior---i.e., their behavior at the instant
the detection process is applied. \ 

\begin{itemize}
\item This will be called the \textit{static U/D} (S-U/D) \textit{model} of
U/D targets.
\end{itemize}

Section 5 of \cite{MahSensors1019Exact} derived explicit formulas for the
multitarget probability distributions \ $f_{k|k}^{d}(X|Z_{1:k})$ \ for
D-targets \cite[Eq. (75)]{MahSensors1019Exact}\ and $\
f_{k|k}^{u}(X|Z_{1:k}) $ \ for U-targets \cite[Eq. (90)]{MahSensors1019Exact}%
\ and their \ corresponding probability generating functionals (p.g.fl.'s) \ 
$G_{k|k}^{d}(X|Z_{1:k})$, \cite[Eq. (5)]{MahSensors1019Exact}\ and \ $%
G_{k|k}^{u}(X|Z_{1:k})$ \ \cite[Eq. (6)]{MahSensors1019Exact},
instantaneously at observation times \ $t_{k}$. \ It did not consider the
following question: \ Can individual targets be identified as U-targets vs.
D-targets? \ 

\subsection{The Dynamic Model for U/D Targets \label{A-Intro-AA-DUD}}

PMBM filters employ a \textit{historical-behavioral}, rather than an
instantaneous-statistical, interpretation of U/D-targets, one which presumes
that this question is answerable. \ Specifically, U/D targets are defined as
(\cite[p. 10]{MahSensors1019Exact}, lines 10-13):

\begin{enumerate}
\item At time \ $t_{k}$ \ a target is a U-target if it has not been detected
at any of the observation times \ $t_{1},...,t_{k}$.

\item If a U-target is detected at time \ $t_{k+1}$ \ then it is a D-target
at this and all subsequent times---i.e., D-targets cannot become U-targets.
\end{enumerate}

Expressed more intuitively: \ A target is presumed conjectural until proven
\textquotedblleft real\textquotedblright\ by the act of generating some
measurement \ $z\in \mathbb{Z}$ \ in the sensor measurement space \ $\mathbb{%
Z}$.

However, in \ Item 3 of \cite[p. 13]{MahSensors1019Exact} and Section 9 of 
\cite[p. 3]{Mah-TRFS-arXiv2024} it was noted that:

\begin{enumerate}
\item the PMBM filter's U/D approach appears to violate the two-step
structure of the MRBF; and

\item this incompatibility is due to the seriously erroneous physical
assumption that all measurements arise from newly-detected---and therefore
newly-appearing---targets. \ 
\end{enumerate}

That is, they occur only when U-targets transition to D-targets. \ 

\begin{itemize}
\item This will be referred to as the \textit{U-birth model}\ for
measurement generation.
\end{itemize}

\subsection{The Five Purposes of this Paper \label{A-Intro-AA-Purpose}}

\subsubsection{First Purpose}

In Sections \ref{A-TwoStepUD} and \ref{A-1StepU/D} it will be shown that,
respectively, the PMBM U/D approach:

\begin{enumerate}
\item cannot be rigorously formulated using either the two-step MRBF or the
conventional single-step \textquotedblleft joint\textquotedblright\ version
of that filter (i.e., in which the two update steps are merged into a single
step, see Eqs. (\ref{eq-BF-4},\ref{eq-Multitarg})); but \ 

\item can be rigorously reformulated using an \textit{unconventional}
single-step MRBF---i.e., one that employs a novel Bayes recursive transition
function (see Eqs. (\ref{eq-BFUD-1}-\ref{eq-BFUD-5},\ref{eq-BFUD-6}-\ref%
{eq-BFUD-8},\ref{eq-NUD2})).
\end{enumerate}

\begin{itemize}
\item This will be called the \textit{dynamic U/D} (D-U/D) \textit{model} of
U/D\ targets. \ 
\end{itemize}

The D-U/D model consists of two parts: \ a) the definition of the U/D
single-target state space (Eq. (\ref{eq-StateSpace})); and b) the definition
of the novel Bayesian single-target recursive joint transition function
(Eqs. (\ref{eq-BFUD-1}-\ref{eq-BFUD-5})). \ The multitarget case will be
addressed in Section \ref{A-Multi}.

The D-U/D reformulation does not obviate the erroneousness of the U-birth
model. \ For, as pointed out in \cite[pp. 53-55]{Mah-Artech}, if measurement
models are seriously inaccurate then, as signal-to-noise (SNR) decreases, a
multitarget filter must increasingly \textquotedblleft
waste\textquotedblright\ data trying (and eventually failing) to overcome
the mismatch between model and reality. \ Most obviously, measurements can
also arise from clutter/false alarms. \ Thus when clutter is dense (SNR is
small) the U-birth model will---at every observation time---result in the
mandatory creation of a large number of non-existent targets.

\subsubsection{Second Purpose}

Sections \ref{A-1StepU/D}-\ref{A-Multi} and Section \ref{A-Comp} will,
respectively,\ a)\ provide a theoretically rigorous basis for the novel
D-U/D model and b) compare it to the S-U/D model. \ The main results are as
follows, assuming that the two approaches have been \textquotedblleft
aligned,\textquotedblright\ in the sense of Section \ref{A-Comp AA-Align},\
to permit an apples-to-apples comparison:

\begin{enumerate}
\item \textit{D-Targets}: \ The respective p.g.fl.'s for the S-U/D vs. D-U/D
models are (see Eqs. (\ref{eq-Sensors-D},\ref{eq-Main-D})): 
\begin{eqnarray}
G_{k|k}^{d}[h] &\propto &\int f_{k}^{\ast }(Z_{k}|X)\,(hp_{D})^{X}\,\frac{%
\delta G_{k|k-1}}{\delta X}[p_{D}^{c}]\delta X  \label{eq-MR-D-1} \\
\breve{G}_{k|k}^{d}[\breve{h}] &\propto &\int \breve{f}_{k}^{\ast }(Z_{k}|%
\breve{X})\,(\breve{h}|_{1}p_{D})^{\breve{X}}\,\frac{\delta \breve{G}%
_{k-1|k-1}}{\delta \breve{X}}[\breve{h}^{\mathbb{D}}p_{D}^{c}]\delta \breve{X%
}.  \label{eq-MR-D-2}
\end{eqnarray}

\item \textit{U-Targets}: \ The respective p.g.fl.'s for the S-U/D vs. D-U/D
models are (see Eqs. (\ref{eq-Sensors-U},\ref{eq-Main-U})): \ 
\begin{eqnarray}
G_{k|k}^{u}[h] &\propto &\int f_{k}^{\ast }(Z_{k}|X)\,p_{D}^{X}\,\frac{%
\delta G_{k|k-1}}{\delta X}[hp_{D}^{c}]\delta X  \label{eq-MR-U-1} \\
\breve{G}_{k|k}^{u}[\breve{h}] &\propto &\int \breve{f}_{k}^{\ast }(Z_{k}|%
\breve{X})\,(\breve{h}p_{D})^{\breve{X}}\,\frac{\delta \breve{G}_{k-1|k-1}}{%
\delta \breve{X}}[\breve{h}^{\mathbb{U}}p_{D}^{c}]\delta \breve{X}.
\label{eq-MR-U-2}
\end{eqnarray}
\end{enumerate}

The meaning of Eqs. (\ref{eq-MR-D-1}-\ref{eq-MR-U-2}) is that the D-U/D
approach appears to provide a more complete and nuanced model of U/D targets
than the S-U/D approach. \ In the latter, detection and non-detection are
both either \textquotedblleft strictly on\textquotedblright\ or
\textquotedblleft strictly off\textquotedblright ; whereas in the former
they are \textquotedblleft partially on\textquotedblright\ or
\textquotedblleft partially off\textquotedblright\ to account for the
transition of U-targets to D-targets and for the impossibility of transition
of D-targets to U-targets.

\subsubsection{Third Purpose}

This is to provide a far simpler illustration of these points. \ The
probability hypothesis density (PHD)\ filter \cite[Sec. 15.3]{Mah-Artech}, 
\cite[Sec. 8.4]{Mah-Newbook} is a statistically first-order approximation of
the (unlabeled) MRBF, with linear computational complexity in the number of
targets and the number of measurements. \ 

In Section \ref{A-PHD}, PHD filters will be derived for both the S-U/D and
D-U/D approaches and then compared. \ In both cases, the number and states
of U-targets and the number and states of D-targets can be easily estimated.

\subsubsection{Fourth Purpose}

Section \ref{A-Comp AA-Comp} will employ purely algebraic methods to verify
the statistically-derived formulas for \ $G_{k|k}^{d}(X|Z_{1:k})$ \ \cite[%
Eq. (5)]{MahSensors1019Exact},\ $G_{k|k}^{u}(X|Z_{1:k})$ \ \cite[Eq. (6)]%
{MahSensors1019Exact}, $\ f_{k|k}^{d}(X|Z_{1:k})$ \ \cite[Eq. (75)]%
{MahSensors1019Exact}, and $\ f_{k|k}^{u}(X|Z_{1:k})$ \ \cite[Eq. (90)]%
{MahSensors1019Exact};\ as well as for \ $G_{k|k}(X|Z_{1:k})$ \ \cite[Eq. (4)%
]{MahSensors1019Exact}.\ 

\subsubsection{Fifth Purpose}

Section \ref{A-Indep} will apply the preceding results to further critique a
second problematic aspect of the PMBM U/D approach: \ \textquotedblleft
U/D-parallelism\textquotedblright ---i.e., the claim that PMBM filters
propagate U-targets and D-targets separately in parallel. \ 

Briefly stated, in the S-U/D case, if \ $f_{k|k}(X|Z_{1:k})$ \ is a PMBM
distribution then the U-target distribution \ $f_{k|k}^{u}(X|Z_{1:k})$ \
should be Poisson and the D-target distribution \ $f_{k|k}^{d}(X|Z_{1:k})$ \
should be MBM. \ But neither is necessarily true, and this remains unchanged
in the D-U/D case.

\subsection{Organization of the Paper \label{A-Intro-AA-Org}}

The paper is organized as follows: analysis of the two-step U/D approach
(Section \ref{A-TwoStepUD}); the single-step, single-target recursive Bayes
filter (Section \ \ref{A-RBFSing}); the D-U/D single-step Bayes filter
(Section \ref{A-1StepU/D}); the D-U/D Bernoulli filter (Section \ref{A-Bern}%
); the multitarget U/D filter (Section \ref{A-Multi}); comparison of the
S-U/D and D-U/D approaches (Section \ref{A-Comp}); U/D-parallelism (Section %
\ref{A-Indep}); the S-U/D and D-U/D PHD\ Filters (Section \ref{A-PHD});
mathematical derivations (Section \ref{A-Der}); and Conclusions (Section \ref%
{A-Concl}). \ 

\section{Analysis of the Two-Step U/D Approach \label{A-TwoStepUD}}

The section is organized as follows: \ the two-step single-target recursive
Bayes filter (Section \ref{A-TwoStepUD-AA-2step}); the D-U/D state-space
model (Section \ref{A-TwoStepUD-AA-StateMod}); and analysis of the PMBM
version of U/D targets (Section \ref{A-TwoStepUD-AA-Anal}).

Let \ $\mathbb{X}_{0}$ \ denote the single-target state space and \ $\mathbb{%
Z}$ \ the sensor single-target measurement space.

\subsection{The Two-Step, Single-Target Recursive Bayes Filter \label%
{A-TwoStepUD-AA-2step}}

Assume that a single target is observed by a single sensor with missed
detections but no clutter or false alarms. \ Let \ $Z_{1:k}:Z_{1},...,Z_{k}%
\subseteq \mathbb{Z}$ \ be a time-sequence of finite sets of measurements
where, because of these assumptions, each \ $Z_{1},...,Z_{k}$ \ is either
the empty set \ $\emptyset $ \ or a singleton set \ $\{z\}$. Then the
general form of the single-target, two-step recursive Bayes filter is a
sequence of time-update and measurement-update steps \cite[Eqs. (3.54,3.56)]%
{Mah-Artech}: \ \ 
\begin{eqnarray}
f_{k|k-1}(x|Z_{1:k-1}) &=&\int
f_{k|k-1}(x|x_{k-1},Z_{1:k-1})f_{k-1|k-1}(x_{k-1}|Z_{1:k-1})dx_{k-1}\;\;\;
\label{eq-BF-1} \\
f_{k|k}(x|Z_{1:k}) &=&\frac{f_{k}(Z_{k}|x,Z_{1:k-1})\,f_{k|k-1}(x|Z_{1:k-1})%
}{f_{k}(Z_{k}|Z_{1:k-1})}\text{ \ \ \ where}  \label{eq-BF-2} \\
f_{k}(Z_{k}|Z_{1:k-1}) &=&\int
f_{k}(Z_{k}|x_{k},Z_{1:k-1})\,f_{k|k-1}(x_{k}|Z_{1:k-1})dx_{k}
\label{eq-BF-3}
\end{eqnarray}%
with initial distribution \ $f_{0|0}(x)$; and where \ $%
f_{k|k-1}(x|x_{k-1},Z_{1:k-1})$ \ is the (general) single-target Markov
density and \ $f_{k}(Z_{k}|x,Z_{1:k-1})$ \ is the (general) single-target
measurement density. \ 

\subsection{The D-U/D State-Space Model \label{A-TwoStepUD-AA-StateMod}}

In the PMBM U/D approach, U/D targets are (as with earlier treatments of
them in the multitarget tracking literature\footnote{%
The concept apparently originated in \cite{Reid-MHT}, where U-targets were
called \textquotedblleft unknown targets.\textquotedblright}) addressed in a
heuristic and ad hoc fashion. \ The \textquotedblleft undetected
target\textquotedblright\ concept is rather vague. \ What, for example, is
the difference between it and a nonexistent target? \ Both generate the null
measurement \ $\emptyset $ \ at any given instant, and thus are in this
sense indistinguishable until the undetected target finally (if ever)
generates some \ $z\in \mathbb{Z}$.\footnote{%
The S-U/D model can distinguish nonexistent from undetected targets. \
Whereas the former can never generate measurements \ $z\in \mathbb{Z}$, \
the former can; and this difference is statistically visible \cite[p. 21]%
{MahSensors1019Exact}..}

Henceforth, individual targets are endowed with an additional, discrete,
state variable \ $o$, \ such that \ $o=1$ \ or \ $o=0$ \ depending on
whether they are D-targets or U-targets, respectively. \ That is, target
states have the mathematical form%
\begin{equation}
\breve{x}=(x,o)\in \mathbb{X}_{0}\times \{0,1\}=\mathbb{\breve{X}}_{0}.
\label{eq-StateSpace}
\end{equation}%
Denote 
\begin{eqnarray}
\mathbb{D} &=&\mathbb{X}_{0}\times \{1\}\text{ \ (D-targets)} \\
\mathbb{U} &=&\mathbb{X}_{0}\times \{0\}\text{ \ (U-targets)}
\end{eqnarray}%
so that \ $\mathbb{D}\cup \mathbb{U}=\mathbb{\breve{X}}_{0}$ \ and \ $%
\mathbb{D}\cap \mathbb{U}=\emptyset $. \ Note that\ $\ \mathbb{D},\mathbb{U}$
\ are topologically disconnected and thus are both open and closed subsets
of \ $\mathbb{\breve{X}}_{0}$.

The integral of a real-valued density function \ $\breve{f}(\breve{x})$ \ of
\ $\breve{x}\in \mathbb{\breve{X}}_{0}$\ \ is \ 
\begin{equation}
\int \breve{f}(\breve{x})d\breve{x}\overset{_{\text{def.}}}{=}\sum_{o\in
\{0,1\}}\int \breve{f}(x,o)dx=\int \breve{f}(x,0)dx+\int \breve{f}(x,1)dx.
\label{eq-DUD-integral}
\end{equation}

Now because there is a single-target with missed detections but no clutter,
the sensor measurement density has the simple form \ 
\begin{equation}
\breve{f}_{k}(Z|\breve{x})=\left\{ 
\begin{array}{ccc}
1-p_{D}(\breve{x}) & \text{if} & Z=\emptyset \\ 
\breve{p}_{D}(\breve{x})\,\breve{f}_{k}(z|\breve{x}) & \text{if} & Z=\{z\}
\\ 
0 & \text{if} & |Z|\geq 2%
\end{array}%
\right.  \label{eq-Like-0}
\end{equation}%
where it is assumed that \ $\breve{f}_{k}(Z|\breve{x},Z_{1:k-1})=\breve{f}%
_{k}(Z|\breve{x})$. \ Also, $\ \breve{p}_{D}(\breve{x})\overset{\text{abbr.}}%
{=}\breve{p}_{D,k}(\breve{x})$ \ and \ $\breve{f}_{k}(z|\breve{x})\overset{%
\text{abbr.}}{=}\breve{L}_{z}(\breve{x})$ \ are, respectively, the
probability of detection and the measurement density (a.k.a. likelihood
function) of the target \ $\breve{x}=(x,o)$ \ if it is detected. \ If \ $%
Z_{k}=\emptyset $ \ then \ $\breve{x}$ \ was not detected at time \ $t_{k}$;
whereas if \ $Z_{k}=\{z_{k}\}$ \ then it was detected with probability \ $%
\breve{p}_{D}(\breve{x})$ \ and generated measurement \ $z_{k}$ \ with
probability (density)\ \ $\breve{f}_{k}(z|\breve{x})$.

Eq. (\ref{eq-Like-0}) defines a valid measurement density since, for all \ $%
\breve{x}$,\footnote{%
Eq. (\ref{eq-0a}) is a special case of the RFS set integral defined below in
Eq. (\ref{eq-SetInt}).} \ 
\begin{equation}
\int \ \breve{f}_{k}(Z|\breve{x})\delta Z\overset{\text{def.}}{=}\check{f}%
_{k}(\emptyset |\breve{x})+\int \ \breve{f}_{k}(\{z\}|\breve{x})dz=1.
\label{eq-0a}
\end{equation}

\subsection{Analysis of the PMBM U/D Model \label{A-TwoStepUD-AA-Anal}}

Can the PMBM version of U/D-targets be formulated using the two-step MRBF? \
The answer is \textquotedblleft No,\textquotedblright\ as can be seen as
follows. \ 

If the U-target \ $(x,0)$ \ generates some \ $z_{k}\in \mathbb{Z}$ \ at time
\ $t_{k}$ \ then\footnote{%
More precisely, since \ $\breve{f}_{k}(Z|x,0)$ \ is the distribution of the
random measurement-set \ $\Sigma _{k}\subseteq \mathbb{Z}$ \ then \ $\breve{f%
}_{k}(\emptyset |x,0)=\Pr (\Sigma _{k}=\emptyset )$ \ and so \ $1-\breve{f}%
_{k}(\emptyset |x,0)=\Pr (\Sigma _{k}\neq \emptyset )$. \ Thus \ \ $\breve{f}%
_{k}(\emptyset |x,0)=0$ \ is equivalent to \ $\Pr (\Sigma _{k}\neq \emptyset
)=1$, which means that the following event occurs with probability $1$: \
some nonempty measurement is generated.} \ \ 
\begin{equation}
\breve{f}_{k}(\emptyset |x,0)=1-\int f_{k}(\{z\}|x,0)dz=0,  \label{eq-Null}
\end{equation}%
in which case it transitions to the D-target \ $(x,1)$ \ at time \ $t_{k+1}$
\ and thereafter.\footnote{%
It might seem that \ $(x,0)$ \ transitions instantaneously to \ $(x,1)$ \
when it is detected at time \ $t_{k}$. \ However, the detection process has
a small duration time \ $\Delta t>0$. \ Assume that the data rate is slow
enough that \ $\Delta t\ll t_{k+1}-t_{k}$. \ Then the effective detection
time is \ \ $t_{k+1}$. \ } \ 

However, U-targets and D-targets are \textit{observationally identical}. \
That is, they are detectable in exactly the same manner and, if detected,
generate measurements in exactly the same manner. \ That is: 
\begin{eqnarray}
\breve{p}_{D}(x,o) &=&p_{D}(x)  \label{eq-Like-1} \\
\ \breve{f}_{k}(z|x,o) &=&f_{k}(z|x)=L_{z}(x)  \label{eq-Like-2} \\
\ \breve{f}_{k}(Z|x,o) &=&f_{k}(Z|x).  \label{eq-Like-3}
\end{eqnarray}%
D-targets and U-targets are also \textit{kinematically identical}. \ That
is, if \ 
\begin{equation}
f_{k|k-1}(x|x_{k-1})=M_{x}(x_{k-1})
\end{equation}%
is the conventional single-target Markov density then the single-target
D-U/D Markov transition density is \ 
\begin{equation}
\breve{f}_{k|k-1}(x,o|x_{k-1},o_{k-1})=\delta
_{o,o_{k-1}}\,f_{k|k-1}(x|x_{k-1}).  \label{eq-Trans}
\end{equation}%
That is, D-targets transition to D-targets and U-targets transition to
U-targets and transition in exactly the same manner. \ Moreover, \ $\breve{f}%
_{k|k-1}(x,o|x_{k-1},o_{k-1})$ \ is physically unrelated to \ $\breve{f}%
_{k}(Z|x,o)=f_{k}(X|x)$. \ 

This means that the transition of a U-target to a D-target cannot be modeled
using \ $\breve{f}_{k|k-1}(\breve{x}|\breve{x}_{k-1})$ \ or its general form
\ $\breve{f}_{k|k-1}(\breve{x}|\breve{x}_{k-1},Z_{1:k-1})$, because neither
involves the detection process at time \ $t_{k}$. \ Furthermore, both \ $%
\breve{f}_{k}(Z|\breve{x})$ and its general form \ $\breve{f}_{k}(Z|\breve{x}%
,Z_{1:k-1})$\ \ do involve the detection process at time \ $t_{k}$ \ but
neither can model target transitions. \ 

Indeed, the U-birth model seems to implicitly have either a Markov density
of the logically impossible form $\breve{f}_{k|k-1}(\breve{x}|\breve{x}%
_{k-1},Z_{k})$ \ or a measurement density of the equally questionable form \ 
$\breve{f}_{k}(Z,\breve{x}|\breve{x}_{k})$.

It is therefore perplexing that (as noted in Section 9 of \cite%
{Mah-TRFS-arXiv2024}) the PMBM\ approach is simultaneously claimed to a)
encompass U/D targets and b) be based on the two-step MRBF.

To resolve this conundrum, note that it follows from Eqs. (\ref{eq-Like-0},%
\ref{eq-Like-1}-\ref{eq-Like-3}) that%
\begin{equation}
\breve{f}_{k}(\emptyset |x,0)=f_{k}(\emptyset |x)=1-p_{D}(x),
\end{equation}%
which, according to Eq. (\ref{eq-Null}), must vanish if \ $(x,0)$ \ is to\
transition to a D-target. \ But it vanishes\ if and only if \ $p_{D}(x)=1$.
\ That is: \ U-targets are detectable if and only if all targets are
perfectly detectable---which is\ true only for very high-SNR scenarios.

The following conclusions can therefore be drawn: \ 

\begin{enumerate}
\item The U-birth model violates the two-step structure of the MRBF and,
consequently, the PMBM approach to U/D targets is theoretically erroneous.

\item The single-step MRBF must be investigated to determine if it can be
employed to salvage the approach.
\end{enumerate}

As is now demonstrated in Section \ref{A-RBFSing}, no such salvage operation
will succeed unless a single-step MRBF using a novel transition function $%
\breve{f}_{k|k-1}(Z,\breve{x}|\breve{x}_{k-1})$ \ is employed. \
Specification of this MRBF in Section \ref{A-1StepU/D} will complete the
definition of the full D-U/B approach.

\section{The Single-Step, Single-Target Bayes Filter \label{A-RBFSing} \ }

For the moment, ignore the D-U/D state variable \ $o$. \ Substituting Eq. (%
\ref{eq-BF-1}) into Eq. (\ref{eq-BF-2}) we get the \textit{single-step}
recursive Bayes filter: \ \ 
\begin{equation}
f_{k|k}(x|Z_{1:k})=\frac{\int \tilde{f}_{k|k-1}(Z_{k},x|x_{k-1},Z_{1:k-1})%
\,f_{k-1|k-1}(x_{k-1}|Z_{1:k-1})dx_{k-1}}{f_{k}(Z_{k}|Z_{1:k-1})}
\label{eq-BF-4}
\end{equation}%
with initial distribution \ $f_{0|0}(x)$ \ and $Z_{1:0}\overset{_{\text{def.}%
}}{=}Z_{1}$, where the single-step \textit{joint transition function} (JTF)
is%
\begin{equation}
\tilde{f}_{k|k-1}(Z_{k},x|x_{k-1},Z_{1:k-1})=f_{k}(Z_{k}|x,Z_{1:k-1})%
\,f_{k|k-1}(x|x_{k-1},Z_{1:k-1}).
\end{equation}

\begin{remark}
\label{Rem-Valid}Note that Eq. (\ref{eq-BF-4}) follows directly from Bayes'
rule and the total probability theorem. \ It is, therefore, a generally
valid Bayesian recursion formula regardless of the specific algebraic form
of \ $\tilde{f}_{k|k-1}(Z,x|x_{k-1},Z_{1:k-1})$.
\end{remark}

Hereafter it is assumed (as is typical in multitarget tracking) that the
following independence assumptions apply: \ 
\begin{eqnarray}
f_{k}(Z_{k}|x,Z_{1:k-1}) &=&f_{k}(Z_{k}|x) \\
f_{k|k-1}(x|x_{k-1},Z_{1:k-1}) &=&f_{k|k-1}(x|x_{k-1}).
\end{eqnarray}

Also define the \textquotedblleft conventional joint transition
function\textquotedblright\ (C-JTF) to be: \ \ 
\begin{equation}
f_{k|k-1}^{\text{con}}(Z,x|x_{k-1})=f_{k}(Z|x)\,f_{k|k-1}(x|x_{k-1})
\label{eq-Like-5}
\end{equation}%
where%
\begin{equation}
f_{k}(Z|x)=\left\{ 
\begin{array}{ccc}
1-p_{D}(x) & \text{if} & Z=\emptyset \\ 
p_{D}(x)\,f_{k}(z|x) & \text{if} & Z=\{z\} \\ 
0 & \text{if} & |Z|\geq 2%
\end{array}%
\right. .  \label{eq-MeasDen}
\end{equation}%
This mathematically models the following physical process: \ The target with
state \ $x_{k-1}$ \ at time \ $t_{k-1}$ \ transitions to the target with
state \ $x$ \ at time \ $t_{k}$, \ which then generates the (empty or
singleton) measurement-set \ $Z$. \ 

The analogous C-JTF for the D-U/D case (CUD-JTF) is \ \ 
\begin{equation}
\breve{f}_{k|k-1}^{\text{con}}(Z,x,o|x_{k-1},o_{k-1})=\delta
_{o,o_{k-1}}\,f_{k}(Z|x)\,f_{k|k-1}(x|x_{k-1})  \label{eq-LIke-6}
\end{equation}%
where the factor \ $\delta _{o,o_{k-1}}$ \ is due to Eq. (\ref{eq-Trans}). \
Note that \ 
\begin{eqnarray}
\breve{f}_{k|k-1}^{\text{con}}(Z,x|x_{k-1},o_{k-1}) &=&\sum_{o\in \{0,1\}}%
\breve{f}_{k|k-1}^{\text{con}}(Z,x,o|x_{k-1},o_{k-1}) \\
&=&f_{k|k-1}^{\text{con}}(Z,x|x_{k-1})  \label{eq-Comp}
\end{eqnarray}%
for all \ $Z,x,x_{k-1},o_{k-1}$. \ 

\section{The D-U/D Single-Step Bayes Filter \label{A-1StepU/D}}

The D-U/D analog of Eq.(\ref{eq-BF-4}) is \ 
\begin{equation}
\breve{f}_{k|k}(\breve{x}|Z_{1:k})=\frac{\int \breve{f}_{k|k-1}(Z_{k},\breve{%
x}|\breve{x}_{k-1})\,\breve{f}_{k-1|k-1}(\breve{x}_{k-1}|Z_{1:k-1})d\breve{x}%
_{k-1}}{f_{k}(Z_{k}|Z_{1:k-1})}
\end{equation}%
with initial distribution \ $\breve{f}_{0|0}(\breve{x})$ \ and \ $Z_{1:0}%
\overset{_{\text{def.}}}{=}Z_{1}$\ and where \ $\breve{x}=(x,o)$ \ and \ $%
\breve{x}_{k-1}=(x_{k-1},o_{k-1})$.

The \textit{novel} \textit{UD-JTF} (NUD-JTF) \ $\breve{f}_{k|k-1}(Z_{k},%
\breve{x}|\breve{x}_{k-1})$ \ is defined as: \ \ \ \ \ \ 
\begin{eqnarray}
\breve{f}_{k|k-1}(Z,x,0|x_{k-1},1) &=&0  \label{eq-BFUD-1} \\
\breve{f}_{k|k-1}(\{z\},x,0|x_{k-1},0) &=&0  \label{eq-BFUD-2} \\
\breve{f}_{k|k-1}(\{z\},x,1|x_{k-1},o)
&=&p_{D}(x)\,f_{k}(z|x)\,f_{k|k-1}(x|x_{k-1})  \label{eq-BFUD-3} \\
\breve{f}_{k|k-1}(\emptyset ,x,o|x_{k-1},o)
&=&(1-p_{D}(x))\,f_{k|k-1}(x|x_{k-1})  \label{eq-BFUD-4} \\
\breve{f}_{k|k-1}(\emptyset ,x,1|x_{k-1},0) &=&0  \label{eq-BFUD-5}
\end{eqnarray}%
where \ $o=0,1$ \ and \ $Z=\emptyset ,\{z\}$.

Eqs. (\ref{eq-BFUD-1}-\ref{eq-BFUD-5}) are justifiable for the following
reasons:

\begin{enumerate}
\item Eq. (\ref{eq-BFUD-1}): \ A D-target \ $(x_{k-1},1)$ \ cannot
transition to a U-target \ $(x,0)$. \ 

\item Eq. (\ref{eq-BFUD-2}): \ A U-target $(x_{k-1},0)$ \ at time \ $t_{k}$
\ cannot, at the same time \ $t_{k}$, generate a nonempty measurement \ $%
\{z\}$. \ 

\item Eq. (\ref{eq-BFUD-3}) with \ $o=0$: \ A U-target \ $(x_{k-1},0)$ \ at
time \ $t_{k-1}$ \ transitions to a D-target \ $(x,1)$ \ at time \ $t_{k}$ \
with generated measurement $\{z\}$.

\item Eq. (\ref{eq-BFUD-3}) with \ $o=1$: \ A D-target \ $(x_{k-1},1)$ \ at
time \ $t_{k-1}$ \ transitions to a D-target\ \ $(x,1)$ \ at time \ $t_{k}$
\ with generated measurement \ $\{z\}$.

\item Eq. (\ref{eq-BFUD-4}) with \ $o=0$: \ A U-target \ $(x_{k-1},0)$ \ at
time \ $t_{k-1}$ \ transitions to a U-target \ $(x,0)$ \ at time \ $t_{k}$ \
(which therefore was not detected and so generated measurement \ $\emptyset $%
).

\item Eq. (\ref{eq-BFUD-4}) with \ $o=1$: \ A D-target \ $(x_{k-1},1)$ \ at
time \ $t_{k-1}$ \ transitions to a D-target \ $(x,1)$ \ at time \ $t_{k}$,
which is not detected.

\item Eq. (\ref{eq-BFUD-5}): \ If a U-target\ \ $(x_{k-1},0)$ \ at time \ $%
t_{k-1}$ \ transitions to a detected target \ $(x,1)$\ \ at time \ $t_{k}$ \
with generated measurement \ $\emptyset $, \ then the latter must actually
be a U-target, not a D-target.
\end{enumerate}

Eq. (\ref{eq-BFUD-3}) with \ $o=0$ \ is especially significant in that it
statistically models the transition of U-targets to D-targets. \ 

Compare Eqs. (\ref{eq-BFUD-1}-\ref{eq-BFUD-5}) with the CUD-JTF in Eq. (\ref%
{eq-LIke-6}) which is, recall,%
\begin{equation}
\breve{f}_{k|k-1}^{\text{con}}(Z,x,o|x_{k-1},o_{k-1})=\delta
_{o,o_{k-1}}\,f_{k}(Z|x)\,f_{k|k-1}(x|x_{k-1}),  \label{eq-Conventiional}
\end{equation}%
or, equivalently, 
\begin{eqnarray}
\breve{f}_{k|k-1}^{\text{con}}(\{z\},x,o|x_{k-1},o_{k-1}) &=&\delta
_{o,o_{k-1}}\,p_{D}(x)\,f_{k}(z|x)\,f_{k|k-1}(x|x_{k-1}) \\
\breve{f}_{k|k-1}^{\text{con}}(\emptyset ,x,o|x_{k-1},o_{k-1}) &=&\delta
_{o,o_{k-1}}\,(1-p_{D}(x))\,f_{k|k-1}(x|x_{k-1}).
\end{eqnarray}

The NUD-JTF differs from the CUD-JTF in only two respects: a) Eq. (\ref%
{eq-BFUD-2}) and b) Eq. (\ref{eq-BFUD-3}) with \ $o=0$ (which have been
interchanged with each other):%
\begin{eqnarray}
\breve{f}_{k|k-1}(\{z\},x,0|x_{k-1},0) &=&0  \label{eq-Differ-1} \\
&\neq &p_{D}(x)\,f_{k}(z|x)\,\,f_{k|k-1}(x|x_{k-1}) \\
&=&\breve{f}_{k|k-1}^{\text{con}}(\{z\},x,0|x_{k-1},0)
\end{eqnarray}%
\begin{eqnarray}
\breve{f}_{k|k-1}(\{z\},x,1|x_{k-1},0)
&=&p_{D}(x)\,f_{k}(z|x)\,f_{k|k-1}(x|x_{k-1})  \label{eq-Differ-2} \\
&\neq &0 \\
&=&\breve{f}_{k|k-1}^{\text{con}}(\{z\},x,1|x_{k-1},0).
\end{eqnarray}%
Eq. (\ref{eq-Differ-1}) indicates that when a U-target transitions to a
U-target, the latter cannot generate an empty measurement---since then it
should actually be a D-target. \ Eq. (\ref{eq-Differ-2}), on the other hand,
allows U-targets to transition to D-targets.

It must be verified that \ $\breve{f}_{k|k-1}(Z,\breve{x}|\breve{x}_{k-1})$
\ is a well-defined state-transition density function---i.e., that: 
\begin{eqnarray}
\int \breve{f}_{k|k-1}(Z,\breve{x}|\breve{x}_{k-1})\delta Zd\breve{x}
&=&\sum_{o\in \{0,1\}}\int \breve{f}_{k|k-1}(Z,x,o|\breve{x}_{k-1})\delta
Zdx\;\;  \label{eq-Norm} \\
&=&1
\end{eqnarray}%
for all \ $\breve{x}_{k-1}=(x_{k-1},o_{k-1})$. \ From Eqs. (\ref{eq-BFUD-1}-%
\ref{eq-BFUD-5}) it is easily seen that the NUD-JTF, similarly to the
CUD-JTF, satisfies Eq. (\ref{eq-Comp}): \ 
\begin{eqnarray}
\sum_{o\in \{0,1\}}\breve{f}_{k|k-1}(Z,x,o|x_{k-1},o_{k-1})
&=&f_{k}(Z|x)\,f_{k|k-1}(x|x_{k-1}) \\
&=&f_{k|k-1}^{\text{con}}(Z,x|x_{k-1})
\end{eqnarray}%
for all \ $Z,x,x_{k-1},o_{k-1}.$ \ Eq. (\ref{eq-Norm}) immediately follows
from this.

Note that Eqs. (\ref{eq-BFUD-1}-\ref{eq-BFUD-5}) can be written more
succinctly as 
\begin{eqnarray}
\breve{f}_{k|k-1}(Z,x,o|x_{k-1},1) &=&\delta _{o,1}\,f_{k|k-1}^{\text{con}%
}(Z,x|x_{k-1})  \label{eq-BFUD-6} \\
\breve{f}_{k|k-1}(Z,x,0|x_{k-1},0) &=&\delta _{|Z|,0}\,f_{k|k-1}^{\text{con}%
}(Z,x|x_{k-1})  \label{eq-BFUD-7} \\
\breve{f}_{k|k-1}(Z,x,1|x_{k-1},0) &=&(1-\delta _{|Z|,0})\,f_{k|k-1}^{\text{%
con}}(Z,x|x_{k-1}),  \label{eq-BFUD-8}
\end{eqnarray}%
or even more succinctly as%
\begin{equation}
\breve{f}_{k|k-1}(Z,x,o|x_{k-1},o_{k-1})=\left( \delta _{o,1}+(-1)^{o}\delta
_{o_{k-1},0}\delta _{|Z|,0}\right) \,f_{k|k-1}^{\text{con}}(Z,x|x_{k-1})
\label{eq-BFUD-9}
\end{equation}%
where \ $|Z|$ \ denotes the number of elements in \ $Z$ \ and where \ $%
f_{k|k-1}^{\text{con}}(Z,x|x_{k-1})$ \ was defined in Eq. (\ref{eq-Like-5}).
\ Thus \ $\breve{f}_{k|k-1}^{\text{con}}(Z,\breve{x}|\breve{x}_{k-1})$ \ of
Eq. (\ref{eq-LIke-6}) results from replacing \ $\delta _{o,1}+(-1)^{o}\delta
_{o_{k-1},0}\delta _{|Z|,0}$ \ with \ $\delta _{o,o_{k-1}}$. \ \ 

\section{The D-U/D Bernoulli filter \label{A-Bern}}

The section is organized as follows: \ the Bernoulli filter (Section \ref%
{A-Bern-AA-Bern}) and the D-U/D\ Bernoulli filter (Section \ref{A-Bern-AA-UD}%
).

\subsection{The Bernoulli Filter \label{A-Bern-AA-Bern}}

The Bernoulli filter \cite{BaTuong-Thesis}, \cite[pp. 514-528]{Mah-Artech}, 
\cite{RisticTutorial2013}, \cite{MahWiley2025} is an algebraically exact
closed-form special case of the MRBF (in the sense of \cite%
{MahSensors1019Exact}), assuming that a) the target and clutter generation
processes are independent, b) the multi-object distribution \ $\kappa
_{k}(Z) $ \ of the clutter process is known, c), the detection profile \ $%
p_{D,k|k-1}(x)$ \ is known, and d) the number of targets does not exceed
one. \ 

Its most familiar version is a two-step filter. \ However, there is also a
single-step version \cite[Eq. (111)]{MahAccess2020-PMM}:\footnote{%
Eq. (\ref{eq-PHD-single}) is slightly different than \cite[Eq. (111)]%
{MahAccess2020-PMM} in that it consolidates\ \ $\hat{L}_{Z}$\ $\ $and \ $%
M_{x}$ $\ $into \ $M_{Z,x}$ (see Eq. (\ref{eq-Bern-Cons})).\ } \ \ 
\begin{equation}
D_{k|k}(x)=\frac{\hat{L}_{Z_{k}}(x)%
\,B_{k|k-1}(x)(1-D_{k-1|k-1}[1])+D_{k-1|k-1}[p_{S}M_{Z_{k},x}]}{%
(1-B_{k|k-1}[1]+B_{k|k-1}[\hat{L}_{Z_{k}}])(1-D_{k-1|k-1}[1])+D_{k-1|k-1}[%
\hat{R}_{Z_{k}}]}.  \label{eq-PHD-single}
\end{equation}

Here, a) \ $B_{k|k-1}(x)\geq 0$ \ with \ $\int B_{k|k-1}(x)dx\leq 1$ \ is
the target-appearance PHD; b)\ \ $D_{k|k-1}(x)\geq 0$ \ and \ $%
D_{k|k}(x)\geq 0$\ \ are density functions with \ \thinspace $\int
D_{k|k}(x)dx\leq 1$ and \ $\int D_{k|k-1}(x)dx\leq 1$ \ for all \ $k\geq 1$;
c) where, if \ $f_{k}(Z|X)$ \ with \ $|X|\leq 1$ \ is the measurement
density, \ \ \ \ 
\begin{eqnarray}
\hat{L}_{Z}(x) &=&f_{k}(Z|\{x\})/\kappa _{k}(Z)  \label{eq-Bern-0} \\
&=&1-p_{D}(x)+p_{D}(x)\sum_{z\in Z}\frac{f_{k}(z|x)\,\kappa _{k}(Z-\{z\})}{%
\kappa _{k}(Z)}  \label{eq-Bern0a}
\end{eqnarray}%
\begin{eqnarray}
M_{Z,x}(x_{k-1}) &=&\hat{L}_{Z}(x)\,M_{x}(x_{k-1})  \label{eq-Bern-Cons} \\
&=&\frac{f_{k}(Z|\{x\})\,f_{k|k-1}(x|x_{k-1})}{\kappa _{k}(Z)} \\
&=&\frac{\breve{f}_{k|k-1}^{\text{Bern}}(Z,x|x_{k-1})}{\kappa _{k}(Z)}\text{
\ \ where}  \label{eq-Bern-1} \\
\breve{f}_{k|k-1}^{\text{Bern}}(Z,x|x_{k-1})
&=&f_{k}(Z|\{x\})\,f_{k|k-1}(x|x_{k-1})  \label{eq-Bern1a}
\end{eqnarray}%
\begin{eqnarray}
\hat{R}_{Z}(x_{k-1}) &=&1-p_{S}(x_{k-1})  \label{eq-Bern-2} \\
&&+p_{S}(x_{k-1})\int \hat{L}_{Z}(x)\,f_{k|k-1}(x|x_{k-1})dx  \notag
\end{eqnarray}%
\begin{equation}
=1-p_{S}(x_{k-1})+p_{S}(x_{k-1})\,\frac{\int \breve{f}_{k|k-1}^{\text{Bern}%
}(Z,x|x_{k-1})dx}{\kappa _{k}(Z)}  \label{eq-Bern-3}
\end{equation}%
d) $p_{S}(x_{k-1})\overset{_{\text{abbr.}}}{=}p_{S,k|k-1}(x_{k-1})$ \ is the
probability of target survival; and e), if \ $0\leq h(x)\leq 1$ \ is a
unitless function then \ 

\begin{equation}
D[h]=\int h(x)\,D(x)dx\text{, \ \ \ }B[h]=\int h(x)\,B(x)dx.
\end{equation}

\subsection{The D-U/D Bernoulli Filter \label{A-Bern-AA-UD}}

Eq. (\ref{eq-PHD-single}) is easily extended to the U/D case by generalizing
the NUD-JTF of Eqs. (\ref{eq-BFUD-6}-\ref{eq-BFUD-8}) to the following
NUD-JTF for Bernoulli filters: \ 
\begin{eqnarray}
\breve{f}_{k|k-1}(Z,x,0|x_{k-1},1) &=&0  \label{eq-E} \\
\breve{f}_{k|k-1}(Z,x,1|x_{k-1},1) &=&\breve{f}_{k|k-1}^{\text{Bern}%
}(Z,x|x_{k-1})  \label{eq-F} \\
\breve{f}_{k|k-1}(Z,x,0|x_{k-1},0) &=&\delta _{|Z|,0}\,\breve{f}_{k|k-1}^{%
\text{Bern}}(Z,x|x_{k-1})  \label{eq-G} \\
\breve{f}_{k|k-1}(Z,x,1|x_{k-1},0) &=&(1-\delta _{|Z|,0})\,\breve{f}%
_{k|k-1}^{\text{Bern}}(Z,x|x_{k-1})  \label{eq-H}
\end{eqnarray}%
where \ $\breve{f}_{k|k-1}^{\text{Bern}}(Z,x|x_{k-1})$ \ was defined in Eq. (%
\ref{eq-Bern1a}).

This NUD-JTF is well-defined. \ For, since \ $\int f_{k}(Z|\{x\})\delta Z=1$%
, then 
\begin{eqnarray}
\int \breve{f}_{k|k-1}(Z,\breve{x}|\breve{x}_{k-1})\delta Zd\tilde{x}
&=&\sum_{oe\{0,1\}}\int \breve{f}_{k|k-1}(Z,x,o|\breve{x}_{k-1})\delta Zdx \\
&=&1
\end{eqnarray}%
for all \ $\breve{x}_{k-1}=(x_{k-1},o_{k-1})$. \ 

Given this, the D-U/D analog of Eq. (\ref{eq-PHD-single}) is \ 
\begin{equation}
\breve{D}_{k|k}(\breve{x})=\frac{\breve{L}_{Z_{k}}(\breve{x})\,\breve{B}%
_{k|k-1}(\breve{x})(1-\breve{D}_{k-1|k-1}[1])+\breve{D}_{k-1|k-1}[p_{S}%
\breve{M}_{Z_{k},\breve{x}}]}{(1-\breve{B}_{k|k-1}[1]+\breve{B}_{k|k-1}[%
\breve{L}_{Z_{k}}])(1-\breve{D}_{k-1|k-1}[1])+\breve{D}_{k-1|k-1}[\breve{R}%
_{Z_{k}}]}  \label{eq-PHD-UD}
\end{equation}%
where the obvious analogs of Eqs. (\ref{eq-Bern-0}-\ref{eq-Bern-2}) are: 
\begin{equation}
\breve{L}_{Z}(x,o)=\hat{L}_{Z}(x)
\end{equation}%
\begin{equation}
\breve{M}_{Z,\breve{x}}(\breve{x}_{k-1})=\frac{\breve{f}_{k|k-1}(Z,\breve{x}|%
\breve{x}_{k-1})}{\kappa _{k}(Z)}
\end{equation}%
\begin{eqnarray}
\breve{R}_{Z}(x_{k-1},o_{k-1}) &=&1-p_{S}(x_{k-1}) \\
&&+p_{S}(x_{k-1})\,\frac{\breve{f}_{k|k-1}(Z|x_{k-1},o_{k-1})}{\kappa _{k}(Z)%
}  \notag
\end{eqnarray}%
\begin{eqnarray}
\breve{f}_{k|k-1}(Z|\breve{x}_{k-1}) &=&\int \breve{f}_{k|k-1}(Z,\breve{x}|%
\breve{x}_{k-1})d\breve{x} \\
&=&\sum_{o\in \{0,1\}}\int \breve{f}_{k|k-1}(Z,x,o|\breve{x}_{k-1})dx.
\end{eqnarray}%
Also, since newly-appearing targets have not yet been detected and thus must
be U-targets, we have%
\begin{equation}
\breve{B}_{k|k-1}(x,o)=\delta _{o,0}\,B_{k|k-1}(x)
\end{equation}%
for some \ $B_{k|k-1}(x)\geq 0$ \ with \ $\int B_{k|k-1}(x)dx\leq 1$, from
which follows 
\begin{eqnarray}
\breve{B}_{k|k-1}[\breve{L}_{Z_{k}}]) &=&\sum_{o\in \{0,1\}}\int \hat{L}%
_{Z_{k}}(x)\,\breve{B}_{k|k-1}(x,o)dx \\
&=&\int \hat{L}_{Z_{k}}(x)\,B_{k|k-1}(x)dx=B_{k|k-1}[\hat{L}_{Z_{k}}] \\
\breve{B}_{k|k-1}[1] &=&\sum_{o\in \{0,1\}}\int \breve{B}_{k|k-1}(x,o)dx=%
\int B_{k|k-1}(x)dx.
\end{eqnarray}%
Note that from Eqs. (\ref{eq-E}-\ref{eq-H}),%
\begin{eqnarray}
\breve{f}_{k|k-1}(Z|x_{k-1},o_{k-1}) &=&\int \breve{f}_{k|k-1}(Z,\breve{x}%
|x_{k-1},o_{k-1})d\breve{x} \\
&=&\sum_{o\in \{0,1\}}\int \breve{f}_{k|k-1}(Z,x,o|x_{k-1},o_{k-1})dx \\
&=&\int f_{k}(Z|\{x\})\,f_{k|k-1}(x|x_{k-1})dx \\
&=&\int \breve{f}_{k}^{\text{Bern}}(Z,x|x_{k-1})dx.
\end{eqnarray}

\section{The D-U/D Multitarget Filter \label{A-Multi}}

This section demonstrates that the preceding analyses can be extended to the
general multitarget case.

The most obvious approach would be to begin with the multitarget
generalization of Eq. (\ref{eq-BF-4}), \ 
\begin{equation}
\breve{f}_{k|k}(\breve{X}|Z_{1:k})=\frac{\int \breve{f}_{k|k-1}(Z_{k},\breve{%
X}|\breve{X}_{k-1})\,\breve{f}_{k-1|k-1}(\breve{X}_{k-1}|Z_{1:k-1})\delta 
\breve{X}_{k-1}}{f_{k}(Z_{k}|Z_{1:k-1})},  \label{eq-Multitarg}
\end{equation}%
with initial distribution \ $\breve{f}_{0|0}(\breve{X})$; \ where the set
integral \ $\int \cdot \delta \breve{X}_{k-1}$ \ will be defined in Eq. (\ref%
{eq-SetInt}); and\ where \ $\breve{f}_{k|k-1}(Z,\breve{X}|\breve{X}%
_{k-1},Z_{1:k-1})$ \ is some D-U/D multitarget NUD-JTF, assuming the usual
independence assumption \ 
\begin{equation}
\breve{f}_{k|k-1}(Z,\breve{X}|\breve{X}_{k-1},Z_{1:k-1})=\breve{f}_{k|k-1}(Z,%
\breve{X}|\breve{X}_{k-1}).
\end{equation}%
However, there is an obvious difficulty: \ What is \ $\breve{f}_{k|k-1}(Z,%
\breve{X}|\breve{X}_{k-1})$? \ This issue is resolvable will be investigated
further in Section \ref{A-Comp AA-NUD}. \ 

The most concise and conceptually streamlined approach, however, is to begin
with the RFS framework's \textit{probability generating functional }%
(p.g.fl.) technique \cite[Sec. 14.8]{Mah-Artech}.

\begin{remark}
\label{Rem-LRFS}The analysis in this section can be extended to labeled RFS
(LRFS) theory \cite{Vo-TSP-Overview2024}, \cite{VoVoTSPconjugate}, \cite%
{Beard2020}, \cite[Chpt. 15]{Mah-Newbook}. \ This possibility will not be
considered here since the most recent PMBM incarnation---the trajectory PMBM
filter---does not employ (and in fact, prohibits) LRFS labels \cite%
{Mah-TRFS-arXiv2024}. \ 
\end{remark}

\begin{remark}
\label{Rem-BernWarn} The D-U/D Bernoulli filter of Eq. (\ref{eq-PHD-UD}) is
not a special case of the approach described in this section. \ This is
because the latter assumes that the target motion and target appearance
processes are independent, whereas the former specifies, because of the
restriction \ $|X|\leq 1$, that they must be synchronized in a particular
manner.
\end{remark}

The section is organized as follows: \ the p.g.fl. framework (Section \ref%
{A-Multi-AA-PGFL}); the general form of the non-U/D p.g.fl. two-step Bayes
filter (Section \ref{A-Multi-AA-2StepGen}); the non-U/D p.g.fl. two-step
Bayes filter for standard models (Section \ref{A-Multi-AA-2StepStd}); the
non-U/D p.g.fl. one-step Bayes filter for standard models (Section \ref%
{A-Multi-AA-1Step}); and the p.g.fl. one-step Bayes filter for the\ D-U/D
case (Section \ref{A-Multi-AA-1StepU/D}).

\subsection{The p.g.fl. Framework \label{A-Multi-AA-PGFL}}

For the moment ignore the D-U/D variable \ $o$. \ Let \ $f(X)$ \ be a
multitarget probability density function defined on finite subsets \ $%
X\subseteq \mathbb{X}_{0}$. \ Then its p.g.fl. \ $0\leq G_{f}[h]\leq 1$ \ is
defined as%
\begin{equation}
G_{f}[h]=\int h^{X}\,f(X)\delta X  \label{eq-PGFL}
\end{equation}%
where a) $0\leq h(x)\leq 1$ \ are unitless \textquotedblleft test
functions\textquotedblright\ for \ $x\in \mathbb{X}_{0}$; b)\ $%
h^{X}=\emptyset $ \ if \ $X=\emptyset $ \ and \ $h^{X}=\prod_{x\in X}h(x)$ \
if otherwise; and c) 
\begin{equation}
\int f(X)\delta X\ =f(\emptyset )+\sum_{n\geq 1}\frac{1}{n!}\int
f(\{x_{1},...,x_{n}\})dx_{1}\cdots dx_{n}  \label{eq-SetInt}
\end{equation}%
is the RFS set integral.

The multitarget density \ $f(X)$ \ can be recovered from\ $G_{f}[h]$ \ using
the RFS functional derivative \ $\delta /\delta X$ \cite[Sect. 11.4.1]%
{Mah-Artech}: \ 
\begin{equation}
f(X)=\left[ \frac{\delta G_{f}}{\delta X}[h]\right] _{h=0}=\frac{\delta G_{f}%
}{\delta X}[0]  \label{eq-Recover}
\end{equation}%
with \ $(\delta G_{f}/\delta \emptyset )[h]\overset{_{\text{def.}}}{=}%
G_{f}[h]$ \ if \ $X=\emptyset $. \ If \ $X_{(n)}=\{x_{1},...,x_{n}\}$ \ with
\ $|X|=n\geq 1$ \ and \ $X_{(n-1)}=X-\{x_{n}\}$ \ then this derivative is
recursively defined as \ 
\begin{equation}
\frac{\delta G}{\delta X_{(n)}}[h]=\frac{\delta }{\delta x_{n}}\frac{\delta G%
}{\delta X_{(n-1)}}[h]  \label{eq-FuncDer-1}
\end{equation}%
where the \textit{intuitive} definition\footnote{%
Intuitive rather than rigorous, because \ $\delta _{x}$ \ is not a valid
test function.} of \ $\delta /\delta x$ \ is, if \ $\delta _{x}$ \ is the
Dirac delta function concentrated at \ $x$, \ 
\begin{equation}
\frac{\delta G}{\delta x}[h]=\lim_{\varepsilon \searrow 0}\frac{%
G[h+\varepsilon \delta _{x}]-G[h]}{\varepsilon }.  \label{eq-FuncDer-2}
\end{equation}

The p.g.fl. multitarget Bayes filter, to be described shortly in Eqs. (\ref%
{eq-PGFL-P1}-\ref{eq-PGFL-B1}), is a p.g.fl. version of the MRBF and is
central to the RFS approach that follows.\ 

If \ $f(x)$ \ is a density function and \ $h(x)$ \ a unitless function,
define\footnote{%
The notation \ $f[h]$ \ is employed rather than the often-used \ $\langle
f,h\rangle $. \ This is in part because the latter, which is the preferred
notation for a scalar product, is misleading since \ $f$ \ and \ $h$ \
belong to different function spaces.}%
\begin{equation}
f[h]=\int h(x)\,f(x)dx.  \label{eq-FuncNotation}
\end{equation}%
If \ $f(x)$ \ is a probability distribution then \ $f[h]$ \ is the simplest
nontrivial p.g.fl.

The p.g.fl. and probability distribution of a \textit{Poisson RFS} are,
respectively,%
\begin{equation}
G[h]=e^{D[h-1]}\text{, \ \ \ \ }f(X)=e^{-D[1]}D^{X}
\end{equation}%
where \ $D(x)\geq 0$ \ is a density function.

The section is organized as follows: \ The general p.g.fl. two-step Bayes
filter (Section \ref{A-Multi-AA-2StepGen}); the p.g.fl. two-step Bayes
filter \ for \textquotedblleft standard\textquotedblright\ multitarget
models (Section \ref{A-Multi-AA-2StepStd}); the p.g.fl. one-step Bayes
filter for standard multitarget models (Section \ref{A-Multi-AA-1Step}); and
the D-U/D p.g.fl. one-step Bayes filter (Section \ref{A-Multi-AA-1StepU/D}).

\subsection{P.g.fl. Two-Step Bayes Filter (General) \label%
{A-Multi-AA-2StepGen}}

This consists of two main parts: \ time-update and measurement-update.

\subsubsection{P.g.fl. Time-Update}

Suppose that, at time \ $t_{k-1}$, the posterior \ $f_{k-1|k-1}(X)\overset{_{%
\text{abbr.}}}{=}f_{k-1|k-1}(X|Z_{1:k-1})$ \ and its p.g.fl. \ $%
G_{k-1|k-1}[h]$ \ are given. \ Then the predicted p.g.fl. at time \ $t_{k}$
\ is given by \cite[Eqs. (14.265,14.266)]{Mah-Artech}: \ 
\begin{eqnarray}
G_{k|k-1}[h] &=&\int G_{k|k-1}[h|X_{k-1}]\,\,f_{k-1|k-1}(X_{k-1})\delta
X_{k-1}  \label{eq-PGFL-P1} \\
G_{k|k-1}[h|X_{k-1}] &=&\int h^{X}\,f_{k|k-1}(X|X_{k-1})\delta X;
\label{eq-PGFL-P2}
\end{eqnarray}%
where \ $\,f_{k|k-1}(X|X_{k-1})$ \ is the multitarget Markov density at time
\ $t_{k}$.

\subsubsection{P.g.fl Measurement-Update}

Suppose that, at time \ $t_{k-1}$, the distribution \ $f_{k|k-1}(X)\overset{%
_{\text{abbr.}}}{=}f_{k|k-1}(X|Z_{1:k-1})$ \ and its p.g.fl. \ $G_{k|k-1}[h]$
\ are given. \ Then the measurement-updated p.g.fl. at time \ $t_{k}$ \ is 
\cite[Eq. (14.280)]{Mah-Artech}

\begin{equation}
G_{k|k}[h]=\frac{\frac{\delta F_{k}}{\delta Z_{k}}[0,h]}{\frac{\delta F_{k}}{%
\delta Z_{k}}[0,1]}\propto \frac{\delta F_{k}}{\delta Z_{k}}[0,h]
\label{eq-PGFL-B1}
\end{equation}%
where a) \cite[Eqs. (14.281,14.282)]{Mah-Artech} \ \ \ \ \ \ 
\begin{eqnarray}
F_{k}[g,h] &=&\int h^{X}\,G_{k}[g|X]\,f_{k|k-1}(X)\delta X\text{ \ \ \ where}
\label{eq-PGFL-B2} \\
G_{k}[g|X] &=&\int g^{Z}\,f_{k}(Z|X)\delta Z  \label{eq-PGFL-B3}
\end{eqnarray}%
is the \textquotedblleft $F$-functional\textquotedblright\ at time \ $t_{k}$%
; b) \ $f_{k}(Z|X)$ \ is the multitarget measurement density; c) $\delta
F_{k}/\delta Z$ \ denotes the RFS functional derivative of \ $F_{k}[g,h]$ \
with respect to $\ g$; and d) $0\leq g(z)\leq 1$ \ are unitless test
functions for \ $z\in \mathbb{Z}$. \ 

\subsection{P.g.fl. Two-Step Bayes Filter (\textquotedblleft
Standard\textquotedblright\ Models) \label{A-Multi-AA-2StepStd}}

\subsubsection{P.g.fl. Time-Update (\textquotedblleft
Standard\textquotedblright\ Multitarget Motion\ Model)}

This is given by \cite[Eq. (14.273)]{Mah-Artech}: \ \ 
\begin{eqnarray}
G_{k|k-1}[h] &=&e^{B[h-1]}\,G_{k-1|k-1}[1-p_{S}+p_{S}M_{h}]\text{ \ \ \ where%
}  \label{eq-Std-P1} \\
M_{h}(x_{k-1}) &=&\int h(x)\,f_{k|k-1}(x|x_{k-1})dx=\int
h(x)\,M_{x}(x_{k-1})dx;  \label{eq-Std-P2}
\end{eqnarray}%
and where \ $f_{k|k-1}(x|x_{k-1})$ \ is the single-target Markov density; \ $%
p_{S}(x_{k-1})\overset{_{\text{abbr.}}}{=}p_{S,k|k-1}(x_{k-1})$ \ is the
probability of target survival; \ $B(x)\overset{_{\text{abbr.}}}{=}%
B_{k|k-1}(x)$ \ is the intensity function of the Poisson target-appearance
process at time \ $t_{k}$; and \ $B[h]=\int h(x)\,B_{k\left\vert
k-1\right\vert }(x)dx$.

\subsubsection{P.g.fl. Measurement-Update (\textquotedblleft
Standard\textquotedblright\ Multitarget Measurement\ Model)}

This is given by\ \cite[Eqs. (12.151,12.145, 14.290)]{Mah-Artech}: \ \ \ \ 
\begin{eqnarray}
G_{k}[g|X] &=&e^{\kappa \lbrack g-1]}\,\left( 1-p_{D}+p_{D}L_{g}\right) ^{X}%
\text{ \ \ where}  \label{eq-Std-B1} \\
L_{g}(x) &=&\int g(z)\,f_{k}(z|x)dz=\int g(z)\,L_{z}(x)dz  \label{eq-Std-B2}
\\
F_{k}[g,h] &=&e^{\kappa \lbrack g-1]}\,G_{k|k-1}[h(1-p_{D}+p_{D}L_{g})];
\label{eq-Std-B3}
\end{eqnarray}%
and where \ $f_{k}(z|x)$ \ is the single-target measurement density; \ $%
p_{D}(x)\overset{_{\text{abbr.}}}{=}p_{D,k|k-1}(x)$ \ is the target
probability of detection; \ $\kappa (z)\overset{_{\text{abbr.}}}{=}\kappa
_{k}(z)$ \ is the intensity function of the Poisson target-appearance
process at time \ $t_{k}$; and \ $\kappa \lbrack g]=\int g(z)\,\kappa
_{k}(z)dz$.

\subsection{P.g.fl. One-Step Bayes Filter (Standard Models) \label%
{A-Multi-AA-1Step}}

Substitute the formula for \ $G_{k}[g|X]$, \ Eq. (\ref{eq-Std-B1}), into Eq.
(\ref{eq-PGFL-B2}) to get:%
\begin{equation}
F_{k}[g,h]=e^{\kappa \lbrack g-1]}\int (h\left( 1-p_{D}+p_{D}L_{g}\right)
^{X}\,f_{k|k-1}(X)\delta X.
\end{equation}%
Then the definition of a p.g.fl., Eq. (\ref{eq-PGFL}), yields: \ 
\begin{equation}
F_{k}[g,h]=e^{\kappa \lbrack g-1]}\,G_{k|k-1}[h(1-p_{D}+p_{D}L_{g})].
\label{eq-Std-B-0}
\end{equation}%
Finally, from Eq. (\ref{eq-Std-P1}) we get: \ \ 
\begin{eqnarray}
F_{k}[g,h] &=&e^{\kappa \lbrack g-1]}\,e^{B[h(1-p_{D}+p_{D}L_{g})-1]}\,
\label{eq-Std-B-F} \\
&&\cdot G_{k-1|k-1}[1-p_{S}+p_{S}M_{h(1-p_{D}+p_{D}L_{g})}\,]  \notag
\end{eqnarray}%
from which \ $G_{k|k}[h]\propto (\delta F_{k}/\delta Z_{k})[0,h]$ \ can be
derived using Eq. (\ref{eq-PGFL-B1}). \ 

Abbreviate%
\begin{equation}
M_{g,h}=M_{h(1-p_{D}+p_{D}L_{g})}.  \label{eq-Abbr}
\end{equation}%
Then \ 
\begin{equation}
M_{g,h}(x_{k-1})=\int_{Z:|Z|\leq 1}h(x)\,g^{Z}\,f_{k|k-1}^{\text{con}%
}(Z,x|x_{k-1})\delta Zdx.  \label{eq-Core}
\end{equation}%
For, from Eqs. (\ref{eq-Std-P2},\ref{eq-Std-B2}), note that 
\begin{eqnarray}
&&M_{g,h}(x_{k-1})  \label{eq-Core-0} \\
&=&\int h(x)\,(1-p_{D}(x))\,f_{k|k-1}(x|x_{k-1})dx  \notag \\
&&+\int h(x)\,g(z)\,p_{D}(x)\,f_{k}(z|x)\,f_{k|k-1}(x|x_{k-1})dxdz  \notag
\end{eqnarray}%
\begin{eqnarray}
&=&\int h(x)\,f_{k|k-1}^{\text{con}}(\emptyset ,x|x_{k-1})dx \\
&&+\int h(x)\,g^{\{z\}}\,f_{k|k-1}^{\text{con}}(\{z\},x|x_{k-1})dzdx  \notag
\\
&=&\int_{Z:|Z|\leq 1}h(x)\,g^{Z}\,f_{k|k-1}^{\text{con}}(Z,x|x_{k-1})\delta
Zdx.  \label{eq-Core-1}
\end{eqnarray}

It then follows from Eq. (\ref{eq-Std-B-F}) that the one-step p.g.fl. Bayes
filter is completely determined by the following functions: \ $\kappa
_{k}(z) $, $B_{k|k-1}(x)$, \ $p_{S,k|k-1}(x_{k-1})$ \ and the C-JTF \ $%
f_{k|k-1}^{\text{con}}(Z,x|x_{k-1})$. \ \ 

\subsection{P.g.fl. One-Step Bayes filter: \ D-U/D Case \label%
{A-Multi-AA-1StepU/D}}

Now apply these results to the U/D state space \ $\mathbb{\breve{X}}_{0}$. \
In this case test functions on \ $\mathbb{\breve{X}}_{0}$ \ have the form \ $%
\breve{h}(x,o)$. \ The set-indicator functions\ of \ $\mathbb{D}$ \ and \ $%
\mathbb{U}$ \ are test functions of particular interest: 
\begin{eqnarray}
\mathbf{1}_{\mathbb{D}}(x,o) &=&\delta _{o,1}  \label{eq-Test-1} \\
\mathbf{1}_{\mathbb{U}}(x,o) &=&\delta _{o,0}.  \label{eq-Test-2}
\end{eqnarray}%
So are the test functions \ \ 
\begin{eqnarray}
&&\breve{h}|_{1}(x)\overset{_{\text{abbr.}}}{=}\breve{h}|_{1}(x,o)\overset{_{%
\text{def.}}}{=}\breve{h}(x,1)  \label{eq-Test-3} \\
&&\breve{h}|_{0}(x)\overset{_{\text{abbr.}}}{=}\breve{h}|_{0}(x,o)\overset{_{%
\text{def.}}}{=}\breve{h}(x,0)
\end{eqnarray}%
and \ 
\begin{eqnarray}
\breve{h}^{\mathbb{D}}(x,o) &=&\left\{ 
\begin{array}{ccc}
\breve{h}|_{1}(x) & \text{if} & o=1 \\ 
1 & \text{if} & o=0%
\end{array}%
\right.  \label{eq-Test-5} \\
\breve{h}^{\mathbb{U}}(x,o) &=&\left\{ 
\begin{array}{ccc}
1 & \text{if} & o=1 \\ 
\breve{h}|_{0}(x) & \text{if} & o=0%
\end{array}%
\right.  \label{eq-Test-6}
\end{eqnarray}%
or, equivalently, 
\begin{equation}
\breve{h}^{\mathbb{D}}=\mathbf{1}_{\mathbb{U}}+\mathbf{1}_{\mathbb{D}}\breve{%
h}\text{, \ \ }\breve{h}^{\mathbb{U}}=\mathbf{1}_{\mathbb{D}}+\mathbf{1}_{%
\mathbb{U}}\breve{h}.  \label{eq-Test-7}
\end{equation}

It is shown in Section \ref{A-Der-AA-Diff} that \ 
\begin{eqnarray}
\text{\ }\frac{\delta }{\delta \breve{x}}\breve{h} &=&\delta _{\breve{x}}%
\text{, \ \ \ \ \ \ }\frac{\delta }{\delta \breve{x}}\breve{h}|_{1}=\delta _{%
\breve{x}}^{d}\text{,}  \label{eq-Diff-1} \\
\frac{\delta }{\delta \breve{x}}\breve{h}^{\mathbb{D}} &=&\mathbf{1}_{%
\mathbb{D}}\,\delta _{\breve{x}}\text{, \ \ }\frac{\delta }{\delta \breve{x}}%
\breve{h}^{\mathbb{U}}=\mathbf{1}_{\mathbb{U}}\,\delta _{\breve{x}}
\label{eq-Diff-2}
\end{eqnarray}%
where \ 
\begin{eqnarray}
&&\delta _{(x,o)}(x^{\prime },o^{\prime })\overset{_{\text{def.}}}{=}\delta
_{o,o^{\prime }}\,\delta _{x_{1}}(x^{\prime })  \label{eq-DIff-3} \\
&&\delta _{(x,o)}^{d}(x^{\prime },o^{\prime })\overset{_{\text{def.}}}{=}%
\delta _{o,1}\,\delta _{x}(x^{\prime }).  \label{eq-Diff-4}
\end{eqnarray}%
\ \ 

From Eqs. (\ref{eq-Core-0}-\ref{eq-Core-1}) it follows that Eq. (\ref%
{eq-Core}) should be replaced by its D-U/D analog: \ 
\begin{eqnarray}
&&\breve{M}_{g,\breve{h}}(\breve{x}_{k-1})  \label{eq-AA} \\
&=&\int_{|Z|\leq 1}\breve{h}(\breve{x})\,g^{Z}\,\breve{f}_{k|k-1}(Z,\breve{x}%
,|\breve{x}_{k-1})\delta Zd\breve{x}  \notag \\
&=&\sum_{o\in \{0,1\}}\int_{|Z|\leq 1}\breve{h}(x,o)\,g^{Z}\,\breve{f}%
_{k|k-1}(Z,x,o|\breve{x}_{k-1})\delta Zdx
\end{eqnarray}%
\ 
\begin{eqnarray}
&=&\int_{|Z|\leq 1}\breve{h}(x,0)\,g^{Z}\,\breve{f}_{k|k-1}(Z,x,0|\breve{x}%
_{k-1})\delta Zdx  \label{eq-BB} \\
&&+\int_{Z:|Z|\leq 1}\breve{h}(x,1)\,g^{Z}\,\breve{f}_{k|k-1}(Z,x,1|\breve{x}%
_{k-1})\delta Zdx.  \notag
\end{eqnarray}%
This results in the replacement of Eq. (\ref{eq-Std-B-F}) by its D-U/D\
analog: \ 
\begin{equation}
\breve{F}_{k}[g,\breve{h}]=e^{\kappa \lbrack g-1]}\,e^{\breve{B}[\breve{h}%
(1-p_{D}+p_{D}L_{g})-1]}\,\breve{G}_{k-1|k-1}[1-p_{S}+p_{S}\breve{M}_{g,%
\breve{h}}]  \label{eq-CC}
\end{equation}%
where, since newly-appearing targets have not yet been detected and thus are
U-targets,%
\begin{equation}
\breve{B}(x,o)\overset{_{\text{abbr.}}}{=}\breve{B}_{k|k-1}(x,o)=\delta
_{o,0}\,B_{k|k-1}(x)
\end{equation}%
and thus%
\begin{equation}
\breve{B}[\breve{h}]=\sum_{o\in \{0,1\}}\int \breve{h}(x,o)\,\delta
_{o,0}\,B(x)dx=\int \breve{h}|_{0}(x)\,B(x)dx.
\end{equation}

In Section \ref{A-Der-AA-&&&1} it is demonstrated, using Eqs. (\ref{eq-AA},%
\ref{eq-BB},\ref{eq-BFUD-1}-\ref{eq-BFUD-5}), that 
\begin{equation}
\breve{M}_{g,\breve{h}}=\breve{M}_{\breve{h}(1-p_{D})+\breve{h}%
|_{1}p_{D}L_{g}},  \label{eq-Core-UD}
\end{equation}%
where \ $\breve{h}|_{1}$ \ was defined in Eq. (\ref{eq-Test-3}) and where,
using Eq. (\ref{eq-Trans}),%
\begin{eqnarray}
&&\breve{M}_{\breve{h}}(x_{k-1},o_{k-1})\overset{_{\text{def.}}}{=}\int 
\breve{h}(\breve{x})\,\,\breve{f}_{k|k-1}(\breve{x}|x_{k-1},o_{k-1})d\breve{x%
}  \label{eq-Trans-2} \\
&=&\int \breve{h}(x,o_{k-1})\,M_{x}(x_{k-1})dx=M_{\breve{h}%
|_{o_{k-1}}}(x_{k-1}).  \label{eq-Trans-3}
\end{eqnarray}%
In this case Eq. (\ref{eq-CC}) becomes%
\begin{eqnarray}
\breve{F}_{k}[g,\breve{h}] &=&e^{\kappa \lbrack g-1]}e^{\breve{B}[\breve{h}%
(1-p_{D})+\breve{h}|_{1}p_{D}L_{g})-1]}  \label{eq-DD} \\
&&\cdot \breve{G}_{k-1|k-1}[1-p_{S}+p_{S}\breve{M}_{\breve{h}(1-p_{D})+%
\breve{h}|_{1}p_{D}L_{g}}].  \notag
\end{eqnarray}

\section{Comparison of U/D Approaches \label{A-Comp}}

As previously noted, Section 5 of \cite{MahSensors1019Exact} presented a
S-U/D model of U/D targets. \ The purpose of this section is to compare it
to the D-U/D model. \ It is organized as follows: \ alignment (Section \ref%
{A-Comp AA-Align}); censored RFSs (Section \ref{A-Comp AA-Censor}); the
p.g.fl.'s of U/D targets (Section \ref{A-Comp AA-UDFGFL}); comparison of the
approaches (Section \ref{A-Comp AA-Comp}); and discussion of multitarget
NUD-JTFs (Section \ref{A-Comp AA-NUD}).

\subsection{Alignment of S-U/D and D-U/D \label{A-Comp AA-Align} \ }

A direct apples-with-apples comparison is clearly impossible since S-U/D
considers target detection at single instants whereas D-U/D considers it
over a span of time. \ To permit comparison, we must first align the two
approaches by neutralizing the time-prediction part of the NUD-JTF. \
Specifically, the following assumptions must be made: \ a)\ no target
appearances, b) no target disappearances, and c) no target motion. \ Or,
expressed mathematically, \ 
\begin{eqnarray}
B_{k|k-1}(x) &=&0\text{ \ identically}  \label{eq-Align-1} \\
p_{S,k|k-1}(x_{k-1}) &=&1\text{ \ identically}  \label{eq-Align-2} \\
f_{k|k-1}(x|x_{k-1}) &=&\delta _{x_{k-1}}(x)  \label{eq-Align-3}
\end{eqnarray}%
where \ $\delta _{x_{k-1}}$ \ is the Dirac delta function concentrated at \ $%
x_{k-1}$.

Given this, it is demonstrated in Section \ref{A-Der-AA-&&&3} that Eq. (\ref%
{eq-DD}) becomes%
\begin{equation}
\breve{F}_{k}[g,\breve{h}]=e^{\kappa \lbrack g-1]}\,\breve{G}_{k-1|k-1}[%
\breve{h}p_{D}^{c}+\breve{h}|_{1}p_{D}L_{g}]  \label{eq-FF}
\end{equation}%
where \ $\breve{h}|_{1}$ \ was defined in Eq. (\ref{eq-Test-3}) and where 
\begin{equation}
\ p_{D}^{c}=1-p_{D}.
\end{equation}%
Thus by Eq. (\ref{eq-PGFL-B1}),\ the corresponding measurement-updated
p.g.fl. is \ 
\begin{equation}
\breve{G}_{k|k}[\breve{h}]\propto \frac{\delta \breve{F}_{k}}{\delta Z_{k}}%
[0,\breve{h}].  \label{eq-FF-PGFL}
\end{equation}

\subsection{Censored RFSs \label{A-Comp AA-Censor}}

Statistical analysis of the D-U/D approach requires formulas for the
respective distributions (or, equivalently, the p.g.fl.'s) of D-targets and
U-targets. \ This, in turn, requires the concept of a \textquotedblleft
censored\ RFS.\textquotedblright\ \ 

Let $\ \breve{\Xi}\subseteq \mathbb{\breve{X}}_{0}$ \ be an RFS and \ $%
\breve{O}\subseteq \mathbb{\breve{X}}_{0}$ \ be \textit{open} in \ $\mathbb{%
\breve{X}}_{0}$. \ Then \ $\breve{\Xi}\cap \breve{O}\subseteq \mathbb{\breve{%
X}}_{0}$ \ is the RFS for which any elements of \ $\breve{O}^{c}$ \ have
been removed from \ $\breve{\Xi}$. \ 

From \cite[pp. 165]{GMN}, Prop. 23, the \textquotedblleft belief
measure\textquotedblright\ of this RFS\ is%
\begin{equation}
\breve{\beta}_{\breve{\Xi}\cap \breve{O}}(\breve{S})=\breve{\beta}_{\breve{%
\Xi}}(\breve{S}\cup \breve{O}^{c})
\end{equation}%
for all closed \ $\breve{S}\subseteq \mathbb{\breve{X}}_{0}$. \ Note that
since \ $\breve{O}^{c}=\mathbb{\breve{X}}_{0}-\breve{O}$ \ is closed, $%
\breve{\beta}_{\breve{\Xi}\cap \breve{O}}(\breve{S})$ \ is mathematically
well-defined.\footnote{\textit{Errata}: \ A version of \cite{GMN}, Prop. 23
appeared as Exercise 43 in \cite[p. 397]{Mah-Artech}. \ There,
\textquotedblleft $T$\textquotedblright\ \ should have been (as had been
stipulated in \cite[pp. 165]{GMN}) an \textit{open} (and not, as
inadvertently stated in \cite[p. 397]{Mah-Artech}, a closed) subset. \ \ } \
In particular, note that $\mathbb{D},\mathbb{U}\subseteq \mathbb{\breve{X}}%
_{0}$ \ are open (as well as closed) subsets of \ $\mathbb{\breve{X}}_{0}$,
so that \ $\breve{\beta}_{\breve{\Xi}\cap \mathbb{D}}(\breve{S})$ \ and \ $%
\breve{\beta}_{\breve{\Xi}\cap \mathbb{U}}(\breve{S})$ \ are well-defined.

P.g.fl.'s are generalized belief measures in the sense that \cite[Eq.
(11.167)]{Mah-Artech} 
\begin{equation}
\breve{\beta}_{\breve{\Xi}}(\breve{S})\overset{_{\text{def.}}}{=}\Pr (\breve{%
\Xi}\subseteq \breve{S})=\int \mathbf{1}_{\breve{S}}^{\breve{X}}\,f_{\breve{%
\Xi}}(\breve{X})\delta \breve{X}=\breve{G}_{\breve{\Xi}}[\mathbf{1}_{\breve{S%
}}]
\end{equation}%
where \ $\mathbf{1}_{\breve{S}}(\breve{x})\overset{_{\text{def.}}}{=}\max_{%
\breve{y}\in \breve{S}}\delta _{\breve{x},\breve{y}}$ \ is the set-indicator
function of \ $\breve{S}$ \ and \ $\delta _{\breve{x},\breve{y}}$ \ is a
Kronecker delta.

In particular, the algebraically simple p.g.fl. 
\begin{equation}
\breve{G}_{\breve{\Xi}\cap \breve{O}}[\breve{h}]=\breve{G}_{\breve{\Xi}%
}[1-(1-\breve{h})\mathbf{1}_{\breve{O}}]  \label{eq-Censor-PGFL}
\end{equation}%
generalizes $\ \breve{\beta}_{\breve{\Xi}\cap \breve{O}}(\breve{S})$ \ since%
\begin{equation}
\breve{G}_{\breve{\Xi}\cap \breve{O}}[\mathbf{1}_{\breve{S}}]=\breve{\beta}_{%
\breve{\Xi}\cap \breve{O}}(\breve{S}).
\end{equation}%
For, if \ $\breve{S}\subseteq \mathbb{\breve{X}}_{0}$ \ is closed then \ 
\begin{eqnarray}
\breve{G}_{\breve{\Xi}}[1-(1-\mathbf{1}_{\breve{S}})\mathbf{1}_{\breve{O}}]
&=&\breve{G}_{\breve{\Xi}}[1-\mathbf{1}_{\breve{S}^{c}}\mathbf{1}_{\breve{O}%
}] \\
&=&\breve{G}_{\breve{\Xi}}[1-\mathbf{1}_{\breve{S}^{c}\cap \breve{O}}] \\
&=&\breve{G}_{\breve{\Xi}}[\mathbf{1}_{(\breve{S}^{c}\cap \breve{O})^{c}}]=%
\breve{G}_{\breve{\Xi}}[\mathbf{1}_{\breve{S}\cup \breve{O}^{c}}] \\
&=&\breve{\beta}_{\breve{\Xi}}(\breve{S}\cup \breve{O}^{c})=\breve{\beta}_{%
\breve{\Xi}\cap \breve{O}}(\breve{S}).
\end{eqnarray}

\begin{remark}
\label{Rem-Censored}Eq. (\ref{eq-Censor-PGFL}) is a valid p.g.fl. since its
corresponding probability distribution \ $\breve{f}_{\breve{\Xi}\cap \breve{O%
}}(\breve{X})$ \ exists. \ Specifically, the following p.g.fl. version of
Prop. 23 of \cite[p. 164]{GMN} is easily verified using repeated application
of the chain rule \cite[Eq. (11.285)]{Mah-Artech} for functional
derivatives: 
\begin{eqnarray}
\breve{f}_{\breve{\Xi}\cap \breve{O}}(\breve{X}) &=&\mathbf{1}_{\breve{O}}^{%
\breve{X}}\,\frac{\,\delta \breve{G}_{\breve{\Xi}}}{\delta \breve{X}}[%
\mathbf{1}_{\breve{O}^{c}}] \\
&=&\left\{ 
\begin{array}{ccc}
\frac{\,\delta \breve{G}_{\breve{\Xi}}}{\delta \breve{X}}[\mathbf{1}_{\breve{%
O}^{c}}] & \text{if} & \breve{X}\subseteq \breve{O} \\ 
0 & \text{if} & \text{otherwise}%
\end{array}%
\right. .
\end{eqnarray}
\end{remark}

Eq. (\ref{eq-Censor-PGFL}) arises from the \textquotedblleft
prodsum\textquotedblright\ fuzzy logic \cite[pp. 122, 130]{Mah-Artech}: \ 
\begin{eqnarray}
\breve{h}_{1}\wedge \breve{h}_{2} &=&\breve{h}_{1}\breve{h}_{2} \\
\breve{h}_{1}\vee \breve{h}_{2} &=&\breve{h}_{1}+\breve{h}_{2}-\breve{h}_{1}%
\breve{h}_{2} \\
\breve{h}^{c} &=&1-\breve{h}.\ 
\end{eqnarray}%
It is easily shown that \ $\mathbf{1}_{\breve{T}_{1}}\wedge \mathbf{1}_{%
\breve{T}_{2}}=\mathbf{1}_{\breve{T}_{1}\cap \breve{T}_{2}}$, \ $\mathbf{1}_{%
\breve{T}_{1}}\vee \mathbf{1}_{\breve{T}_{2}}=\mathbf{1}_{\breve{T}_{1}\cup 
\breve{T}_{2}}$,\ and \ $\mathbf{1}_{\breve{T}_{1}}^{c}=\mathbf{1}_{\breve{T}%
_{1}^{c}}$. \ Given this, the prodsum analog of \ $\breve{S}\cup \breve{T}%
^{c}$ \ is \ 
\begin{eqnarray}
\breve{h}_{1}\vee \breve{h}_{2}^{c} &=&\breve{h}_{1}+1-\breve{h}_{2}-\breve{h%
}_{1}(1-\breve{h}_{2}) \\
&=&1-(1-\breve{h}_{1})\breve{h}_{2}=(\breve{h}_{1}^{c}\breve{h}_{2})^{c}.
\label{eq-Abbr2}
\end{eqnarray}%
Notations such as \ $\breve{h}^{c}=1-\breve{h}$, \ $p_{D}^{c}=1-p_{D}$ \
will be used extensively.

\subsection{The D-U/D p.g.fl.'s \label{A-Comp AA-UDFGFL}}

If \ $\breve{\Xi}_{k|k}\subseteq \mathbb{\breve{X}}_{0}$ \ is the
multitarget RFS at time \ $t_{k}$ \ with p.g.fl. \ $\breve{G}_{k|k}[\breve{h}%
]=\breve{G}_{\breve{\Xi}_{k|k}}[\breve{h}]$ \ then the corresponding
D-target and U-target RFSs are, respectively, \ $\breve{\Xi}\cap \mathbb{D}$
\ and \ $\breve{\Xi}\cap \mathbb{U}$. \ From Eqs. (\ref{eq-Censor-PGFL},\ref%
{eq-Abbr2}) we know that their respective p.g.fl.'s are 
\begin{eqnarray}
\breve{G}_{k|k}^{d}[\breve{h}] &=&\breve{G}_{\breve{\Xi}_{k|k}\cap \mathbb{D}%
}[\breve{h}]=\breve{G}_{k|k}[(\breve{h}^{c}\mathbf{1}_{\mathbb{D}})^{c}]
\label{eq-MM-1} \\
\breve{G}_{k|k}^{u}[\breve{h}] &=&\breve{G}_{\breve{\Xi}_{k|k}\cap \mathbb{U}%
}[\breve{h}]=\breve{G}_{k|k}[(\breve{h}^{c}\mathbf{1}_{\mathbb{U}})^{c}]
\label{eq-MM-2}
\end{eqnarray}%
where \ $\breve{G}_{k|k}[\breve{h}]$ \ is derived, using Eq. (\ref%
{eq-PGFL-B1}), from the $F$-functional of Eq. (\ref{eq-FF}), which, recall,
is:%
\begin{equation}
\breve{F}_{k}[g,\breve{h}]=e^{\kappa \lbrack g-1]}\,\breve{G}_{k-1|k-1}[%
\breve{h}p_{D}^{c}+\breve{h}|_{1}p_{D}L_{g}].
\end{equation}%
For the standard models, as given in Eq. (\ref{eq-Core-UD}), Eqs. (\ref%
{eq-MM-1},\ref{eq-MM-2}) are respectively derived from the $F$-functionals 
\begin{eqnarray}
\breve{F}_{k}^{d}[g,\breve{h}] &=&e^{\kappa \lbrack g-1]}\breve{G}%
_{k-1|k-1}[(\breve{h}^{c}\mathbf{1}_{\mathbb{D}})^{c}p_{D}^{c}+(\breve{h}^{c}%
\mathbf{1}_{\mathbb{D}})^{c}|_{1}p_{D}L_{g}]  \label{eq-LL-D} \\
\breve{F}_{k}^{u}[g,\breve{h}] &=&e^{\kappa \lbrack g-1]}\breve{G}%
_{k-1|k-1}[(\breve{h}^{c}\mathbf{1}_{\mathbb{U}})^{c}p_{D}^{c}+(\breve{h}^{c}%
\mathbf{1}_{\mathbb{U}})^{c}|_{1}p_{D}L_{g}]  \label{eq-LL-U}
\end{eqnarray}%
again using Eq. (\ref{eq-PGFL-B1}).

Eqs. (\ref{eq-LL-D},\ref{eq-LL-U}) can be simplified to

\begin{eqnarray}
\breve{F}_{k}^{d}[g,\breve{h}] &=&e^{\kappa \lbrack g-1]}\breve{G}_{k-1|k-1}[%
\breve{h}^{\mathbb{D}}p_{D}^{c}+\breve{h}_{1}p_{D}L_{g}]  \label{eq-Ffunc-D}
\\
\breve{F}_{k}^{u}[g,\breve{h}] &=&e^{\kappa \lbrack g-1]}\breve{G}_{k-1|k-1}[%
\breve{h}^{\mathbb{U}}p_{D}^{c}+p_{D}L_{g}]  \label{eq-Ffunc-U}
\end{eqnarray}%
where \ $\breve{h}|_{1}$ was defined in Eq. (\ref{eq-Test-3}) and \ $\breve{h%
}^{\mathbb{D}}$ \ and \ $\breve{h}^{\mathbb{U}}$ in Eqs. (\ref{eq-Test-5},%
\ref{eq-Test-6}).

For Eq. (\ref{eq-Ffunc-D}): \ From Eqs. (\ref{eq-Test-7},\ref{eq-Abbr2}) we
get \ $(\breve{h}^{c}\mathbf{1}_{\mathbb{D}})^{c}=\breve{h}^{\mathbb{D}}$ \
and hence \ $(\breve{h}^{c}\mathbf{1}_{\mathbb{D}})^{c}|_{1}=\breve{h}^{%
\mathbb{D}}|_{1}=\breve{h}|_{1}$. \ Thus 
\begin{equation}
(\breve{h}^{c}\mathbf{1}_{\mathbb{D}})^{c}p_{D}^{c}+(\breve{h}^{c}\mathbf{1}%
_{\mathbb{D}})^{c}|_{1}p_{D}L_{g}=\breve{h}^{\mathbb{D}}p_{D}^{c}+\breve{h}%
|_{1}p_{D}L_{g}
\end{equation}%
and so, as claimed, 
\begin{equation}
\breve{F}_{k}^{d}[g,\breve{h}]=e^{\kappa \lbrack g-1]}\breve{G}_{k-1|k-1}[%
\breve{h}^{\mathbb{D}}p_{D}^{c}+\breve{h}_{1}p_{D}L_{g}].
\end{equation}

For Eq. (\ref{eq-Ffunc-U}): \ Again from Eqs. (\ref{eq-Test-7},\ref{eq-Abbr2}%
) we get \ $(\breve{h}^{c}\mathbf{1}_{\mathbb{U}})^{c}=\breve{h}^{\mathbb{U}%
} $ \ and hence \ $(\breve{h}^{c}\mathbf{1}_{\mathbb{U}})^{c}|_{1}=\breve{h}%
^{\mathbb{U}}|_{1}=1$. \ Thus%
\begin{equation}
(\breve{h}^{c}\mathbf{1}_{\mathbb{U}})^{c}p_{D}^{c}+(\breve{h}^{c}\mathbf{1}%
_{\mathbb{U}})^{c}|_{1}p_{D}L_{g}=\breve{h}^{\mathbb{U}}p_{D}^{c}+p_{D}L_{g}
\end{equation}%
and so, as claimed, \ 
\begin{equation}
\breve{F}_{k}^{u}[g,\breve{h}]=e^{\kappa \lbrack g-1]}\breve{G}_{k-1|k-1}[%
\breve{h}^{\mathbb{U}}p_{D}^{c}+p_{D}L_{g}].
\end{equation}%
\ \ \ \ \ \ \ \ 

\subsection{U/D Approaches Compared \label{A-Comp AA-Comp}}

Assume that the clutter process \ $\kappa _{k}(Z)\propto \kappa _{k}^{Z}$ \
is Poisson with intensity function \ $\kappa _{k}(z)$. \ Consider the S-U/D
formula for the posterior p.g.fl. in terms of the predicted p.g.fl. \cite[%
Eq. (4)]{MahSensors1019Exact}: 
\begin{equation}
G_{k|k}[h]\propto \int f_{k}^{\ast }(Z_{k}|X)\,(hp_{D})^{X}\,\frac{\delta
G_{k|k-1}}{\delta X}[hp_{D}^{c}]\delta X.  \label{eq-Orig-1}
\end{equation}%
Here, \ $f_{k}^{\ast }(Z|X)=0$ \ if \ $|X|>|Z|$ \ and, if otherwise\ \cite[%
Eqs. (55,58)]{MahSensors1019Exact},\footnote{%
Thus the set integral \ $\int \cdot \delta X$ \ in Eq. (\ref{eq-Orig-1}) is
actually cardinality-bounded: \ $\int_{X:|X|\leq |Z_{k}|}\cdot \delta X$.\ } 
\begin{equation}
f_{k}^{\ast }(Z|X)\overset{_{\text{def.}}}{=}\left[ f_{k}(Z|X)\right]
_{p_{D}=1}=\kappa _{k}(Z)\sum_{\tau }\prod_{i=1}^{n}\frac{L_{z_{\tau
(i)}}(x_{i})}{\kappa _{k}(z_{\tau (i)})};  \label{eq-fStar}
\end{equation}%
where \ $f_{k}(Z|X)$ \ is defined below in Eq. (\ref{eq-MTA-2}); where \ $%
X=\{x_{1},...,x_{n}\}$ with \ $|X|=n$ \ and \ $Z=\{z_{1},...,z_{m}\}$ \ with
\ $|Z|=m$; and where the summation is taken over the set of one-to-one
functions \ $\tau :\{1,...,n\}\hookrightarrow \{1,...,m\}$ \ (i.e., \ $\tau
(i_{1})=\tau (i_{2})$ \ implies \ $i_{1}=i_{2}$).

Eq. (\ref{eq-Orig-1}) was originally derived as Eq. (4) of \cite%
{MahSensors1019Exact} using intricate statistical reasoning---see \cite[Sec.
6.5]{MahSensors1019Exact}.\footnote{\textit{Errata}: \ In the line following 
\cite[Eq. (141)]{MahSensors1019Exact}, the phrase \textquotedblleft
Substituting Eq. (69)...\textquotedblright\ should be \textquotedblleft
Substituting Eqs. (69,70)...\textquotedblright .} \ It cleanly separates the
detection process for targets from the measurement-generation process for
targets; and also cleanly separates detection from non-detection.

In Section \ref{A-Der AA-altF}, however, Eq. (\ref{eq-Orig-1}) is derived in
a more direct fashion using the \textquotedblleft
turn-the-crank\textquotedblright\ rules for the RFS functional derivative. \
The purpose of this derivation is to facilitate a comparison of the S-U/D
and D-U/D approaches by contrasting their respective p.g.fl.'s. \ 

\subsubsection{Direct Derivation of Eq. (\protect\ref{eq-Orig-1}) \label%
{A-Comp AA-Comp-AAA-NewDer}\ }

This derivation can be summarized as follows. \ From Eq. (\ref{eq-Std-B-0})
we know that the $F$-functional for the standard multitarget measurement
model is%
\begin{equation}
F_{k}[g,h]=e^{\kappa \lbrack g-1]}G_{k|k-1}[\widehat{hp_{D}^{c}}+\overbrace{%
hp_{D}}L_{g}].  \label{eq-TT}
\end{equation}%
Here, the notations \textquotedblleft $\,\widehat{--}\,$\textquotedblright\
and \textquotedblleft $\overbrace{--}$\textquotedblright\ \ respectively
highlight and distinguish the nondetection-functional $%
H_{k}^{u}[h]=hp_{D}^{c}$ \ and detection-functional $\ H_{k}^{d}[h]=hp_{D}$
\ parts of the total bifunctional measurement model \ $%
h(p_{D}^{c}+p_{D}L_{g})$. \ 

The derivation in Section \ref{A-Der AA-altF} is simply a specific
implementation of the p.g.fl. measurement-update formula, Eq. (\ref%
{eq-PGFL-B1}). \ Intuitively speaking, it \textquotedblleft
tracks\textquotedblright\ the \textquotedblleft algebraic
motions\textquotedblright\ of \ $\widehat{hp_{D}^{c}}$ \ and \ $\overbrace{%
hp_{D}}$ \ beginning with Eq. (\ref{eq-TT}) and ending with Eq. (\ref%
{eq-Orig-1}), the latter of which is, in \textquotedblleft algebraic
motion\textquotedblright\ notation, \ \ 
\begin{equation}
G_{k|k}[h]\propto \int f_{k}^{\ast }(Z_{k}|X)\,(\overbrace{hp_{D}})^{X}\,%
\frac{\delta G_{k|k-1}}{\delta X}[\widehat{hp_{D}^{c}}]\delta X,
\label{eq-Sensors-Tot}
\end{equation}%
and which verifies the statistically-derived S-U/D formula \cite[Eq. (4)]%
{MahSensors1019Exact}.

Suppose instead that the $F$-functional is\ \ 
\begin{equation}
F_{k}^{d}[g,h]=e^{\kappa \lbrack g-1]}G_{k|k-1}[\widehat{p_{D}^{c}}+%
\overbrace{hp_{D}}L_{g}].  \label{eq-Ffunc-D-new}
\end{equation}%
Then inspection of the derivation in Section \ref{A-Der AA-altF} immediately
leads to 
\begin{equation}
G_{k|k}^{d}[h]\propto \int f_{k}^{\ast }(Z_{k}|X)\,(\overbrace{hp_{D}})^{X}\,%
\frac{\delta G_{k|k-1}}{\delta X}[\widehat{p_{D}^{c}}]\delta X,
\label{eq-Sensors-D}
\end{equation}%
which verifies the statistically-derived S-U/D formula \cite[Eq. (5)]%
{MahSensors1019Exact} for the p.g.fl. of D-targets. \ 

Comparing the definition of a p.g.fl., Eq. (\ref{eq-PGFL}), with this
equation it is easily seen by inspection that the formula for the S-U/D
distribution of D-targets is:%
\begin{equation}
f_{k|k}^{d}(X|Z_{1:k})\propto f_{k}^{\ast }(Z_{k}|X)\,(\overbrace{p_{D}}%
)^{X}\,\frac{\delta G_{k|k-1}}{\delta X}[\widehat{p_{D}^{c}}],
\end{equation}%
which verifies the statistically-derived S-U/D formula \cite[Eq. (75)]%
{MahSensors1019Exact}. \ (The formula for the $F$-functional for D-targets,
Eq. (\ref{eq-Ffunc-D-new}), is a new result, however.)\ 

Similarly, suppose that \ \ 
\begin{equation}
F_{k}^{u}[g,h]=e^{\kappa \lbrack g-1]}G_{k|k-1}[\widehat{hp_{D}^{c}}+%
\overbrace{p_{D}}L_{g}].  \label{eq-Ffunc-U-new}
\end{equation}%
Then inspection of the derivation in Section \ref{A-Der AA-altF} immediately
leads to \ 
\begin{equation}
G_{k|k}^{u}[h]\propto \int f_{k}^{\ast }(Z_{k}|X)\,(\overbrace{p_{D}})^{X}\,%
\frac{\delta G_{k|k-1}}{\delta X}[\widehat{hp_{D}^{c}}]\delta X,
\label{eq-Sensors-U}
\end{equation}%
which verifies the statistically-derived S-U/D formula \cite[Eq. (5)]%
{MahSensors1019Exact} for the p.g.fl. of U-targets. \ \ 

Now apply \ $\delta /\delta Y$ \ to both sides of Eq. (\ref{eq-Sensors-U}):%
\begin{equation}
\frac{\delta G_{k|k}^{u}}{\delta Y}[h]\propto \int f_{k}^{\ast }(Z_{k}|X)\,(%
\overbrace{p_{D}})^{X}\,\frac{\delta }{\delta Y}\frac{\delta G_{k|k-1}}{%
\delta X}[\widehat{hp_{D}^{c}}]\delta X
\end{equation}%
and set \ $h=0$ \ to get, from Eq. (\ref{eq-Recover}),%
\begin{equation}
f_{k|k}^{u}(Y)=\frac{\delta G_{k|k}^{u}}{\delta Y}[0].
\end{equation}%
Then in Section \ref{A-Der-AA-UtargDistr} it is shown that%
\begin{equation}
f_{k|k}^{u}(X|Z_{1:k})\propto (\widehat{p_{D}^{c}})^{X}\int f_{k}^{\ast
}(Z_{k}|Y)\,(\overbrace{p_{D}})^{Y}\,f_{k|k-1}(Y\cup X)\delta Y,
\label{eq-Udistr}
\end{equation}%
which verifies the S-U/D formula \cite[Eq. (90)]{MahSensors1019Exact} for
the distribution of U-targets. \ (The formula for the $F$-functional for
U-targets, Eq. (\ref{eq-Ffunc-U-new}), is a new result.)\ 

\subsubsection{Comparison of S-U/D and D-U/D \label{A-Comp AA-Comp-AAA-Comp}}

The same argument can be directly applied to the D-U/D case. \ From Eq. (\ref%
{eq-Ffunc-D}) we know that the D-U/D $F$-functional for D-targets is \ \ 
\begin{equation}
\breve{F}_{k}^{d}[g,\breve{h}]=e^{\kappa \lbrack g-1]}\breve{G}_{k-1|k-1}[%
\widehat{\breve{h}^{\mathbb{D}}p_{D}^{c}}+\overbrace{\breve{h}|_{1}p_{D}}%
L_{g}]  \label{eq-Main-D-F}
\end{equation}%
where \ $\breve{h}^{\mathbb{D}}$ \ was defined in Eq. (\ref{eq-Test-5}). \
Inspection of the derivation in Section \ref{A-Der AA-altF} immediately
leads us to the following D-U/D formula for the p.g.fl. of D-targets: 
\begin{equation}
\breve{G}_{k|k}^{d}[\breve{h}]\propto \int \breve{f}_{k}^{\ast }(Z_{k}|%
\breve{X})\,(\overbrace{\breve{h}|_{1}p_{D}})^{\breve{X}}\,\frac{\delta 
\breve{G}_{k|k-1}}{\delta \breve{X}}[\widehat{\breve{h}^{\mathbb{D}}p_{D}^{c}%
}]\delta \breve{X}  \label{eq-Main-D}
\end{equation}%
where \ $\breve{f}_{k}^{\ast }(Z|\breve{X})\,$\ is identical to \ $%
f_{k}^{\ast }(Z|X)\,$\ as defined in \ Eq. (\ref{eq-fStar}) except that \ $X$
\ is replaced by \ $\breve{X}=\{(x_{1},o_{1}),...,(x_{n},o_{n})\}$ \ with \ $%
|\breve{X}|=n$: \ 
\begin{equation}
f_{k}^{\ast }(Z|\breve{X})=\kappa _{k}(Z)\sum_{\tau }\prod_{i=1}^{n}\frac{%
L_{z_{\tau (i)}}(x_{i})}{\kappa _{k}(z_{\tau (i)})}.  \label{eq-fstar}
\end{equation}%
\ 

Likewise, from From Eq. (\ref{eq-Ffunc-U}) the D-U/D $F$-functional for
U-targets is%
\begin{equation}
\breve{F}_{k}^{u}[g,\breve{h}]=e^{\kappa \lbrack g-1]}\breve{G}_{k-1|k-1}[%
\widehat{\breve{h}^{\mathbb{U}}p_{D}^{c}}+\overbrace{\breve{h}p_{D}}L_{g}]
\label{eq-Main-U-F}
\end{equation}%
where \ $\breve{h}^{\mathbb{U}}$ \ is as in Eq. (\ref{eq-Test-6}). \
Inspection of the derivation in Section \ref{A-Der AA-altF} immediately
leads us to the following D-U/D formula for the p.g.fl. of U-targets: \ 
\begin{equation}
\breve{G}_{k|k}^{u}[\breve{h}]\propto \int \breve{f}_{k}^{\ast }(Z_{k}|%
\breve{X})\,(\overbrace{\breve{h}p_{D}})^{\breve{X}}\,\frac{\delta \breve{G}%
_{k|k-1}}{\delta \breve{X}}[\widehat{\breve{h}^{\mathbb{U}}p_{D}^{c}}]\delta 
\breve{X}.  \label{eq-Main-U}
\end{equation}

For the sake of completeness, recall from Eq. (\ref{eq-FF}) that the $F$%
-functional for general D-U/D targets is \ 
\begin{equation}
\breve{F}_{k}[g,\breve{h}]=e^{\kappa \lbrack g-1]}\breve{G}_{k-1|k-1}[%
\widehat{\breve{h}p_{D}^{c}}+\overbrace{\breve{h}|_{1}p_{D}}L_{g}].
\label{eq-Main-UD-F}
\end{equation}%
Thus the p.g.fl. for general D-U/D targets is \ 
\begin{equation}
\breve{G}_{k|k}[\breve{h}]\propto \int \breve{f}_{k}^{\ast }(Z_{k}|\breve{X}%
)\,(\overbrace{\breve{h}p_{D}})^{\breve{X}}\,\frac{\delta \breve{G}_{k|k-1}}{%
\delta \breve{X}}[\widehat{\breve{h}|_{1}p_{D}^{c}}]\delta \breve{X}.
\label{eq-Main-UD}
\end{equation}

\subsection{Multitarget NUD-JTFs \label{A-Comp AA-NUD}}

The D-U/D single-step MRBF recursion was described in Eq. (\ref{eq-Multitarg}%
), along with the concept of a D-U/D multitarget NUD-JTF \ $\breve{f}%
_{k|k-1}(Z,\breve{X}|\breve{X}_{k-1})$. \ Unlike the earlier NUD-JTFs, the
formula for \ $\breve{f}_{k|k-1}(Z,\breve{X}|\breve{X}_{k-1})$ \ is not only
algebraically complicated but is a generalized function.\footnote{%
Dirac delta functions have a rigorous mathematical basis, using either L.
Schwartz's \textquotedblleft distribution\textquotedblright\ (linear
functional) theory \cite{Schwartz} or M. Lighthill's contemporaneous
\textquotedblleft generalized function\textquotedblright\ (equivalence
classes of function sequences) theory \cite{Lighthill}. \ Like physicists,
engineers employ a less formal and more intuitive approach: \ the Dirac
delta is the Radon-Nikod\'{y}m derivative of the Dirac measure \ $\Delta
_{x}(S)\overset{_{\text{def.}}}{=}\mathbf{1}_{S}(x)$---i.e., \ $\Delta
_{x}(S)=\int_{S}\delta _{x}(y)dy$. \ }\ \ This is because of the
complexifying effect of the factor \ $\breve{h}|_{1}$\ in the U/D-target $F$%
-functional, Eq. (\ref{eq-FF}):\footnote{%
See Footnote 18.} 
\begin{equation}
\breve{F}_{k}[g,\breve{h}]=e^{\kappa \lbrack g-1]}\,\breve{G}_{k-1|k-1}[%
\breve{h}p_{D}^{c}+\breve{h}|_{1}p_{D}L_{g}].
\end{equation}

However, if one assumes the aligned case, Eqs. (\ref{eq-Align-1}-\ref%
{eq-Align-3}), it is possible to derive a formula for \ $\breve{f}_{k|k-1}(Z,%
\breve{X}|\breve{X}_{k-1})$ \ similar in form to that for the standard-model
\ $f_{k}(Z|X)$, which is \cite[Eq. (12.140)]{Mah-Artech}:%
\begin{eqnarray}
&&f_{k}(Z|X)  \label{eq-MTA} \\
&=&e^{-\kappa _{k}[1]}\kappa _{k}^{Z}(1-p_{D})^{X}\sum_{\alpha
}\prod_{i:\alpha (i)>0}\frac{p_{D}(x_{i})\,L_{z_{\alpha (i)}}(x_{i})}{%
(1-p_{D}(x_{i}))\,\kappa _{k}(z_{\alpha (i)})}  \notag \\
&=&e^{-\kappa _{k}[1]}\kappa _{k}^{Z}\sum_{\alpha }\left( \prod_{i:\alpha
(i)=0}p_{D}^{c}(x_{i})\right) \left( \prod_{i:\alpha (i)>0}\frac{%
p_{D}(x_{i})\,L_{z_{\alpha (i)}}(x_{i})}{\kappa _{k}(z_{\alpha (i)})}\right)
\label{eq-MTA-2}
\end{eqnarray}%
where \ $X=\{x_{1},...,x_{n}\}$ \ with \ $|X|=n$; \ where \ $%
Z=\{z_{1},...,z_{m}\}$ \ with \ $|Z|=m$; and where \ $\alpha
:\{1,...,n\}\rightarrow \{0,1,...,m\}$ \ is a measurement-to-track
association\ (MTA)---i.e., \ $\alpha (i_{1})=\alpha (i_{2})>0$ \ implies \ $%
i_{1}=i_{2}$.

The demonstration of this claim has two parts.

\textit{Part 1}: In Section \ref{A-Der-AA-NUD1} it is shown that the partial
p.g.fl. of \ $\breve{f}_{k|k-1}(Z,\breve{X}|\breve{X}_{k-1})$ \ with respect
to \ $g$ \ is\footnote{%
Note that \ $\int \breve{L}_{\breve{x},1}(\breve{x}^{\prime })d\breve{x}=1$
\ for all \ $\breve{x}^{\prime }\in \mathbb{\breve{X}}_{0}$ \ and thus \ $%
\int \breve{G}_{k|k-1}[1,\breve{X}|\breve{X}_{k-1}]\delta \breve{X}=1$ \ for
all finite \ $\breve{X}_{k-1}\subseteq \mathbb{\breve{X}}_{0}$. \ Also note
that while Eq. (\ref{eq-Lg}) depends on the Dirac deltas \ $\delta _{\breve{x%
}}\,$ and \ $\delta _{\breve{x}}^{d}$, the MRBF in Eq. (\ref{eq-Multitarg})
does not. \ } \ 
\begin{eqnarray}
\breve{G}_{k|k-1}[g,\breve{X}|\breve{X}_{k-1}]\overset{_{\text{def.}}}{=}
&&\int g^{Z}\,\breve{f}_{k|k-1}(Z,\breve{X}|\breve{X}_{k-1})\delta Z \\
&=&\delta _{|\breve{X}|,|\breve{X}^{\prime }|}\,e^{\kappa \lbrack
g-1]}\sum_{\pi }\prod_{i=1}^{n}\breve{L}_{\breve{x}_{\pi (i)},g}(\breve{x}%
_{i}^{\prime }).  \label{eq-NUD1}
\end{eqnarray}%
Here, a) $\ \breve{X}=\{\breve{x}_{1},...,\breve{x}_{n}\}$ \ and \ $\breve{X}%
_{k-1}=\{\breve{x}_{1}^{\prime },...,\breve{x}_{n}^{\prime }\}$ \ with \ $|%
\breve{X}|=|\breve{X}_{n-1}|=n$ b) $\ \pi $ \ is a permutation on \ $1,...,n$%
; \ and c) if \ $\delta _{\breve{x}}\,,\delta _{\breve{x}}^{d}$ \ are as
defined in Eqs. (\ref{eq-DIff-3},\ref{eq-Diff-4}) then \ \ 
\begin{equation}
\breve{L}_{\breve{x},g}=\delta _{\breve{x}}\,p_{D}^{c}+\delta _{\breve{x}%
}^{d}\,p_{D}L_{g}.  \label{eq-Lg}
\end{equation}

\textit{Part 2}: \ Using Eq. (\ref{eq-NUD1}), it is shown in Section \ref%
{A-Der-AA-NUD2} that%
\begin{eqnarray}
&&\breve{f}_{k|k-1}(Z,\breve{X}|\breve{X}_{k-1})  \label{eq-NUD2} \\
&=&e^{-\kappa _{k}[1]}\kappa _{k}^{Z}\delta _{|\breve{X}|,|\breve{X}^{\prime
}|}\sum_{\pi }\sum_{\alpha }\left( \prod_{i:\alpha (i)=0}\left( \delta _{%
\breve{x}_{\pi (i)}}(\breve{x}_{i}^{\prime })\,p_{D}^{c}(\breve{x}%
_{i}^{\prime })\right) \right)  \notag \\
&&\cdot \left( \prod_{i:\alpha (i)>0}\frac{\delta _{\breve{x}_{\pi (i)}}^{d}(%
\breve{x}_{i}^{\prime })\,p_{D}(\breve{x}_{i}^{\prime })\,L_{z_{\alpha (i)}}(%
\breve{x}_{i}^{\prime })}{\kappa _{k}(z_{\alpha (i)})}\right) .  \notag
\end{eqnarray}

\section{The U/D-Parallelism Assumption \label{A-Indep} \ }

The U-birth model is not the only perplexing aspect of the PMBM formulation
of U/D targets. \ The PMBM filter propagates PMBM RFSs, which are
set-theoretic unions of multi-Bernoulli mixture RFSs (the \textquotedblleft
MBM\textquotedblright\ part of \textquotedblleft PMBM\textquotedblright
\textquotedblright ) and Poisson RFSs (the \textquotedblleft
P\textquotedblright\ part). \ It is claimed that the MBM part propagates the
D-targets whereas the Poisson part propagates the U-targets. \ This is what
was earlier described as \textquotedblleft
U/D-parallelism.\textquotedblright\ \ 

In the case of the third (\textquotedblleft hybrid
labeled-unlabeled\textquotedblright )\ version of the PMBM filter \cite[%
Sect. 4.5]{MahSensors1019Exact}, an even stronger claim was made via the
statistical independence formula \cite[Eq. (49)]{MahSensors1019Exact}%
\begin{equation}
f_{k|k}(X|Z_{1:k})=f_{k|k}^{d}(X^{d}|Z_{1:k})\,f_{k|k}^{u}(X^{u}|Z_{1:k}),
\label{eq-U/D-error}
\end{equation}%
where \ $X=X^{d}\uplus X^{u}$ (disjoint union); where \ $%
f_{k|k}^{d}(X^{u}|Z_{1:k})$ \ is the Poisson U-target distribution; and
where $\ f_{k|k}^{d}(X^{d}|Z_{1:k})$ \ is the labeled MBM (LMBM) D-target
distribution. \ This implies that D-targets and U-targets can be
independently \textquotedblleft ...propagated in parallel, in both cases by
carrying out a prediction step and an update step...\textquotedblright\
using separate filters.\ \ 

Eq. (\ref{eq-U/D-error}) was shown to be erroneous. \ For, in the very same
paper that Eq. (\ref{eq-U/D-error}) appeared in, it was (accurately) claimed
that \textquotedblleft ...the update step for \ $X^{d}$ \ [i.e., at time \ $%
t_{k}$] involves the prediction results for both \ $X^{d}$ \ and \ $X^{u}$
[i.e., at time \ $t_{k-1}$]...\textquotedblright\ \ But if so, the D-target
and U-target processes will be statistically correlated at time \ $t_{k}$. \ 

The questionable nature of U/D-parallelism can be further demonstrated using
the results of the previous sections. \ If it were true, one would expect
that, if \ $f_{k|k}(X|Z_{1:k})$ \ is a PMBM distribution (i.e., the
distribution of a PMBM RFS), then the corresponding S-U/D distribution \ $%
f_{k|k}^{u}(X|Z_{1:k})$ for U-targets\ should be Poisson. \ But this is not
necessarily the case, as was demonstrated by the counterexample in \cite[%
Sect. 5.9]{MahSensors1019Exact}.

That counterexample can be generalized as follows. \ On the one hand,
application of Eq. (\ref{eq-MR-U-1}) shows that if \ $%
G_{k|k-1}[h]=e^{D[h-1]} $ \ is Poisson then \ $G_{k|k}[h]$ \ is PMBM but not
Poisson, and yet (see Section \ref{A-Der-AA-Parallel-1}) $\ $%
\begin{equation}
G_{k|k}^{u}[h]=e^{D[p_{D}^{c}(h-1)]}  \label{eq-Parallel-1}
\end{equation}%
is Poisson, which supports the claim. \ But on the other hand, if \ $%
G_{k|k-1}[h]=1-q+q\,s[h]$ \ is Bernoulli then so are both \ 
\begin{equation}
G_{k|k}[h]\propto 1-q+q\,s[\hat{L}_{Z_{k}}h]\   \label{eq-Parallel-2}
\end{equation}%
(see Section \ref{A-Der-AA-Parallel-2}) and, this time contrary to claim, 
\begin{equation}
G_{k|k}^{u}[h]\propto 1-q+q\,s[\hat{L}_{Z_{k}}+p_{D}^{c}(h-1)]
\label{eq-Parallel-3}
\end{equation}
(see Section \ref{A-Der-AA-Parallel-3}), where \ $\hat{L}_{Z}$ \ is as
defined in Eqs. (\ref{eq-Bern-0},\ref{eq-Bern0a}).\footnote{%
Eq. (\ref{eq-Parallel-3}) is intuitively reasonable. \ If \ $p_{D}=1$ \ (all
targets are D-targets) then \ $G_{k|k}^{u}[h]=1=f_{k|k}^{u}(\emptyset
|Z_{1:k})$ \ (there are no U-targets). \ And if \ $p_{D}=0$ \ (i.e., all
targets are U-targets and \ $\hat{L}_{Z_{k}}=1$), then \ $G_{k|k}^{u}[h]=$\ $%
G_{k|k-1}h]$ \ since target detections are impossible.}

This discrepancy is even more pronounced in the D-U/D case. \ For,
application of Eq. (\ref{eq-MR-U-2}) shows that if \ $\breve{G}_{k|k-1}[%
\breve{h}]=e^{\breve{D}\breve{[}\breve{h}-1]}$ \ is Poisson then \ $\breve{G}%
_{k|k}^{u}[\breve{h}]$ \ is PMBM but, again contrary to claim, not Poisson.

\section{The S-U/D and D-U/D PHD\ Filters \label{A-PHD}}

The section is organized as follows: the PHD\ filter (Section \ref%
{A-PHD-AA-PHD}); the S-U/D PHD\ filter (Section \ref{A-PHD-AA-SUD}); the
D-U/D PHD\ filter (Section \ref{A-PHD-AA-DUD}); and estimation using the
D-U/D PHD\ filter (Section \ref{A-PHD-AA-DUDE}).

\subsection{The PHD\ Filter \label{A-PHD-AA-PHD}}

Let \ $f_{\Xi }(X)$ \ be the distribution of the RFS \ $\Xi \subseteq 
\mathbb{X}_{0}$. \ Its probability hypothesis density (PHD), a density
function \ $D_{\Xi }(x)\geq 0$, is a statistical first-order
factorial-moment of \ $f_{\Xi }(X)$ \ in the sense that%
\begin{equation}
D_{\Xi }(x)=\frac{\delta G_{\Xi }}{\delta x}[1]=\int f_{\Xi }(\{x\}\cup
X)\delta X.  \label{eq-PHD-basic}
\end{equation}%
It has the properties that a) \ $D_{\Xi }(x)$ \ is the target density at \ $%
x $ \ and b) $\ N_{\Xi }(S)=\int_{S}D_{\Xi }(x)dx$ \ is the expected number
of targets in the region \ $S\subseteq \mathbb{X}_{0}$. \ 

The PHD\ filter \cite[Sec. 15.3]{Mah-Artech}, \cite[Sec. 8.4]{Mah-Newbook},
employs the first-order Poisson approximation \ $f_{\Xi }(X)\cong e^{-D_{\Xi
}[1]}D_{\Xi }^{X}$. \ Its time-update and measurement-update equations are%
\begin{eqnarray}
D_{k|k-1}(x) &=&B_{k|k-1}(x)+D_{k-1|k-1}[p_{S}M_{x}]  \label{eq-PHD-P} \\
D_{k|k}(x) &\cong &\left( p_{D}^{c}(x)+\sum_{z\in Z_{k}}\frac{%
p_{D}(x)\,L_{z}(x)}{\kappa _{k}(z)+D_{k|k-1}[p_{D}L_{z}]}\right)
\,D_{k|k-1}(x).  \label{eq-PHD-B}
\end{eqnarray}%
Substituting Eq. (\ref{eq-PHD-P}) into Eq. (\ref{eq-PHD-B}) results in its
single-step version: 
\begin{eqnarray}
D_{k|k}(x) &\cong &p_{D}^{c}(x)\,\left(
B_{k|k-1}(x)+D_{k-1|k-1}[p_{S}M_{x}]\right)  \label{eq-PHD-SingleStep} \\
&&+\sum_{z\in Z_{k}}\frac{p_{D}(x)\,L_{z}(x)\,\left(
B_{k|k-1}(x)+D_{k-1|k-1}[p_{S}M_{x}]\right) }{\kappa
_{k}(z)+B_{k|k-1}[p_{D}L_{z}]+D_{k-1|k-1}[p_{S}M_{p_{D}L_{z}}]}.  \notag
\end{eqnarray}

The multitarget state is estimated by taking the nearest integer \ $\hat{n}%
_{k|k}$ \ of \ $N_{k|k}=\int D_{k|k}(x)dx$ and then determining (if it
exists) the finite subset \ 
\begin{equation}
\hat{X}_{k|k}=\{\hat{x}_{1},...,\hat{x}_{\hat{n}_{k|k}}\}\subseteq \mathbb{X}%
_{0}\   \label{eq-PHD-E}
\end{equation}%
such that \ $D_{k|k}(\hat{x}_{1})>...>D_{k|k}(\hat{x}_{\hat{n}_{k|k}})$ \
are the $\ \hat{n}_{k|k}$ \ largest maxima of \ $D_{k|k}(x)$.

\begin{remark}
\label{Rem-PHD}When the PHD filter was introduced in 2000, computing power
was primitive by current standards. \ Prior to the labeled RFS theoretical
and computational innovations introduced in 2011 and 2013 in \cite%
{Vo-ISSNIP12-Conjugate}, \cite{VoVoTSPconjugate}, heuristically labeled RFS
filters on \ $\mathbb{X}_{0}$ \ were necessary as a stopgap to avoid
computational intractability. \ In particular, the PHD filter is not
theoretically rigorous because it is based on the Poisson approximation and
Poisson RFSs are not LRFSs. \ Indeed, Poisson RFSs \textit{cannot model any
multitarget reality.\footnote{%
For more detail, see the first paragraphs of \cite[Sec. 3]%
{Mah-TRFS-arXiv2024} or \cite[Sec. III-A]{Mah20220arXiv-TRFS}.}} \
Nevertheless, PHD\ filters are still of interest in some practical
applications because of their algebraic simplicity, computational
efficiency, and surprisingly good multitarget tracking performance (given
the inaccuracy of the Poisson approximation).\footnote{%
The PHD\ filter was originally proposed as a target-cluster tracker, not a
multitarget tracker---see \cite[p. 566]{Mah-Artech}, Remark 27.} \ 
\end{remark}

\subsection{The S-U/D PHD\ Filter \label{A-PHD-AA-SUD}}

This is based on variations of the $F$-functional of Eq. (\ref{eq-Std-B-F}):
\ 
\begin{eqnarray}
F_{k}[g,h] &=&e^{\kappa
_{k}[g-1]}e^{B_{k|k-1}[h(p_{D}^{c}+p_{D}L_{g})-1]}%
\,G_{k-1}[p_{S}^{c}+p_{S}M_{h(p_{D}^{c}+p_{D}L_{g})}\,]\;\;\;\;
\label{eq-PHD-F1} \\
F_{k}^{d}[g,h] &=&e^{\kappa
_{k}[g-1]}e^{B_{k|k-1}[p_{D}^{c}+hp_{D}L_{g}-1]}%
\,G_{k-1}[p_{S}^{c}+p_{S}M_{p_{D}^{c}+hp_{D}L_{g}}\,]  \label{eq-PHD-F2} \\
F_{k}^{u}[g,h] &=&e^{\kappa
_{k}[g-1]}e^{B_{k|k-1}[hp_{D}^{c}+p_{D}L_{g}-1]}%
\,G_{k-1}[p_{S}^{c}+p_{S}M_{hp_{D}^{c}+p_{D}L_{g}}\,]  \label{eq-PHD-F3}
\end{eqnarray}%
where \ $G_{k-1}[h]\overset{_{\text{abbr.}}}{=}G_{k-1|k-1}[h]$. \ 

In the case of Eq. (\ref{eq-PHD-F1}), the corresponding PHD\ filter is
derived by combining Eq. (\ref{eq-PHD-basic}) and Eq. (\ref{eq-PGFL-B1}) as
follows, \ 
\begin{equation}
D_{k|k}(x)=\frac{\frac{\delta F_{k}}{^{\delta Z_{k}\delta x}}[0,1]}{\frac{%
\delta F_{k}}{^{\delta Z_{k}}}[0,1]},  \label{eq-PHD-SUD0}
\end{equation}%
and---as is demonstrated in Section \ref{A-Der-AA-Equiv}---results (as it
should) in the single-step PHD\ filter, Eq. (\ref{eq-PHD-SingleStep}).

In the case of Eqs. (\ref{eq-PHD-F2},\ref{eq-PHD-F3}), Eqs. (\ref%
{eq-PHD-basic},\ \ref{eq-PGFL-B1}) result (using the same approach as in
Section \ref{A-Der-AA-Equiv}) in the D-target and U-target recursions \ 
\begin{eqnarray}
D_{k}^{d}(x) &\cong &\sum_{z\in Z_{k}}\frac{p_{D}(x)\,L_{z}(x)\,\left(
B_{k|k-1}(x)+D_{k-1}[p_{S}M_{x}]\right) }{\kappa
_{k}(z)+B_{k|k-1}[p_{D}L_{z}]+D_{k-1}[p_{S}M_{p_{D}L_{z}}]}
\label{eq-PHD-SUD1} \\
D_{k}^{u}(x) &\cong &p_{D}^{c}(x)\,\left(
B_{k|k-1}(x)+D_{k-1}[p_{S}M_{x}]\right)  \label{eq-PHD-SUD2} \\
D_{k}(x) &=&D_{k}^{d}(x)+D_{k}^{u}(x)  \label{eq-PHD-SUD3} \\
D_{k-1}(x) &=&D_{k-1}^{d}(x)+D_{k-}^{u}(x)  \label{eq-PHD-SUD4}
\end{eqnarray}%
where \ $D_{k}^{d}(x)\overset{_{\text{abbr.}}}{=}D_{k|k}^{d}(x)$, \ $%
D_{k}^{u}(x)\overset{_{\text{abbr.}}}{=}D_{k|k}^{u}(x)$, and \ $D_{k}(x)%
\overset{_{\text{abbr.}}}{=}D_{k|k}(x)$ \ for \ $k\geq 1$. \ 

\subsection{The D-U/D PHD\ filter \label{A-PHD-AA-DUD}}

This is based on the $F$-functional of Eq. (\ref{eq-DD}):%
\begin{equation}
\breve{F}_{k}[g,\breve{h}]=e^{\kappa _{k}[g-1]}e^{\breve{B}_{k|k-1}[\breve{h}%
p_{D}^{c}+\breve{h}|_{1}p_{D}L_{g})-1]}\,\breve{G}_{k-1|k-1}[1+p_{S}\breve{M}%
_{\breve{h}p_{D}^{c}+\breve{h}|_{1}p_{D}L_{g}-1}].  \label{eq-PHD-F4}
\end{equation}%
Application of Eqs. (\ref{eq-PHD-basic},\ \ref{eq-PGFL-B1}) results in the
recursion formulas 
\begin{eqnarray}
\breve{D}_{k}^{d}(x) &\cong &p_{D}^{c}(x)\,\breve{D}_{k-1}^{d}[p_{S}M_{x}]
\label{eq-PHD-DUD1} \\
&&+\sum_{z\in Z_{k}}\frac{p_{D}(x)\,L_{z}(x)\left( B_{k|k-1}^{u}(x)+\breve{D}%
_{k-1}^{d/u}[p_{S}M_{x}]\right) }{\kappa _{k}(z)+B_{k|k-1}^{u}[p_{D}L_{z}]+%
\breve{D}_{k-1}^{d/u}[p_{S}M_{p_{D}L_{z}}]}  \notag \\
\breve{D}_{k}^{u}(x) &\cong &p_{D}^{c}(x)\,\left( B_{k|k-1}^{u}(x)+\breve{D}%
_{k-1}^{u}[p_{S}M_{x}]\right)  \label{eq-PHD-DUD2} \\
\breve{D}_{k}^{d/u}(x) &=&\breve{D}_{k}^{d}(x)+\breve{D}_{k}^{u}(x)
\label{eq-PHD-DUD3} \\
\breve{D}_{k-1}^{d/u}(x) &=&\breve{D}_{k-1}^{d}(x)+\breve{D}_{k-1}^{u}(x)
\label{eq-PHD-DUD4} \\
\breve{D}_{k}(x,o) &=&\delta _{o,1}\breve{D}_{k}^{d}(x)+\delta _{o,0}\breve{D%
}_{k}^{u}(x)  \label{eq-PHD-DUD5}
\end{eqnarray}%
where \ $\breve{D}_{k}^{d}(x)\overset{_{\text{abbr.}}}{=}\breve{D}_{k}(x,1)$%
, \ $\breve{D}_{k}^{u}(x)\overset{_{\text{abbr.}}}{=}\breve{D}_{k}(x,0)$, $%
\breve{D}_{k-1}^{d}(x)\overset{_{\text{abbr.}}}{=}\breve{D}_{k-1}(x,1)$, \ $%
\breve{D}_{k-1}^{u}(x)\overset{_{\text{abbr.}}}{=}\breve{D}_{k-1}(x,0)$, and
\ $B_{k|k-1}^{u}(x)\overset{_{\text{abbr.}}}{=}\breve{B}_{k|k-1}(x,0)$ \
where $\breve{B}_{k|k-1}(x,1)=0$ (because newly-appearing targets have not
been detected yet and thus are U-targets).

The derivation of these equations is algebraically intricate and can be
found in Section \ref{A-Der-AA-DUD}.\ 

\subsection{Estimation Using the D-U/D PHD\ filter \label{A-PHD-AA-DUDE}}

Perhaps the most interesting aspect of the S-U/D and D-U/D PHD\ filters is
that they both seem to allow the number and states of the U-targets to be
easily estimated.

For the S-U/D PHD\ filter,this is accomplished by applying the PHD state
estimator in Eq. (\ref{eq-PHD-E}) to \ $D_{k}^{u}(x)$ \ in Eq. (\ref%
{eq-PHD-SUD2}):%
\begin{eqnarray}
D_{k}^{u}(x) &\cong &p_{D}^{c}(x)\,\left(
B_{k|k-1}(x)+D_{k-1}[p_{S}M_{x}]\right) \\
&=&p_{D}^{c}(x)\,\left(
B_{k|k-1}(x)+D_{k-1}^{d}[p_{S}M_{x}]+D_{k-1}^{u}[p_{S}M_{x}]\right) .
\label{eq-PHD-E2}
\end{eqnarray}

For the D-U/D PHD filter, it is accomplished by applying the PHD\ state
estimator to the D-U/D PHD \ $\breve{D}_{k}(x,o)$ \ in Eqs. (\ref%
{eq-PHD-DUD1}-\ref{eq-PHD-DUD5}). \ That is, take the nearest integer \ $%
\hat{n}_{k}$ \ of \ $\breve{N}_{k}=\int \breve{D}_{k}(\breve{x})d\breve{x}$
\ and determine 
\begin{equation}
\hat{X}_{k}=\{(\hat{x}_{1},\hat{o}_{1}),...,(\hat{x}_{\hat{n}_{k}},\hat{o}_{%
\hat{n}_{k}})\}\subseteq \mathbb{\breve{X}}_{0}
\end{equation}%
such that \ $\breve{D}_{k}(\hat{x}_{1},\hat{o}_{1})>...>\breve{D}_{k}(\hat{x}%
_{\hat{n}_{k}},\hat{o}_{\hat{n}_{k}})$ \ are the $\ \hat{n}_{k}$ \ largest
maxima of \ $\breve{D}_{k}(x,o)$. \ Note that \ $\breve{x}_{i}=(\hat{x}_{i},%
\hat{o}_{i})$ \ is the state of a D-target if \ $\hat{o}_{i}=1$ \ and a
U-target if $\hat{o}_{i}=0$.

\section{Mathematical Derivations \label{A-Der}}

\subsection{Derivation of Eqs. (\protect\ref{eq-Diff-1},\protect\ref%
{eq-Diff-2}) \label{A-Der-AA-Diff}}

The Dirac delta function \ $\delta _{\breve{x}}(\breve{x}^{\prime })$\ was
defined in Eq. (\ref{eq-DIff-3}). \ So begin with \ $\breve{h}|_{1}$. \ From
Eq. (\ref{eq-FuncDer-2}) we have%
\begin{eqnarray}
&&\frac{\delta }{\delta (x_{1},o_{1})}\breve{h}|_{1}(x_{1}^{\prime
},o_{1}^{\prime }) \\
&=&\lim_{\varepsilon \searrow 0}\frac{\breve{h}|_{1}(x_{1}^{\prime
},o_{1}^{\prime })+\varepsilon \,\left( \delta _{(x_{1},o_{1})}\right)
|_{1}(x_{1}^{\prime },o_{1}^{\prime })-\breve{h}|_{1}(x_{1}^{\prime
},o_{1}^{\prime })}{\varepsilon }  \notag
\end{eqnarray}%
\begin{eqnarray}
&=&\lim_{\varepsilon \searrow 0}\frac{\breve{h}(x_{1}^{\prime
},1)+\varepsilon \,\delta _{o_{1},1}\delta _{x_{1}}(x_{1}^{\prime })-\breve{h%
}(x_{1}^{\prime },1)}{\varepsilon } \\
&=&\delta _{o_{1},1}\,\delta _{x_{1}}(x_{1}^{\prime })\overset{_{\text{def.}}%
}{=}\delta _{\breve{x}_{1}}^{d}(\breve{x}_{1}^{\prime }).
\end{eqnarray}

Now consider \ $\breve{h}^{\mathbb{D}}$. \ Again from Eq. (\ref{eq-FuncDer-2}%
) we have \ \ \ 
\begin{equation}
\frac{\delta }{\delta \breve{x}_{1}}\breve{h}^{\mathbb{D}}=\lim_{\varepsilon
\searrow 0}\frac{(\breve{h}+\varepsilon \delta _{\breve{x}_{1}})^{\mathbb{D}%
}-\breve{h}^{\mathbb{D}}}{\varepsilon }
\end{equation}%
\begin{eqnarray}
&=&\lim_{\varepsilon \searrow 0}\frac{\mathbf{1}_{\mathbb{U}}+\mathbf{1}_{%
\mathbb{D}}(\breve{h}+\varepsilon \delta _{\breve{x}_{1}})-\mathbf{1}_{%
\mathbb{U}}-\mathbf{1}_{\mathbb{D}}\breve{h}}{\varepsilon } \\
&=&\mathbf{1}_{\mathbb{D}}\delta _{\breve{x}_{1}}.
\end{eqnarray}%
\ 

\subsection{Derivation of Eq. (\protect\ref{eq-Core-UD}) \label%
{A-Der-AA-&&&1}}

Two cases must be considered, \ $o_{k-1}=0$ \ and $o_{k-1}=1$.

\textit{Case 1} ($o_{k-1}=0$): \ By Eqs. (\ref{eq-AA},\ref{eq-BB}) we have 
\begin{eqnarray}
&&\breve{M}_{g,h}(x_{k-1},0) \\
&=&\int_{Z:|Z|\leq 1}\breve{h}(x,0)\,g^{Z}\,\breve{f}%
_{k|k-1}(Z,x,0|x_{k-1},0)\delta Zdx  \notag \\
&&+\int_{Z:|Z|\leq 1}\breve{h}(x,1)\,g^{Z}\,\breve{f}%
_{k|k-1}(Z,x,1|x_{k-1},0)\delta Zdx.  \notag
\end{eqnarray}%
\begin{eqnarray}
&=&\int \breve{h}(x,0)\,\,\breve{f}_{k|k-1}(\emptyset ,x,0|x_{k-1},0)dx \\
&&+\int \breve{h}(x,0)\,g(z)\,\breve{f}_{k|k-1}(\{z\},x,0|x_{k-1},0)dzdx 
\notag
\end{eqnarray}%
\begin{eqnarray}
&&+\int \breve{h}(x,1)\,\,\breve{f}_{k|k-1}(\emptyset ,x,1|x_{k-1},0)dx 
\notag \\
&&+\int \breve{h}(x,1)\,g(z)\,\breve{f}_{k|k-1}(\{z\},x,1|x_{k-1},0)dzdx 
\notag
\end{eqnarray}%
\begin{eqnarray}
&=&\int \breve{h}(x,0)\,\,(1-p_{D}(x))\,f_{k|k-1}(x|x_{k-1})dx \\
&&+0+0  \notag \\
&&+\int \breve{h}(x,1)\,g(z)\,p_{D}(x)\,f_{k}(z|x)\,f_{k|k-1}(x|x_{k-1})dzdx
\notag
\end{eqnarray}%
\begin{equation}
=M_{\breve{h}|_{0}(1-p_{D})}(x_{k-1})+M_{\breve{h}|_{1}p_{D}L_{g}}(x_{k-1})
\end{equation}%
and using Eq. (\ref{eq-Trans-3}),%
\begin{eqnarray}
&=&\breve{M}_{\breve{h}(1-p_{D})}(x_{k-1},0)+\breve{M}_{\breve{h}%
|_{1}p_{D}L_{g}}(x_{k-1},0) \\
&=&\breve{M}_{\breve{h}(1-p_{D})+\breve{h}|_{1}p_{D}L_{g}}(x_{k-1},0)
\end{eqnarray}%
as claimed.

\textit{Case 2} ($o_{k-1}=1$): \ Again by Eqs. (\ref{eq-AA},\ref{eq-BB}), we
have 
\begin{eqnarray}
&&\breve{M}_{g,\breve{h}}(x_{k-1},1) \\
&=&\int_{Z:|Z|\leq 1}\breve{h}(x,0)\,g^{Z}\,\breve{f}%
_{k|k-1}(Z,x,0|x_{k-1},1)\delta Zdx  \notag \\
&&+\int_{Z:|Z|\leq 1}\breve{h}(x,1)\,g^{Z}\,\breve{f}%
_{k|k-1}(Z,x,1|x_{k-1},1)\delta Zdx.  \notag
\end{eqnarray}%
\begin{equation}
=0+\int_{Z:|Z|\leq 1}\breve{h}(x,1)\,g^{Z}\,\breve{f}%
_{k|k-1}(Z,x,1|x_{k-1},1)\delta Zdx
\end{equation}%
\begin{eqnarray}
&=&\int \breve{h}|_{1}(x)\,\breve{f}_{k|k-1}(\emptyset ,x,1|x_{k-1},1)dx \\
&&+\int \breve{h}|_{1}(x)\,g(z)\,\breve{f}_{k|k-1}(\{z\},x,1|x_{k-1},1)dzdx.
\notag
\end{eqnarray}%
\ 
\begin{eqnarray}
&=&\int \breve{h}|_{1}(x)\,(1-p_{D}(x))\,f_{k|k-1}(x|x_{k-1})dx \\
&&+\int \breve{h}|_{1}(x)\,g(z)\,p_{D}(x)\,f_{k}(z|x)%
\,f_{k|k-1}(x|x_{k-1})dzdx  \notag
\end{eqnarray}%
and once again using Eq. (\ref{eq-Trans-3}),%
\begin{eqnarray}
&=&M_{\breve{h}|_{1}(1-p_{D})}(x_{k-1})+M_{\breve{h}|_{1}p_{D}L_{g}}(x_{k-1})
\\
&=&M_{\breve{h}(1-p_{D})}(x_{k-1},1)+M_{\breve{h}|_{1}p_{D}L_{g}}(x_{k-1},1)
\\
&=&\breve{M}_{\breve{h}(1-p_{D})+\breve{h}|_{1}p_{D}L_{g}}(x_{k-1},1)
\end{eqnarray}%
as claimed.

\subsection{Derivation of Eq. (\protect\ref{eq-FF}) \label{A-Der-AA-&&&3}}

First apply Eqs. (\ref{eq-Align-1},\ref{eq-Align-2}) to Eq. (\ref{eq-DD}),
in which case the D-U/D $F$-functional becomes:%
\begin{equation}
\breve{F}_{k}[g,\breve{h}]=e^{\kappa \lbrack g-1]}\,\breve{G}_{k-1|k-1}[%
\breve{M}_{\breve{h}p_{D}^{c}+\breve{h}|_{1}p_{D}L_{g}}].  \label{eq-EE}
\end{equation}%
Now apply Eq. (\ref{eq-Align-3}) to Eq. (\ref{eq-Core-UD}) to get:%
\begin{eqnarray}
\breve{M}_{g,\breve{h}}(x_{k-1},o_{k-1}) &=&\int \breve{h}%
(x,o_{k-1})\,p_{D}^{c}(x)\,f_{k|k-1}(x|x_{k-1})dx \\
&&+\int \breve{h}|_{1}(x)\,\,p_{D}(x)\,L_{g}(x)\,f_{k|k-1}(x|x_{k-1})dx 
\notag
\end{eqnarray}%
\begin{eqnarray}
&=&\int \breve{h}(x,o_{k-1})\,p_{D}^{c}(x)\,\delta _{x_{k-1}}(x)dx \\
&&+\int \breve{h}|_{1}(x)\,\,p_{D}(x)\,L_{g}(x)\,\delta _{x_{k-1}}(x)dx 
\notag
\end{eqnarray}%
\begin{equation}
=\breve{h}(x_{k-1},o_{k-1})\,p_{D}^{c}(x_{k-1})+\breve{h}|_{1}(x_{k-1})\,%
\,p_{D}(x_{k-1})\,L_{g}(x_{k-1})
\end{equation}%
and thus%
\begin{equation}
\breve{M}_{g,\breve{h}}=\breve{h}p_{D}^{c}+\breve{h}|_{1}p_{D}L_{g}.
\label{eq-ZZ}
\end{equation}

Eq. (\ref{eq-EE}) then, as claimed, becomes \ 
\begin{equation}
\breve{F}_{k}[g,\breve{h}]=e^{\kappa \lbrack g-1]}\,\breve{G}_{k-1|k-1}[%
\breve{h}p_{D}^{c}+\breve{h}|_{1}p_{D}L_{g}]
\end{equation}%
where%
\begin{equation}
\breve{G}_{k-1|k-1}[\breve{h}]=\breve{G}_{k|k-1}[\breve{h}].
\end{equation}

\subsection{Functional-Derivative Derivation of $G_{k|k}[h]$ \ \label{A-Der
AA-altF}}

As previously noted, the formula \cite[Eq. (4)]{MahSensors1019Exact} was
derived using statistical methods. \ Here we present a more direct, purely
algebraic derivation using the RFS functional calculus \cite[Chpt. 11]%
{Mah-Artech}. \ The derivation is divided into two parts: \ without clutter
(Section \ref{A-Der AA-altF-AA-Without}) and with clutter (Section \ref%
{A-Der AA-altF-AA-With}).\ 

\subsubsection{Case 1: \ Without clutter \label{A-Der AA-altF-AA-Without}}

In this case, from Eq. (\ref{eq-PGFL-B2}) the $F$-functional is%
\begin{equation}
F_{k}[g,h]=G_{k|k-1}[\widehat{hp_{D}^{c}}+\overbrace{hp_{D}}L_{g}]
\end{equation}%
where here the notations \textquotedblleft $\widehat{--}$\textquotedblright\
and \textquotedblleft $\overbrace{--}$\textquotedblright\ highlight and
distinguish the non-detection $hp_{D}^{c}$ \ and detection \ $hp_{D}$ \
parts of the total single-target bifunctional measurement model \ $%
h(p_{D}^{c}+p_{D}L_{g})$. \ 

First apply the chain rule \cite[Eq. (11.285)]{Mah-Artech} for the
functional derivative at \ $z_{1}\in \mathbb{Z}$ \ with respect to the
variable \ \ $g$ \ in \ $F_{k}[g,h]$: 
\begin{eqnarray}
\frac{\delta F_{k}}{\delta z_{1}}[g,h] &=&\int_{\mathbb{X}_{0}}\frac{\delta 
}{\delta z_{1}}(hp_{D}^{c}+hp_{D}L_{g})(y)\,\frac{\delta G_{k|k-1}}{\delta y}%
[\widehat{hp_{D}^{c}}+\overbrace{hp_{D}}L_{g}]dy \\
&=&\int_{\mathbb{X}_{0}}(hp_{D}L_{z_{1}})^{\{y\}}\,\frac{\delta G_{k|k-1}}{%
\delta y}[\widehat{hp_{D}^{c}}+\overbrace{hp_{D}}L_{g}]dy
\end{eqnarray}%
From Eq. (\ref{eq-PGFL-B1}) this results, after setting \ $g=0$, 
\begin{equation}
G_{k|k}[h]\propto \frac{\delta F_{k}}{\delta z_{1}}[0,h]=\int_{\mathbb{X}%
_{0}}L_{z_{1}}^{\{y_{1}\}}(\overbrace{hp_{D}})^{\{y_{1}\}}\,\frac{\delta
G_{k|k-1}}{\delta y_{1}}[\widehat{hp_{D}^{c}}]dy_{1}.
\end{equation}%
Apply the chain rule again at\ \ $z_{2}$ \ and set \ $g=0$: 
\begin{equation}
\frac{\delta ^{2}F_{k}}{\delta z_{1}\delta z_{2}}[0,h]=\int_{\mathbb{X}%
_{0}^{2}}L_{z_{1}}^{\{y_{1}\}}L_{z_{1}}^{\{y_{2}\}}(\overbrace{hp_{D}}%
)^{\{y_{1},y_{3}\}}\frac{\delta ^{2}G_{k|k-1}}{\delta y_{1}\delta y_{2}}[%
\widehat{hp_{D}^{c}}]dy_{1}dy_{1}
\end{equation}%
It is inductively clear by inspection that for general measurement-sets \ $%
Z=\{z_{1},...,z_{m}\}\subseteq \mathbb{Z}$ \ with \ $m=|Z|$ \ we get:%
\begin{equation}
\frac{\delta F_{k}}{\delta Z}[0,h]=\int_{\mathbb{X}_{0}^{m}}\left(
\prod_{j=1}^{m}L_{z_{j}}^{\{y_{j}\}}\right) (\overbrace{hp_{D}}%
)^{\{y_{1},...,y_{m}\}}\frac{\delta ^{m}G_{k|k-1}}{\delta y_{1}\cdots \delta
y_{m}}[\widehat{hp_{D}^{c}}]dy_{1}\cdots dy_{m}.
\end{equation}%
Note that this formula is valid regardless of the ordering of \ $%
z_{1},...,z_{m}$, \ so that%
\begin{equation}
m!\frac{\delta F_{k}}{\delta Z}[0,h]=\int_{\mathbb{X}_{0}^{m}}\left(
\sum_{\pi }\prod_{j=1}^{m}L_{z_{\pi (j)}}^{\{y_{j}\}}\right) (\overbrace{%
hp_{D}})^{\{y_{1},...,y_{m}\}}\frac{\delta ^{m}G_{k|k-1}}{\delta y_{1}\cdots
\delta y_{m}}[\widehat{hp_{D}^{c}}]dy_{1}\cdots dy_{m}
\end{equation}%
where \ $\pi $ \ denotes the \ $m!$ \ permutations on \ $1,...,m$. \ 

Multiplying both sides by \ $\kappa _{k}(Z)/m!$ \ we get:%
\begin{eqnarray}
&&\kappa _{k}(Z)\frac{\delta F_{k}}{\delta Z}[0,h] \\
&=&\frac{\kappa _{k}(Z)}{m!}\int_{\mathbb{X}_{0}^{m}}\left( \sum_{\pi
}\prod_{j=1}^{m}\frac{L_{z_{\pi (j)}}(y_{j})}{\kappa (z_{\pi (j)})}\right)
\,(\overbrace{hp_{D}})^{\{y_{1},...,y_{m}\}}  \notag
\end{eqnarray}%
\begin{eqnarray}
&&\cdot \frac{\delta ^{m}G_{k|k-1}}{\delta y_{1}\cdots \delta y_{m}}[%
\widehat{hp_{D}^{c}}]dy_{1}\cdots dy_{m}  \notag \\
&\propto &\int \delta _{|Z|,|Y|}\,\hat{f}_{k}^{\ast }(Z|Y)\,(\overbrace{%
hp_{D}})^{Y}\,\frac{\delta G_{k|k-1}}{\delta Y}[\widehat{hp_{D}^{c}}]\delta Y
\end{eqnarray}%
where \ 
\begin{equation}
\hat{f}_{k}^{\ast }(Z|Y)=\kappa _{k}(Z)\sum_{\pi }\prod_{j=1}^{m}\frac{%
L_{z_{\pi (j)}}(y_{k})}{\kappa (z_{\pi (j)})}.  \label{eq-fStar-2}
\end{equation}

\subsubsection{Case 2: \ With clutter \label{A-Der AA-altF-AA-With}}

In this case, from Eq. (\ref{eq-PGFL-B2}) the $F$-functional is

\begin{equation}
F_{k}[g,h]=e^{\kappa \lbrack g-1]}G_{k|k-1}[\widehat{hp_{D}^{c}}+\overbrace{%
hp_{D}}L_{g}].
\end{equation}%
According to the product rule \cite[Eq. (11.253)]{Mah-Artech} for the
functional derivative with respect to the variable \ $g$, 
\begin{equation}
\frac{\delta F_{k}}{\delta Z}[g,h]=\sum_{W\subseteq Z}\frac{\delta }{\delta
(Z-W)}e^{\kappa \lbrack g-1]}\cdot \frac{\delta }{\delta W}G_{k|k-1}[%
\widehat{hp_{D}^{c}}+\overbrace{hp_{D}}L_{g}]
\end{equation}%
\begin{eqnarray}
&=&e^{\kappa \lbrack g-1]}\sum_{W\subseteq Z}\kappa ^{Z-W}\,\frac{\delta }{%
\delta W}G_{k|k-1}[\widehat{hp_{D}^{c}}+\overbrace{hp_{D}}L_{g}] \\
&=&\kappa ^{Z}e^{\kappa \lbrack g-1]}\sum_{W\subseteq Z}\frac{1}{\kappa ^{W}}%
\,\frac{\delta }{\delta W}G_{k|k-1}[\widehat{hp_{D}^{c}}+\overbrace{hp_{D}}%
L_{g}].
\end{eqnarray}%
Hence by Eq. (\ref{eq-PGFL-B1})%
\begin{eqnarray}
&&G_{k|k}[h] \\
&\propto &\frac{\delta F_{k}}{\delta Z}[0,h]  \notag \\
&=&\kappa ^{Z}e^{-\kappa \lbrack 1]}\sum_{W\subseteq Z}\int \delta
_{|W|,|Y|}\,\hat{f}_{k}^{\ast }(W|Y)\,(\overbrace{hp_{D}})^{Y}\,\frac{\delta
G_{k|k-1}}{\delta Y}[\widehat{hp_{D}^{c}}]\delta Y
\end{eqnarray}%
\begin{eqnarray}
&=&\kappa ^{Z}e^{-\kappa \lbrack 1]}\int \left( \sum_{W:W\subseteq Z}\delta
_{|W|,|Y|}\hat{f}_{k}^{\ast }(W|Y)\right) (\overbrace{hp_{D}})^{Y}\frac{%
\delta G_{k|k-1}}{\delta Y}[\widehat{hp_{D}^{c}}]\delta Y\;\; \\
&=&\kappa ^{Z}e^{-\kappa \lbrack 1]}\int f_{k}^{\ast }(Z|Y)\,(\overbrace{%
hp_{D}})^{Y}\,\frac{\delta G_{k|k-1}}{\delta Y}[\widehat{hp_{D}^{c}}]\delta Y
\end{eqnarray}%
where%
\begin{equation}
f_{k}^{\ast }(Z|Y)=\sum_{W:W\subseteq Z}\delta _{|W|,|Y|}\,\hat{f}_{k}^{\ast
}(W|Y)=\sum_{W:W\subseteq Z,|W|=|Y|}\,\hat{f}_{k}^{\ast }(W|Y)
\end{equation}%
and where \ $\,\hat{f}_{k}^{\ast }(W|Y)$ \ was defined in Eq. (\ref%
{eq-fStar-2}). \ This summation is equivalent to a summation over all pairs
\ $(V,\tau )$ \ such that \ $V\subseteq Y$ \ with \ $|V|\leq |Z|$\ \ and
where \ $\tau :V\rightarrow Z$ \ is a one-to-one function.\ That is, \ $%
f_{k}^{\ast }(Z|Y)$ \ is the same as defined in both \cite[Eq. (55)]%
{MahSensors1019Exact} and Eq. (\ref{eq-fStar}). \ 

It follows that%
\begin{equation}
G_{k|k}[h]\propto \int f_{k}^{\ast }(Z|Y)\,(hp_{D})^{Y}\,\frac{\delta
G_{k|k-1}}{\delta Y}[hp_{D}^{c}]\delta Y
\end{equation}%
which, as claimed, verifies \cite[Eq. (4)]{MahSensors1019Exact}.

\subsection{Derivation of Eq. (\protect\ref{eq-Udistr}) \label%
{A-Der-AA-UtargDistr}}

Taking the functional derivative \ $\delta /\delta Y$ \ of both sides of Eq.
(\ref{eq-Sensors-U}) yields%
\begin{equation}
\frac{\delta G_{k|k}^{u}}{\delta Y}[h]\propto \int f_{k}^{\ast }(Z_{k}|X)\,(%
\overbrace{p_{D}})^{X}\,\frac{\delta }{\delta Y}\frac{\delta G_{k|k-1}}{%
\delta X}[\widehat{hp_{D}^{c}}]\delta X  \label{eq-101}
\end{equation}
\begin{eqnarray}
&=&\int f_{k}^{\ast }(Z_{k}|X)\,(\overbrace{p_{D}})^{X}\,(\widehat{p_{D}^{c}}%
)^{Y}\frac{\delta G_{k|k-1}}{\delta (Y\cup X)}[\widehat{hp_{D}^{c}}]\delta X
\label{eq-102} \\
&=&(\widehat{p_{D}^{c}})^{Y}\int f_{k}^{\ast }(Z_{k}|X)\,(\overbrace{p_{D}}%
)^{X}\int (\widehat{hp_{D}^{c}})^{U}\,f_{k|k-1}(U\cup Y\cup X)\delta X\delta
U  \label{eq-103}
\end{eqnarray}%
where Eq. (\ref{eq-103}) is due to the RFS Radon-Nikod\'{y}m formula \cite[%
Eq. (11.251)]{Mah-Artech} and Eq. (\ref{eq-102}) follows from%
\begin{equation}
\frac{\delta }{\delta X}G[qh]=q^{X}\,\frac{\delta G}{\delta X}[qh],
\end{equation}%
where \ $0\leq q(x)\leq 1$ \ is a fixed unitless function. \ Setting \ $h=0$
\ and using the definition of the set integral, Eq. (\ref{eq-SetInt}), we
get, as claimed,%
\begin{eqnarray}
f_{k|k}^{u}(Y) &=&\frac{\delta G_{k|k}^{u}}{\delta Y}[0] \\
&\propto &(\widehat{p_{D}^{c}})^{Y}\int f_{k}^{\ast }(Z_{k}|X)\,(\overbrace{%
p_{D}})^{X}\,f_{k|k-1}(Y\cup X)\delta X.
\end{eqnarray}

\subsection{Derivation of Eq. (\protect\ref{eq-NUD1}) \label{A-Der-AA-NUD1}}

We first sketch the derivation. \ 

The partial p.g.fl. with respect to \ $Z$ \ of \ $\breve{f}_{k|k-1}(Z_{k},%
\breve{X}|\breve{X}_{k-1})$ \ is 
\begin{equation}
\breve{G}_{k|k-1}[g,\breve{X}|\breve{X}_{k-1}]=\int g^{Z}\,\breve{f}%
_{k|k-1}(Z,\breve{X}|\breve{X}_{k-1})\delta Z.
\end{equation}%
From Eq. (\ref{eq-Multitarg}) we get%
\begin{eqnarray}
&&\breve{f}_{k|k}(\breve{X}|Z_{1:k}) \\
&\propto &\int \breve{f}_{k|k-1}(Z_{k},\breve{X}|\breve{X}_{k-1})\,\breve{f}%
_{k-1|k-1}(\breve{X}_{k-1}|Z_{1:k-1})\delta \breve{X}_{k-1}  \notag
\end{eqnarray}%
\begin{eqnarray}
&=&\int \frac{\delta \breve{G}_{k|k-1}}{\delta Z_{k}}[0,\breve{X}|\breve{X}%
_{k-1}]\,\breve{f}_{k-1|k-1}(\breve{X}_{k-1}|Z_{1:k-1})\delta \breve{X}_{k-1}
\\
&=&\int \left[ \frac{\delta }{\delta Z_{k}}\breve{G}_{k|k-1}[g,\breve{X}|%
\breve{X}_{k-1}]\right] _{g=0}\,\breve{f}_{k-1|k-1}(\breve{X}%
_{k-1}|Z_{1:k-1})\delta \breve{X}_{k-1}.  \label{eq-RR}
\end{eqnarray}%
\ It will be shown that%
\begin{eqnarray}
&&\breve{f}_{k|k}(\breve{X}|Z_{1:k}) \\
&\propto &\frac{\delta \breve{F}_{k}}{\delta \breve{X}\delta Z_{k}}[0,0] 
\notag \\
&=&\int \left[ \frac{\delta }{\delta Z_{k}}\left[ e^{\kappa \lbrack g-1]}%
\breve{L}_{\breve{X},g}(\breve{X}_{k-1})\right] \right] _{g=0}\,\breve{f}%
_{k|k-1}(\breve{X}_{k-1}|Z_{1:k-1})\delta \breve{X}_{k-1}  \label{eq-SS}
\end{eqnarray}%
where \ $\breve{F}_{k}[g,\breve{h}]$ \ was defined in Eq. (\ref{eq-Main-UD-F}%
),%
\begin{equation}
\breve{F}_{k}[g,\breve{h}]=e^{\kappa \lbrack g-1]}\breve{G}_{k|k-1}[\breve{h}%
p_{D}^{c}+\breve{h}|_{1}p_{D}L_{g}],
\end{equation}%
and where, for \ $\breve{L}_{\breve{x},g}(\breve{x}^{\prime })$ \ as defined
in Eq. (\ref{eq-Lg}), \ 
\begin{equation}
\breve{L}_{\breve{X},g}(\{\breve{x}_{1},...,\breve{x}_{n}\})=\delta _{|%
\breve{X}|,|\breve{X}^{\prime }|}e^{\kappa \lbrack g-1]}\sum_{\pi }\breve{L}%
_{\breve{x}_{\pi (1)},g}(\breve{x}_{1}^{\prime })\cdots \breve{L}_{\breve{x}%
_{\pi (n)},g}(\breve{x}_{n}^{\prime }).
\end{equation}%
Since the right sides of Eqs. (\ref{eq-RR},\ref{eq-SS}) are proportional to
each other for all \ $\breve{f}_{k|k-1}(\breve{X}_{k-1}|Z_{1:k-1})$, \ we
can conclude that%
\begin{equation}
\breve{G}_{k|k-1}[g,\breve{X}|\breve{X}_{k-1}]\propto e^{\kappa \lbrack
g-1]}\,\breve{L}_{\breve{X},g}(\breve{X}_{k-1})
\end{equation}%
and hence, since \ $\int \breve{G}_{k|k-1}[1,\breve{X}|\breve{X}%
_{k-1}]\delta \breve{X}=1$ and \ $\int \breve{L}_{\breve{X},1}(\breve{X}%
_{k-1})\delta \breve{X}=1$, we get the claimed result: 
\begin{equation}
\breve{G}_{k|k-1}[g,\breve{X}|\breve{X}_{k-1}]=e^{\kappa \lbrack g-1]}\,%
\breve{L}_{\breve{X},g}(\breve{X}_{k-1}).
\end{equation}

So we must verify Eq. (\ref{eq-SS}). \ First, note from Eq. (\ref{eq-Diff-1}%
) that 
\begin{equation}
\frac{\delta }{\delta \breve{X}}(\breve{h}p_{D}^{c}+\breve{h}%
|_{1}p_{D}L_{g})=\left\{ 
\begin{array}{ccc}
\breve{h}p_{D}^{c}+\breve{h}|_{1}p_{D}L_{g} & \text{if} & \breve{X}=\emptyset
\\ 
\breve{L}_{\breve{x}_{1},g} & \text{if} & \breve{X}=\{\breve{x}_{1}\} \\ 
0 & \text{if} & |\breve{X}|\geq 2%
\end{array}%
\right.  \label{eq-XX}
\end{equation}%
where%
\begin{equation}
\breve{L}_{\breve{x}_{1},g}=\delta _{\breve{x}_{1}}p_{D}^{c}+\delta _{\breve{%
x}_{1}}^{d}p_{D}L_{g}.
\end{equation}

Second, from Eq. (\ref{eq-Recover}) for \ $\breve{X}=\emptyset $ \ we have: 
\begin{equation}
e^{-\kappa \lbrack g-1]}\breve{F}_{k}[g,0]=\breve{G}_{k|k-1}[0]=\breve{f}%
_{k|k-1}(\emptyset ).
\end{equation}

Third, take the first functional derivative of $\ \breve{F}_{k}[g,\breve{h}]$
\ with respect to \ $\breve{h}$ \ and apply the fourth chain rule \cite[Eq.
(11.285)]{Mah-Artech}. Then:%
\begin{equation}
e^{-\kappa \lbrack g-1]}\frac{\delta \breve{F}_{k}}{\delta (x_{1},o_{1})}[g,%
\breve{h}]=\int \frac{\delta }{\delta \breve{x}_{1}}(\breve{h}p_{D}^{c}+%
\breve{h}|_{1}p_{D}L)^{\{\breve{x}_{1}^{\prime }\}}\,\frac{\delta \breve{G}%
_{k|k-1}}{\delta \breve{x}_{1}^{\prime }}[\breve{h}p_{D}^{c}+\breve{h}%
|_{1}p_{D}L_{g}]d\breve{x}_{1}^{\prime }
\end{equation}%
and thus from Eq. (\ref{eq-Recover}) and Eq. (\ref{eq-XX}) after setting \ $%
\breve{h}=0$, we get\footnote{%
Note that if it were the case that \ $\breve{L}_{\breve{x}_{1},g}=\delta _{%
\breve{x}_{1}}(p_{D}^{c}+p_{D}L_{g})$ \ then Eq. (\ref{eq-LLL}) would
greatly simplify to \ $(p_{D}^{c}+p_{D}L_{g})^{\{\breve{x}_{1}\}}\,\breve{f}%
_{k|k-1}(\{\breve{x}_{1}\})$. \ This illustrates the complexifying effect of
\ $\breve{h}|_{1}$.} 
\begin{equation}
e^{-\kappa \lbrack g-1]}\frac{\delta \breve{F}_{k}}{\delta (x_{1},o_{1})}%
[g,0]=\int \breve{L}_{\breve{x}_{1},g}(\breve{x}_{1}^{\prime })\,\breve{f}%
_{k|k-1}(\{\breve{x}_{1}^{\prime }\})d\breve{x}_{1}^{\prime }.
\label{eq-LLL}
\end{equation}

Fourth, for the second functional derivative we get%
\begin{equation}
e^{-\kappa \lbrack g-1]}\frac{\delta ^{2}\breve{F}_{k}}{\delta \breve{x}%
_{1}\delta \breve{x}_{2}}[g,0]=\int \breve{L}_{\breve{x}_{1},g}(\breve{x}%
_{1}^{\prime })\,\breve{L}_{\breve{x}_{2}^{\prime },g}(\breve{x}_{2}^{\prime
})\,\breve{f}_{k|k-1}(\{\breve{x}_{1}^{\prime },\breve{x}_{2}^{\prime }\})d%
\breve{x}_{1}^{\prime }d\breve{x}_{2}^{\prime }
\end{equation}

Fifth, by induction, the \ $n$'th functional derivative is 
\begin{eqnarray}
&e^{-\kappa \lbrack g-1]}\frac{\delta ^{n}\breve{F}_{k}}{\delta \breve{x}%
_{1}\cdots \delta \breve{x}_{n}}[g,0]=&\int \breve{L}_{\breve{x}_{1},g}(%
\breve{x}_{1}^{\prime })\cdots \breve{L}_{\breve{x}_{n},g}(\breve{x}%
_{n}^{\prime })\, \\
&&\cdot \breve{f}_{k|k-1}(\{\breve{x}_{1}^{\prime },...,\breve{x}%
_{n}^{\prime }\})d\breve{x}_{1}^{\prime }\cdots d\breve{x}_{n}^{\prime }. 
\notag
\end{eqnarray}

Sixth, since the left side of this is invariant with respect to the ordering
of the \ $\breve{x}_{1},...,\breve{x}_{n}$,%
\begin{eqnarray}
&&\frac{\delta ^{n}\breve{F}_{k}}{\delta \breve{x}_{1}\cdots \delta \breve{x}%
_{n}}[g,0] \\
&=&\frac{e^{\kappa \lbrack g-1]}}{n!}\int \sum_{\pi }\breve{L}_{\breve{x}%
_{\pi (1)},g}(\breve{x}_{1}^{\prime })\cdots \breve{L}_{\breve{x}_{\pi
(n)},g}(\breve{x}_{n}^{\prime })\,  \notag
\end{eqnarray}
\begin{eqnarray}
&&\cdot \breve{f}_{k|k-1}(\{\breve{x}_{1}^{\prime },...,\breve{x}%
_{n}^{\prime }\})d\breve{x}_{1}^{\prime }\cdots d\breve{x}_{n}^{\prime } 
\notag \\
&=&\frac{e^{\kappa \lbrack g-1]}}{n!}\int \breve{L}_{\{\breve{x}_{1},...,%
\breve{x}_{n}\},g}(\{\breve{x}_{1}^{\prime },...,\breve{x}_{n}^{\prime }\})\,%
\breve{f}_{k|k-1}(\{\breve{x}_{1}^{\prime },...,\breve{x}_{n}^{\prime }\})d%
\breve{x}_{1}^{\prime }\cdots d\breve{x}_{n}^{\prime }\;\;\;\;
\end{eqnarray}

Thus seventh, from Eq. (\ref{eq-SetInt}),%
\begin{equation}
\frac{\delta \breve{F}_{k}}{\delta \breve{X}}[g,0]=e^{\kappa \lbrack
g-1]}\int \breve{L}_{\breve{X},g}(\breve{X}^{\prime })\,\breve{f}_{k|k-1}(%
\breve{X}^{\prime }|Z_{1:k-1})\delta \breve{X}^{\prime }
\end{equation}%
where%
\begin{equation}
\breve{L}_{\{\breve{x}_{1},...,\breve{x}_{n}\},g}(\{\breve{x}_{1}^{\prime
},...,\breve{x}_{n}^{\prime }\})\overset{_{\text{def.}}}{=}\delta
_{|X|,|X|}\sum_{\pi }\breve{L}_{\breve{x}_{\pi (1)},g}(\breve{x}_{1}^{\prime
})\cdots \breve{L}_{\breve{x}_{\pi (n)},g}(\breve{x}_{n}^{\prime })
\end{equation}%
and so, as claimed, 
\begin{eqnarray}
&&\breve{f}_{k|k}(\breve{X}|Z_{1:k}) \\
&\propto &\frac{\delta \breve{F}_{k}}{\delta \breve{X}\delta Z_{k}}[0,0] 
\notag \\
&=&\int \left[ \frac{\delta }{\delta Z_{k}}\left[ e^{\kappa \lbrack g-1]}%
\breve{L}_{\breve{X},g}(\breve{X}_{k-1})\right] \right] _{g=0}\,\breve{f}%
_{k|k-1}(\breve{X}_{k-1}|Z_{1:k-1})\delta \breve{X}_{k-1}.  \label{eq-YY}
\end{eqnarray}

\subsection{Derivation of Eq. (\protect\ref{eq-NUD2}) \label{A-Der-AA-NUD2}}

The derivation is similar to that presented in \cite[Sec. 9]{Mah2017arXiv-v2}
(which is simpler than the original one in \cite[Sec. G.18]{Mah-Artech}). \
First note that 
\begin{equation}
\left[ \frac{\frac{\delta }{\delta W}\breve{L}_{\breve{x}_{\pi (i)},g}}{%
\breve{L}_{\breve{x}_{\pi (i)},g}}\right] _{g=0}=\left\{ 
\begin{array}{ccc}
1 & \text{if} & W=\emptyset \\ 
\frac{\delta _{\breve{x}_{\pi (i)}}^{d}\,p_{D}L_{z}}{\delta _{\breve{x}_{\pi
(i)}}\,p_{D}^{c}} & \text{if} & W=\{z\} \\ 
0 & \text{if} & |W|>2%
\end{array}%
\right. .  \label{eq-Temp1}
\end{equation}

Then take the functional derivative \ $\delta /\delta Z$ \ of both sides of
Eq. (\ref{eq-NUD1}) with respect to \ $g$ \ and use the general product rule 
\cite[Eq. (11.274)]{Mah-Artech} to get:%
\begin{eqnarray}
&&\frac{\delta \breve{G}_{k|k-1}}{\delta Z}[g,\breve{X}|\breve{X}_{k-1}]
\label{eq-4a} \\
&=&\delta _{|\breve{X}|,|\breve{X}^{\prime }|}\sum_{\pi }\sum_{\breve{W}%
_{0}\uplus \breve{W}_{1}\uplus ...\uplus \breve{W}_{n}=Z}\left( \frac{\delta 
}{\delta \breve{W}_{0}}e^{\kappa _{k}[g-1]}\right) \prod_{i=1}^{n}\left( 
\frac{\delta }{\delta \breve{W}_{i}}\breve{L}_{\breve{x}_{\pi (i)},g}(\breve{%
x}_{i}^{\prime })\right)  \notag
\end{eqnarray}%
where the summation is taken over all mutually disjoint and possibly empty \ 
$\breve{W}_{0},\breve{W}_{1},...,\breve{W}_{n}$ \ such that \ $\breve{W}%
_{0}\cup \breve{W}_{1}\cup ...\cup \breve{W}_{n}=Z$. \ Then \ \ 
\begin{eqnarray}
&&\frac{\delta \breve{G}_{k|k-1}}{\delta Z}[g,\breve{X}|\breve{X}_{k-1}] \\
&=&\delta _{|\breve{X}|,|\breve{X}^{\prime }|}\sum_{\pi }\sum_{\breve{W}%
_{0}\uplus \breve{W}_{1}\uplus ...\uplus \breve{W}_{n}=Z}e^{\kappa
_{k}[g-1]}\kappa _{k}^{Z-(\breve{W}_{1}\uplus ...\uplus \breve{W}%
_{n})}\prod_{i=1}^{n}\left( \frac{\delta }{\delta \breve{W}_{i}}\breve{L}_{%
\breve{x}_{\pi (i)},g}(\breve{x}_{i}^{\prime })\right)  \notag
\end{eqnarray}%
\begin{equation}
=\delta _{|\breve{X}|,|\breve{X}^{\prime }|}e^{\kappa _{k}[g-1]}\kappa
_{k}^{Z}\sum_{\pi }\sum_{\breve{W}_{0}\uplus \breve{W}_{1}\uplus ...\uplus 
\breve{W}_{n}=Z}\prod_{i=1}^{n}\left( \frac{1}{\kappa ^{_{k}W_{i}}}\frac{%
\delta }{\delta \breve{W}_{i}}\breve{L}_{\breve{x}_{\pi (i)},g}(\breve{x}%
_{i}^{\prime })\right)
\end{equation}%
\begin{eqnarray}
&=&\delta _{|\breve{X}|,|\breve{X}^{\prime }|}e^{\kappa _{k}[g-1]}\kappa
_{k}^{Z}\sum_{\pi }\left( \prod_{i=1}^{n}\breve{L}_{\breve{x}_{\pi (i)},g}(%
\breve{x}_{i}^{\prime })\right)  \label{eq-Temp} \\
&&\cdot \sum_{\breve{W}_{0}\uplus \breve{W}_{1}\uplus ...\uplus \breve{W}%
_{n}=Z}\prod_{i=1}^{n}\left( \frac{\frac{\delta }{\delta \breve{W}_{i}}%
\breve{L}_{\breve{x}_{\pi (i)},g}(\breve{x}_{i}^{\prime })}{\kappa
_{k}^{W_{i}}\,\breve{L}_{\breve{x}_{\pi (i)},g}(\breve{x}_{i}^{\prime })}%
\right) .  \notag
\end{eqnarray}%
Because of Eq. (\ref{eq-Temp1}), the only surviving terms in the summation
in Eq. (\ref{eq-Temp}) are those such that \ $W_{1},...,W_{n}$ \ are either
empty or singleton; and, moreover, \ $W_{i}$ \ contributes a factor to the
product only if it\ is a singleton. \ 

For a given choice of $\ W_{1},...,W_{n}$, \ define \ $\alpha
:\{1,...,n\}\rightarrow \{0,1,...,m\}$ \ by \ $\{\mathbf{z}_{\alpha
(i)}\}=W_{i}$ \ if \ $W_{i}\neq \emptyset $ \ and \ $\alpha (i)=0$ \
otherwise. \ Then \ $\alpha $ \ is an MTA in the sense of Eq. (\ref{eq-MTA}%
). \ Conversely, given an MTA \ $\alpha $ \ define \ $W_{i}=\{\mathbf{z}%
_{\alpha (i)}\}$ \ if \ $\alpha (i)>0$ \ and \ $W_{i}=\emptyset $ \ if
otherwise. \ Either way, $\ $%
\begin{equation}
W_{1}\uplus ...\uplus W_{n}=\{\mathbf{z}_{\alpha (i)}:\alpha (i)>0\}.
\end{equation}%
Thus there is a one-to-one correspondence between MTAs \ $\alpha $ \ and
lists \ $W_{1},...,W_{n}$ \ of mutually disjoint empty or singleton subsets
of \ $Z$. \ Furthermore, only those \ $i$'s with \ $\alpha (i)>0$ \
contribute a factor to the product in Eq. (\ref{eq-Temp}).

Consequently, setting \ $g=0$ \ in Eq. (\ref{eq-Temp}) and rearranging
factors using MTA\ notation, we get the claimed result:%
\begin{eqnarray}
&&\breve{f}_{k|k-1}(Z,\breve{X}|\breve{X}_{k-1}) \\
&=&e^{-\kappa _{k}[1]}\kappa _{k}^{Z}\delta _{|\breve{X}|,|\breve{X}^{\prime
}|}\sum_{\pi }\sum_{\alpha }\left( \prod_{i:\alpha (i)=0}\left( \delta _{%
\breve{x}_{\pi (i)}}(\breve{x}_{i}^{\prime })\,p_{D}^{c}(\breve{x}%
_{i}^{\prime })\right) \right)  \notag \\
&&\cdot \left( \prod_{i:\alpha (i)>0}\frac{\delta _{\breve{x}_{\pi (i)}}^{d}(%
\breve{x}_{i}^{\prime })\,p_{D}(\breve{x}_{i}^{\prime })\,L_{\mathbf{z}%
_{\alpha (i)}}(\breve{x}_{i}^{\prime })}{\kappa _{k}(z_{\alpha (i)})}\right)
.  \notag
\end{eqnarray}

\subsection{Derivation of Eq. (\protect\ref{eq-Parallel-1}) \label%
{A-Der-AA-Parallel-1}}

First note that if \ $G_{k|k-1}[h]=e^{D_{k|k-1}[h-1]}$ \ then%
\begin{equation}
\frac{\delta G_{k|k-1}}{\delta X}[h]=e^{D_{k|k-1}[h-1]}D_{k|k-1}^{X}
\end{equation}%
and so%
\begin{equation}
\frac{\delta G_{k|k-1}}{\delta X}%
[hp_{D}^{c}]=e^{D_{k|k-1}[hp_{D}^{c}-1]}D_{k|k-1}^{X}.
\end{equation}%
From Eq. (\ref{eq-Sensors-U}), 
\begin{eqnarray}
G_{k|k}^{u}[h] &\propto &\int f_{k}^{\ast }(Z|X)\,p_{D}^{X}\,\frac{\delta
G_{k|k-1}}{\delta X}[hp_{D}^{c}]\delta X \\
&=&\int f_{k}^{\ast
}(Z_{k}|X)\,p_{D}^{X}\,e^{D_{k|k-1}[hp_{D}^{c}-1]}D_{k|k-1}^{X}\delta X \\
&=&e^{D_{k|k-1}[p_{D}^{c}h-1]}\int f_{k}^{\ast
}(Z_{k}|X)\,(p_{D}D_{k|k-1})^{X}\delta X
\end{eqnarray}%
and so \ $G_{k|k}^{u}[h]\propto e^{D_{k|k-1}[p_{D}^{c}h-1]}$. \ Thus, as
claimed,%
\begin{equation}
G_{k|k}^{u}[h]=\frac{e^{D_{k|k-1}[p_{D}^{c}h-1]}}{e^{D_{k|k-1}[p_{D}^{c}-1]}}%
=e^{D_{k|k-1}[p_{D}^{c}h-p_{D}^{c}]}=e^{D_{k|k-1}[p_{D}^{c}(h-1)]}.
\end{equation}

\subsection{Derivation of Eq. (\protect\ref{eq-Parallel-2}) \label%
{A-Der-AA-Parallel-2}}

First note that if $G_{k|k-1}[h]=q^{c}+q\,s[h]$ then \ 
\begin{equation}
\frac{\delta G_{k|k-1}}{\delta X}[hp_{D}^{c}]=\left\{ 
\begin{array}{ccc}
G_{k|k-1}[hp_{D}^{c}]\, & \text{if} & X=\emptyset \\ 
q\,s(x) & \text{if} & X=\{x\} \\ 
0 & \text{if} & |X|\geq 2%
\end{array}%
\right. .  \label{eq-BernG}
\end{equation}%
Thus from Eqs. (\ref{eq-Sensors-Tot},\ref{eq-SetInt})

\begin{eqnarray}
&&G_{k|k}^{u}[h] \\
&\propto &\int f_{k}^{\ast }(Z_{k}|X)\,(hp_{D})^{X}\,\frac{\delta G_{k|k-1}}{%
\delta X}[hp_{D}^{c}]\delta X  \notag
\end{eqnarray}%
\begin{eqnarray}
&=&f_{k}^{\ast }(Z_{k}|\emptyset )\,G_{k|k-1}[hp_{D}^{c}]\,+\int f_{k}^{\ast
}(Z_{k}|\{x\})\,h(x)\,p_{D}(x)\,qs(x)dx \\
&=&\kappa _{k}(Z_{k})\,(q^{c}+qs[p_{D}^{c}h])+q\int f_{k}^{\ast
}(Z_{k}|\{x\})\,h(x)\,p_{D}(x)\,s(x)dx
\end{eqnarray}%
and then from Eq. (\ref{eq-fStar}) with \ $X=\{x\}$,%
\begin{eqnarray}
&\propto &q^{c}+qs[p_{D}^{c}h]+q\int \frac{f_{k}^{\ast }(Z_{k}|\{x\})}{%
\kappa _{k}(Z_{k})}\,h(x)\,p_{D}(x)\,s(x)dx \\
&=&q^{c}+p_{D}^{c}qs[h]+q\int \sum_{z\in Z_{k}}\frac{L_{z}(x)\,\kappa
_{k}(Z_{k}-\{z\})}{\kappa _{k}(Z_{k})}\,h(x)\,p_{D}(x)\,s(x)dx.
\end{eqnarray}%
Recombining terms and applying Eqs. (\ref{eq-Bern-0},\ref{eq-Bern0a}) we
get, as claimed, 
\begin{equation}
G_{k|k}^{u}[h]\propto q^{c}+\sum_{z\in Z_{k}}\frac{qs[p_{D}L_{z}h]\,\kappa
_{k}(Z_{k}-\{z\})}{\kappa _{k}(Z_{k})}+qs[p_{D}^{c}h]
\end{equation}%
\begin{eqnarray}
&=&q^{c}+q\,s\left[ h\left( p_{D}^{c}+p_{D}\sum_{z\in Z_{k}}\frac{%
L_{z}\,\kappa _{k}(Z_{k}-\{z\})}{\kappa _{k}(Z_{k})}\right) \right] \\
&=&q^{c}+q\,s\left[ \hat{L}_{Z_{k}}h\right] .
\end{eqnarray}

\subsection{Derivation of Eq. (\protect\ref{eq-Parallel-3}) \label%
{A-Der-AA-Parallel-3}}

The derivation is almost identical to that in Section \ref%
{A-Der-AA-Parallel-2}, except that Eq. (\ref{eq-Sensors-U}) is used rather
than Eq. (\ref{eq-Sensors-Tot}). \ From Eqs. (\ref{eq-BernG},\ref%
{eq-Sensors-U}),

\begin{equation}
G_{k|k}^{u}[h]\propto \int f_{k}^{\ast }(Z_{k}|X)\,p_{D}^{X}\,\frac{\delta
G_{k|k-1}}{\delta X}[hp_{D}^{c}]\delta X
\end{equation}%
\begin{eqnarray}
&=&f_{k}^{\ast }(Z_{k}|\emptyset )\,G_{k|k-1}[hp_{D}^{c}]\,+\int f_{k}^{\ast
}(Z_{k}|\{x\})\,p_{D}(x)\,\frac{\delta G_{k|k-1}}{\delta x}[hp_{D}^{c}]dx \\
&=&\kappa _{k}(Z_{k})\,(q^{c}+qs[p_{D}^{c}h])+\int f_{k}^{\ast
}(Z_{k}|\{x\})\,p_{D}(x)\,qs(x)dx
\end{eqnarray}%
and from Eq. (\ref{eq-fStar}) with \ $X=\{x\}$,%
\begin{eqnarray}
&\propto &q^{c}+qs[p_{D}^{c}h]+\int \frac{f_{k}^{\ast }(Z_{k}|\{x\})}{\kappa
_{k}(Z_{k})}\,p_{D}(x)\,qs(x)dx \\
&=&q^{c}+p_{D}^{c}qs[h]+\int \sum_{z\in Z_{k}}\frac{L_{z}(x)\,\kappa
_{k}(Z_{k}-\{z\})}{\kappa _{k}(Z_{k})}\,p_{D}(x)\,qs(x)dx \\
&=&q^{c}+qs[p_{D}^{c}h]+\sum_{z\in Z_{k}}\frac{qs[p_{D}L_{z}]\,\kappa
_{k}(Z_{k}-\{z\})}{\kappa _{k}(Z_{k})}
\end{eqnarray}%
\begin{equation}
=q^{c}+qs\left[ p_{D}^{c}h+p_{D}\sum_{z\in Z_{k}}\frac{L_{z}\kappa
_{k}(Z_{k}-\{z\})}{\kappa _{k}(Z_{k})}\right] \,
\end{equation}%
\begin{eqnarray}
&=&q^{c}+qs\left[ p_{D}^{c}h-p_{D}^{c}+p_{D}^{c}+p_{D}\sum_{z\in Z_{k}}\frac{%
L_{z}\kappa _{k}(Z_{k}-\{z\})}{\kappa _{k}(Z_{k})}\right] \\
&=&q^{c}+qs\left[ p_{D}^{c}h-p_{D}^{c}+\hat{L}_{Z}\right] =q^{c}+qs\left[ 
\hat{L}_{Z_{k}}+p_{D}^{c}(h-1)\right] .
\end{eqnarray}

\subsection{Equivalence of Eqs. (\protect\ref{eq-PHD-SUD0},\protect\ref%
{eq-PHD-SingleStep}) \label{A-Der-AA-Equiv}}

Abbreviate \ $B(x)=B_{k|k-1}(x)$. \ Assume that \ $%
G_{k-1|k-1}[h]=e^{D_{k-1|k-1}]}$ \ and abbreviate \ $%
G_{k-1|k-1}h]=G_{k-1}[h] $, \ $D_{k-1|k-1}(x)=D_{k-1}(x)$.\ \ Then from Eq. (%
\ref{eq-PHD-F1}),

\begin{eqnarray}
F_{k}[g,h] &=&e^{\kappa \lbrack
g-1]}e^{B[h(1+p_{D}L_{g-1})-1]}%
\,G_{k-1}[p_{S}^{c}+p_{S}M_{h(1+p_{D}L_{g-1})}\,] \\
&=&e^{\kappa \lbrack
g-1]+B[h(1+p_{D}L_{g-1})-1]+D_{k-1}[p_{S}^{c}+p_{S}M_{h(1+p_{D}L_{g-1})}]}
\end{eqnarray}%
and so%
\begin{eqnarray}
\frac{\delta F_{k}}{\delta x}[g,h] &=&e^{\kappa \lbrack
g-1]+B[h(1+p_{D}L_{g-1})-1]+D_{k-1}[1-p_{S}+p_{S}M_{h(1+p_{D}L_{g-1})}]} \\
&&\cdot \,\left( B[\delta _{x}(1+p_{D}L_{g-1})]+D_{k-1}[p_{S}M_{\delta
_{x}(1+p_{D}L_{g-1})}]\right)  \notag
\end{eqnarray}%
and%
\begin{equation}
\frac{\delta F_{k}}{\delta x}[g,1]=e^{Q[g]}\,P_{x}[g]\,
\end{equation}%
where%
\begin{eqnarray}
Q[g] &=&\kappa \lbrack g-1]+B[p_{D}L_{g-1}]+D_{k-1}[1+p_{S}M_{p_{D}L_{g-1}}]
\\
P_{x}[g] &=&B[\delta _{x}(1+p_{D}L_{g-1})]+D_{k-1}[p_{S}M_{\delta
_{x}(1+p_{D}L_{g-1})}] \\
P_{x}[0] &=&B[\delta _{x}p_{D}^{c}]+D_{k-1}[p_{S}M_{\delta _{x}p_{D}^{c}}] \\
&=&p_{D}^{c}(x)\,\left( B(x)\,+D_{k-1}[p_{S}M_{x}]\right) .
\end{eqnarray}%
From Eq. (\ \ref{eq-PHD-SUD0}) and the general product rule for functional
derivatives, \cite[Eq. (11.274)]{Mah-Artech},%
\begin{eqnarray}
\frac{\delta F_{k}^{u}}{\delta x\delta Z_{k}}[0,1] &=&\left[
\sum_{W\subseteq Z_{k}}\frac{\delta }{\delta (Z-W)}e^{Q[g]}\,\frac{\delta
P_{x}}{\delta W}[g]\,\right] _{g=0} \\
&=&e^{Q[0]}\chi ^{Z_{k}}\sum_{W\subseteq Z_{k}}\,\frac{1}{\chi ^{W}}\frac{%
\delta P_{x}}{\delta W}[0]  \label{eq-QQ}
\end{eqnarray}%
where%
\begin{equation}
\chi (z)=\kappa (z)+B[p_{D}L_{z}]+D_{k-1}[p_{S}M_{p_{D}L_{z}}].
\end{equation}%
Note that%
\begin{equation}
\frac{\delta P_{x}}{\delta W}[g]=\left\{ 
\begin{array}{ccc}
P_{x}[g] & \text{if} & W=\emptyset \\ 
\frac{\delta P_{x}}{\delta z}[g] & \text{if} & W=\{z\} \\ 
0 & \text{if} & |W|>1%
\end{array}%
\right.
\end{equation}%
where%
\begin{eqnarray}
\frac{\delta P_{x}}{\delta z}[g] &=&B[\delta
_{x}p_{D}L_{z}]+D_{k-1}[p_{S}M_{\delta _{x}p_{D}L_{z}}] \\
&=&p_{D}(x)\,L_{z}(x)\,\left( B(x)+D_{k-1}[p_{S}M_{x}]\right) .
\end{eqnarray}%
Eq. (\ref{eq-QQ}) thus becomes%
\begin{eqnarray}
&&\frac{\delta F_{k}^{u}}{\delta x\delta Z_{k}}[0,1] \\
&=&e^{Q[0]}\chi ^{Z_{k}}\left( P_{x}[0]+\sum_{z\in Z_{k}}\,\frac{1}{\chi (z)}%
\frac{\delta P_{x}}{\delta z}[0]\right)  \notag
\end{eqnarray}%
\begin{equation}
=e^{Q[0]}\chi ^{Z_{k}}\left( 
\begin{array}{c}
p_{D}^{c}(x)\,\left( B(x)\,+D_{k-1}[p_{S}M_{x}]\right) \\ 
+\sum_{z\in Z_{k}}\,\frac{p_{D}(x)\,L_{z}(x)\,\left(
B(x)+D_{k-1}[p_{S}M_{x}]\right) }{\kappa
(z)+B[p_{S}M_{p_{D}L_{z})}]+D_{k-1}[p_{S}M_{p_{D}L_{z}}]}%
\end{array}%
\right) .
\end{equation}%
Since%
\begin{equation}
\frac{\delta F_{k}^{u}}{\delta Z_{k}}[0,1]=e^{Q[0]}\chi ^{Z_{k}},
\end{equation}%
from Eq. (\ref{eq-PHD-SUD0}) we end up with Eq. (\ref{eq-PHD-SingleStep}),
as claimed.

\subsection{Derivation of Eqs. (\protect\ref{eq-PHD-DUD1}-\protect\ref%
{eq-PHD-DUD5}) \label{A-Der-AA-DUD}}

We employ the basic argument in Section \ref{A-Der-AA-Equiv}, but which here
is more algebraically involved. \ Six lemmas must be established first, and
require the D-U/D integral, (Eq. (\ref{eq-DUD-integral}), and the identities
Eqs. (\ref{eq-Trans-3},\ref{eq-Test-3},\ref{eq-Diff-1}-\ref{eq-Diff-4}): 
\begin{eqnarray}
\frac{\delta }{\delta \breve{x}}\breve{h} &=&\delta _{\breve{x}}\text{, \ }\ 
\frac{\delta }{\delta \breve{x}}\breve{h}|_{1}=\delta _{\breve{x}}^{d}\text{%
, \ }\breve{M}_{(x^{\prime \prime },o^{\prime \prime })}(x^{\prime
},o^{\prime })=\delta _{o^{\prime \prime },o^{\prime }}\,M_{x^{\prime \prime
}}(x^{\prime })\text{, \ }\,\text{ } \\
\delta _{(x,o)}(x^{\prime },o^{\prime }) &=&\delta _{o,o^{\prime }}\,\delta
_{x_{1}}(x^{\prime })\text{, \ \ \ \ }\delta _{(x,o)}^{d}(x^{\prime \prime
},o^{\prime \prime })=\delta _{o,1}\,\delta _{x}(x^{\prime \prime })\text{.\ 
}
\end{eqnarray}

\textit{Lemma 1,} $\ \breve{D}_{k-1}[p_{S}\breve{M}_{p_{D}L_{z}}]=\,\breve{D}%
_{k-1}^{d/u}[p_{S}M_{p_{D}L_{z}}]$:

\begin{equation}
\breve{D}_{k-1}[p_{S}\breve{M}_{p_{D}L_{z}}]=\sum_{o^{\prime }\in
\{0,1\}}\int p_{S}(x^{\prime })\,\breve{M}_{p_{D}L_{z}}(x^{\prime
},o^{\prime })\,\breve{D}_{k-1}(x^{\prime },o^{\prime })dx^{\prime }
\end{equation}%
\begin{eqnarray}
&=&\sum_{o^{\prime }\in \{0,1\}}\int p_{S}(x^{\prime })\,\left(
\sum_{o^{\prime \prime }\in \{0,1\}}\int p_{D}(x^{\prime \prime
})\,L_{z}(x^{\prime \prime })\,\delta _{o^{\prime \prime },o^{\prime
}}\,M_{x^{\prime \prime }}(x^{\prime })dx^{\prime \prime }\right) \\
&&\cdot \,\breve{D}_{k-1}(x^{\prime },o^{\prime })dx^{\prime }  \notag
\end{eqnarray}%
\begin{equation}
=\sum_{o^{\prime },o^{\prime \prime }\in \{0,1\}}\int p_{S}(x^{\prime
})\,p_{D}(x^{\prime \prime })\,L_{z}(x^{\prime \prime })\,\delta _{o^{\prime
\prime },o^{\prime }}\,M_{x^{\prime \prime }}(x^{\prime })\,\breve{D}%
_{k-1}(x^{\prime },o^{\prime })dx^{\prime }dx^{\prime \prime }
\end{equation}%
\begin{eqnarray}
&=&\sum_{o^{\prime \prime }\in \{0,1\}}\int p_{S}(x^{\prime
})\,p_{D}(x^{\prime \prime })\,L_{z}(x^{\prime \prime })\,M_{x^{\prime
\prime }}(x^{\prime })\,\breve{D}_{k-1}(x^{\prime },o^{\prime \prime
})dx^{\prime }dx^{\prime \prime } \\
&=&\int p_{S}(x^{\prime })\,p_{D}(x^{\prime \prime })\,L_{z}(x^{\prime
\prime })\,M_{x^{\prime \prime }}(x^{\prime })\sum_{o^{\prime \prime }\in
\{0,1\}}\breve{D}_{k-1}(x^{\prime },o^{\prime \prime })dx^{\prime
}dx^{\prime \prime }
\end{eqnarray}%
\begin{eqnarray}
&=&\int \,p_{D}(x^{\prime \prime })\,L_{z}(x^{\prime \prime
})\,p_{S}(x^{\prime })\,M_{x^{\prime \prime }}(x^{\prime })\,\breve{D}%
_{k-1}^{d/u}(x^{\prime })dx^{\prime }dx^{\prime \prime } \\
&=&\int \,p_{D}(x^{\prime \prime })\,L_{z}(x^{\prime \prime })\,\breve{D}%
_{k-1}^{d/u}[p_{S}M_{x^{\prime \prime }}]dx^{\prime \prime } \\
&=&\,\breve{D}_{k-1}^{d/u}[p_{S}M_{p_{D}L_{z}}].
\end{eqnarray}

\textit{Lemma 2}, $\breve{D}_{k-1}[p_{S}\breve{M}_{\delta
_{(x.o)}^{d}p_{D}L_{z}}]=\delta _{o,1}\,p_{D}(x)\,L_{z}(x)\,\breve{D}%
_{k-1}^{d/u}[p_{S}M_{x}]$:

\begin{equation}
\breve{D}_{k-1}[p_{S}\breve{M}_{\delta
_{(x.o)}^{d}p_{D}L_{z}}]=\sum_{o^{\prime }\in \{0,1\}}\int p_{S}(x^{\prime
})\,\breve{M}_{\delta _{(x,o)}^{d}p_{D}L_{z}}(x^{\prime },o^{\prime })\,%
\breve{D}_{k-1}(x^{\prime },o^{\prime })dx^{\prime }
\end{equation}%
\begin{eqnarray}
&=&\sum_{o^{\prime }\in \{0,1\}}\int p_{S}(x^{\prime })\, \\
&&\cdot \left( \sum_{o^{\prime \prime }\in \{0,1\}}\int \delta
_{(x,o)}^{d}(x^{\prime \prime },o^{\prime \prime })\,\breve{p}_{D}(x^{\prime
\prime })\,L_{z}(x^{\prime \prime })\,\breve{M}_{(x^{\prime \prime
},o^{\prime \prime })}(x^{\prime },o^{\prime })dx^{\prime \prime }\right) 
\notag \\
&&\cdot \,\breve{D}_{k-1}(x^{\prime },o^{\prime })dx^{\prime }  \notag
\end{eqnarray}%
\begin{eqnarray}
&=&\sum_{o^{\prime },o^{\prime \prime }\in \{0,1\}}\int p_{S}(x^{\prime })\,%
\breve{p}_{D}(x^{\prime \prime })\,L_{z}(x^{\prime \prime })\,\breve{M}%
_{(x^{\prime \prime },o^{\prime \prime })}(x^{\prime },o^{\prime
})\,\,\delta _{(x,o)}^{d}(x^{\prime \prime },o^{\prime \prime })\, \\
&&\cdot \breve{D}_{k-1}(x^{\prime },o^{\prime })dx^{\prime \prime
}dx^{\prime }  \notag
\end{eqnarray}%
\begin{eqnarray}
&=&\sum_{o^{\prime },o^{\prime \prime }\in \{0,1\}}\int p_{S}(x^{\prime })\,%
\breve{p}_{D}(x^{\prime \prime })\,L_{z}(x^{\prime \prime })\,\delta
_{o^{\prime \prime },o^{\prime }}\breve{M}_{x^{\prime \prime }}(x^{\prime
})\,\,\delta _{o.1}\delta _{x}(x^{\prime \prime })\, \\
&&\cdot \breve{D}_{k-1}(x^{\prime },o^{\prime })dx^{\prime \prime
}dx^{\prime }  \notag
\end{eqnarray}%
\begin{eqnarray}
&=&\delta _{o,1}\,p_{D}(x)\,L_{z}(x)\sum_{o^{\prime \prime }\in \{0,1\}}\int
p_{S}(x^{\prime })\,M_{x}(x^{\prime })\,\,\breve{D}_{k-1}(x^{\prime
},o^{\prime \prime })dx^{\prime } \\
&=&\delta _{o,1}\,p_{D}(x)\,L_{z}(x)\int p_{S}(x^{\prime })\,M_{x}(x^{\prime
})\,\,\sum_{o^{\prime \prime }\in \{0,1\}}\breve{D}_{k-1}(x^{\prime
},o^{\prime \prime })dx^{\prime }
\end{eqnarray}%
\begin{eqnarray}
&=&\delta _{o,1}\,\breve{p}_{D}(\breve{x})\,L_{z}(\breve{x})\int
p_{S}(x^{\prime })\,M_{x}(x^{\prime })\,\,\breve{D}_{k-1}^{d/u}(\breve{x}%
^{\prime })dx^{\prime } \\
&=&\delta _{o,1}\,p_{D}(x)\,L_{z}(x)\,\breve{D}_{k-1}^{d/u}[p_{S}M_{x}].
\end{eqnarray}

\textit{Lemma 3}, $\breve{B}[\delta _{(x,o)}^{d}p_{D}L_{z}]=\delta
_{o,1}\,p_{D}(x)\,L_{z}(x)\,\breve{B}^{d/u}(x)$:

\begin{equation}
\breve{B}[\delta _{(x.o)}^{d}p_{D}L_{z}]=\sum_{o^{\prime }\in \{0,1\}}\int
\delta _{o,1}\delta _{x}(x^{\prime })\,p_{D}(x^{\prime })\,L_{z}(x^{\prime
})\,\breve{B}(x^{\prime },o^{\prime })dx^{\prime }
\end{equation}%
\begin{equation}
=\delta _{o,1}\sum_{o^{\prime }\in \{0,1\}}p_{D}(x)\,L_{z}(x)\,\breve{B}%
(x,o^{\prime })=\delta _{o,1}\,p_{D}(x)\,L_{z}(x)\,\breve{B}^{d/u}(x)
\end{equation}%
\begin{equation}
=\int p_{S}(x^{\prime })\,M_{x}(x^{\prime })\,\breve{D}_{k-1}|_{o}(x^{\prime
})dx^{\prime }=\breve{D}_{k-1}|_{o}[p_{S}M_{x}].
\end{equation}

\textit{Lemma 4,}\ $\breve{B}[p_{D}L_{z}]=\breve{B}^{d/u}[p_{D}L_{z}]$:

\begin{equation}
\breve{B}[p_{D}L_{z}]=\sum_{o^{\prime }\in \{0,1\}}\int p_{D}(x^{\prime
})\bigskip \,L_{z}(x^{\prime })\,\breve{B}(x^{\prime },o^{\prime
})dx^{\prime }
\end{equation}%
\begin{equation}
=\int p_{D}(x^{\prime })\,L_{z}(x^{\prime })\,\breve{B}(x^{\prime
},0)dx^{\prime }+\int p_{D}(x^{\prime })\,L_{z}(x^{\prime })\,\breve{B}%
(x^{\prime },1)dx^{\prime }
\end{equation}%
\begin{eqnarray}
&=&\int p_{D}(x^{\prime })\,L_{z}(x^{\prime })\,\breve{B}|_{0}(x^{\prime
})dx^{\prime }+\int p_{D}(x^{\prime })\,L_{z}(x^{\prime })\,\breve{B}%
|_{1}(x^{\prime })dx^{\prime } \\
&=&\breve{B}|_{0}[p_{D}L_{z}]+\breve{B}|_{1}[p_{D}L_{z}]=\breve{B}%
^{d/u}[p_{D}L_{z}].
\end{eqnarray}

\textit{Lemma 5, }$\breve{D}_{k-1}[p_{S}\breve{M}_{\delta
_{(x,o)}p_{D}^{c}}]=p_{D}^{c}(x)\,\breve{D}_{k-1}|_{o}[p_{S}M_{x}]$:

\begin{equation}
\breve{D}_{k-1}[p_{S}\breve{M}_{\delta _{(x,o)}p_{D}^{c}}]=\sum_{o^{\prime
}\in \{0,1\}}p_{S}(x^{\prime })\,\breve{M}_{\delta
_{(x,o)}p_{D}^{c}}(x^{\prime })\,\breve{D}_{k-1}(x^{\prime },o^{\prime
})dx^{\prime }
\end{equation}%
\begin{eqnarray}
&=&\sum_{o^{\prime }\in \{0,1\}}p_{S}(x^{\prime })\, \\
&&\cdot \left( \sum_{o^{\prime \prime }\in \{0,1\}}\int \delta _{o,o^{\prime
\prime }}\delta _{x}(x^{\prime \prime })\,p_{D}^{c}(x^{\prime \prime
})\,\delta _{o^{\prime \prime },o^{\prime }}\,M_{x^{\prime \prime
}}(x^{\prime })dx^{\prime \prime }\right)  \notag \\
&&\cdot \,\breve{D}_{k-1}(x^{\prime },o^{\prime })dx^{\prime }  \notag
\end{eqnarray}%
\begin{eqnarray}
&=&\sum_{o^{\prime },o^{\prime \prime }\in \{0,1\}}\int p_{S}(x^{\prime
})\,\delta _{o,o^{\prime \prime }}\delta _{o^{\prime \prime },o^{\prime
}}\,p_{D}^{c}(x)\,M_{x}(x^{\prime })\,\breve{D}_{k-1}(x^{\prime },o^{\prime
})dx^{\prime } \\
&=&p_{D}^{c}(x)\,\sum_{o^{\prime }\in \{0,1\}}\int p_{S}(x^{\prime
})\,\delta _{o,o^{\prime }}M_{x}(x^{\prime })\,\breve{D}_{k-1}(x^{\prime
},o^{\prime })dx^{\prime }
\end{eqnarray}%
\begin{eqnarray}
&=&p_{D}^{c}(x)\,\int p_{S}(x^{\prime })\,M_{x}(x^{\prime })\,\breve{D}%
_{k-1}(x^{\prime },o)dx^{\prime } \\
&=&p_{D}^{c}(x)\,\int p_{S}(x^{\prime })\,M_{x}(x^{\prime })\,\breve{D}%
_{k-1}|_{o}(x^{\prime })dx^{\prime } \\
&=&p_{D}^{c}(x)\,\breve{D}_{k-1}|_{o}[p_{S}M_{x}].
\end{eqnarray}

\textit{Lemma 6, }$\breve{B}[\delta
_{(x.o)}p_{D}^{c}]=p_{D}^{c}(x)\,B|_{o}(x)$:

\begin{eqnarray}
\breve{B}[\delta _{(x.o)}p_{D}^{c}] &=&\sum_{o^{\prime }\in \{0,1\}}\int
\delta _{(x.o)}(x^{\prime },o^{\prime })\,p_{D}^{c}(x^{\prime },o^{\prime
})\,B(x^{\prime },o^{\prime })dx^{\prime } \\
&=&\sum_{o^{\prime }\in \{0,1\}}\int \delta _{o,o^{\prime }}\delta
_{x}(x^{\prime })\,p_{D}^{c}(x^{\prime })\,B(x^{\prime },o^{\prime
})dx^{\prime }
\end{eqnarray}%
\begin{eqnarray}
&=&\sum_{o^{\prime }\in \{0,1\}}\delta _{o,o^{\prime
}}\,p_{D}^{c}(x)\,B(x,o^{\prime }) \\
&=&p_{D}^{c}(x)\,B(x,o)=p_{D}^{c}(x)\,B|_{o}(x).
\end{eqnarray}

\textit{Main Derivation}: 
\begin{eqnarray}
&&\breve{F}_{k}[g,\breve{h}] \\
&=&e^{\kappa \lbrack g-1]}\,e^{\breve{B}[\breve{h}p_{D}^{c}+\breve{h}%
|_{1}p_{D}L_{g}-1]}\,\breve{G}_{k-1|k-1}[1-p_{S}+p_{S}\breve{M}_{\breve{h}%
p_{D}^{c}+\breve{h}|_{1}p_{D}L_{g}}] \\
&=&e^{\kappa \lbrack g-1]+\breve{B}[\breve{h}p_{D}^{c}+\breve{h}%
|_{1}p_{D}L_{g}-1]+\breve{D}_{k-1}[1-p_{S}+p_{S}\breve{M}_{\breve{h}%
p_{D}^{c}+\breve{h}|_{1}p_{D}L_{g}}]}.
\end{eqnarray}%
From Eq. (\ref{eq-PHD-SUD0}),%
\begin{equation}
\breve{D}_{k}(\breve{x})=\frac{\delta \breve{G}_{k|k}}{\delta \breve{x}}[1]=%
\frac{\frac{\delta \breve{F}_{k}}{\delta \breve{x}\delta Z_{k}}[0,1]}{\frac{%
\delta \breve{F}_{k}}{\delta Z_{k}}[0,1]}  \label{eq-MMMMM}
\end{equation}%
where%
\begin{eqnarray}
\frac{\delta \breve{F}_{k}}{\delta \breve{x}}[g,1] &=&e^{\kappa \lbrack g-1]+%
\breve{B}[p_{D}L_{g-1}]+\breve{D}_{k-1}[1+p_{S}\breve{M}_{p_{D}L_{g-1}}]} \\
&&\cdot \left( \breve{B}[\delta _{\breve{x}}p_{D}^{c}+\delta _{\breve{x}%
}^{d}p_{D}L_{g}]+\breve{D}_{k-1}[p_{S}\breve{M}_{\delta _{\breve{x}%
}p_{D}^{c}+\delta _{\breve{x}}^{d}p_{D}L_{g}}]\right)  \notag \\
&=&e^{Q[g]}\,P_{\breve{x}}[g]\text{ }
\end{eqnarray}%
and where%
\begin{eqnarray}
Q[g]\, &=&\kappa \lbrack g-1]+\breve{B}[p_{D}L_{g-1}]+\breve{D}_{k-1}[1+p_{S}%
\breve{M}_{p_{D}L_{g-1}}] \\
P_{\breve{x}}[g] &=&\breve{B}[\delta _{\breve{x}}p_{D}^{c}+\delta _{\breve{x}%
}^{d}p_{D}L_{g}]+\breve{D}_{k-1}[p_{S}\breve{M}_{\delta _{\breve{x}%
}p_{D}^{c}+\delta _{\breve{x}}^{d}p_{D}L_{g}}].
\end{eqnarray}%
Thus by the general product rule for functional derivatives, \cite[Eq.
(11.274)]{Mah-Artech},%
\begin{equation}
\frac{\delta \breve{F}_{k}}{\delta Z\delta \breve{x}}[g,1]=\sum_{W\subseteq
Z}\frac{\delta }{\delta (Z-W)}e^{Q_{\breve{x}}[g]}\,\frac{\delta P_{\breve{x}%
}}{\delta W}[g].
\end{equation}%
On the one hand%
\begin{eqnarray}
\frac{\delta }{\delta Y}e^{Q_{\breve{x}}[g]} &=&\chi ^{Y}e^{Q_{\breve{x}}[g]}%
\text{ \ \ where} \\
\chi (z) &=&\kappa (z)+\breve{B}[p_{D}L_{z}]+\breve{D}_{k-1}[p_{S}\breve{M}%
_{p_{D}L_{z}}].
\end{eqnarray}%
On the other hand,%
\begin{equation}
\frac{\delta P_{\breve{x}}}{\delta W}[g]=\left\{ 
\begin{array}{ccc}
P_{\breve{x}}[g] & \text{if} & W=\emptyset \\ 
\breve{D}_{k-1}[p_{S}\breve{M}_{\delta _{\breve{x}}^{d}p_{D}L_{z}}] & \text{%
if} & W=\{z\} \\ 
0 & \text{if} & |W|>1%
\end{array}%
\right. .
\end{equation}%
from which follows%
\begin{eqnarray}
\frac{\delta \breve{F}_{k}}{\delta Z\delta \breve{x}}[g,1]
&=&\sum_{W\subseteq Z}\chi ^{Z-W}e^{Q_{\breve{x}}[g]}\frac{\delta P_{\breve{x%
}}}{\delta W}[g]=e^{Q_{\breve{x}}[g]}\sum_{W\subseteq Z}\chi ^{Z-W}\frac{%
\delta P_{\breve{x}}}{\delta W}[g] \\
&=&\chi ^{Z}e^{Q_{\breve{x}}[g]}\left( P_{\breve{x}}[g]+\sum_{z\in Z}\frac{1%
}{\chi (z)}\frac{\delta P_{\breve{x}}}{\delta z}[g]\right)
\end{eqnarray}%
and so%
\begin{equation}
\frac{\delta \breve{F}_{k}}{\delta Z\delta \breve{x}}[0,1]=\chi ^{Z}e^{Q_{%
\breve{x}}[0]}\left( P_{\breve{x}}[0]+\sum_{z\in Z}\frac{1}{\chi (z)}\frac{%
\delta P_{\breve{x}}}{\delta z}[0]\right)
\end{equation}%
\begin{eqnarray}
&=&\chi ^{Z}e^{Q_{\breve{x}}[g]} \\
&&\cdot \left( \breve{B}[\delta _{\breve{x}}p_{D}^{c}]+\breve{D}_{k-1}[p_{S}%
\breve{M}_{\delta _{\breve{x}}p_{D}^{c}}]+\sum_{z\in Z}\frac{\breve{B}%
[\delta _{\breve{x}}^{d}p_{D}L_{z}]+\breve{D}_{k-1}[p_{S}\breve{M}_{\delta _{%
\breve{x}}^{d}p_{D}L_{z}}]}{\kappa (z)+\breve{B}[p_{D}L_{z}]+\breve{D}%
_{k-1}[p_{S}\breve{M}_{p_{D}L_{z}}]}\right) .  \notag
\end{eqnarray}%
Now note that%
\begin{equation}
\breve{F}_{k}[g,1]=e^{\kappa \lbrack g-1]+\breve{D}_{k-1}[p_{S}\breve{M}%
_{p_{D}L_{g-1}}]}=e^{Q[g]}
\end{equation}%
and thus%
\begin{equation}
\frac{\delta \breve{F}_{k}}{\delta Z}[g,1]=\chi ^{Z}e^{Q[g]}
\end{equation}%
and thus by Eq. (\ref{eq-MMMMM}), 
\begin{eqnarray}
\breve{D}_{k}(\breve{x}) &=&\breve{B}[\delta _{\breve{x}}p_{D}^{c}]+\breve{D}%
_{k-1}[p_{S}\breve{M}_{\delta _{\breve{x}}p_{D}^{c}}] \\
&&+\sum_{z\in Z_{k}}\frac{\breve{B}[\delta _{\breve{x}}^{d}p_{D}L_{z}]+%
\breve{D}_{k-1}[p_{S}\breve{M}_{\delta _{\breve{x}}^{d}p_{D}L_{z}}]}{\kappa
(z)+\breve{B}[p_{D}L_{z}]+\breve{D}_{k-1}[p_{S}\breve{M}_{p_{D}L_{z}}]}. 
\notag
\end{eqnarray}%
Applying Lemmas 1-6 we get \ \ \ \ \ \ \ 
\begin{eqnarray}
\breve{D}_{k}(x,o) &=&p_{D}^{c}(x)\,\left( B|_{o}(x)+\breve{D}%
_{k-1}|_{o}[p_{S}M_{x}]\right) \\
&&+\sum_{z\in Z_{k}}\frac{\delta _{o,1}\,p_{D}(x)\,L_{z}(x)\,\left( \breve{B}%
^{d/u}(x)+\breve{D}_{k-1}^{d/u}[p_{S}M_{x}]\right) }{\kappa (z)+\breve{B}%
^{d/u}[p_{D}L_{z}]+\breve{D}_{k-1}^{d/u}[p_{S}M_{p_{D}L_{z}}]}.  \notag
\end{eqnarray}%
Setting \ $o=1$ \ and \ $o=0$ \ and using \ $\breve{B}(x,0)=B(x)$ \ and \ $%
\breve{B}(x,1)=0$, we get, as claimed,%
\begin{eqnarray}
\breve{D}_{k}^{d}(x) &=&p_{D}^{c}(x)\,\breve{D}_{k-1}^{d}[p_{S}M_{x}] \\
&&+\sum_{z\in Z_{k}}\frac{\,p_{D}(x)\,L_{z}(x)\,\left( B_{k|k-1}(x)+\breve{D}%
_{k-1}^{d/u}[p_{S}M_{x}]\right) }{\kappa (z)+\breve{B}_{k|k-1}[p_{D}L_{z}]+%
\breve{D}_{k-1}^{d/u}[p_{S}M_{p_{D}L_{z}}]}  \notag \\
\breve{D}_{k}^{u}(x) &=&p_{D}^{c}(x)\,\left( B_{k|k-1}(x)+\breve{D}%
_{k-1}^{u}[p_{S}M_{x}]\right) .
\end{eqnarray}

\section{Conclusions \label{A-Concl}}

This sequel paper to \cite{MahSensors1019Exact}:

\begin{enumerate}
\item showed that the Poisson multi-Bernoulli mixture (PMBM) approach to
detected targets (D-targets) and undetected targets (U-targets) is
theoretically erroneous;

\item but demonstrated that it can be partially salvaged using a novel
single-step \textquotedblleft D-U/D\textquotedblright\ recursive Bayes
filter that captures the dynamic behavior of U/D-targets;

\item compared this approach with the \textquotedblleft static
U/D\textquotedblright\ (S-U/D) formulation of U/D targets presented in \cite[%
Sec. 5]{MahSensors1019Exact};

\item derived probability hypothesis density filters for both the S-U/D and
D-U/D cases, demonstrating that the number and states of the U-targets and
D-targets can be easily estimated;

\item employed purely algebraic methods to verify the validity of key
formulas in \cite[Sec. 5]{MahSensors1019Exact} for the S-U/D approach; and

\item used these results to show that the PMBM assumption that D-targets and
U-targets can be propagated separately in parallel (\textquotedblleft
U/D-parallelism\textquotedblright ) appears to be erroneous.
\end{enumerate}

\end{document}